\documentclass[aps,prx,twocolumn,
			   groupedaddress,superscriptaddress,
			   amsfonts,amssymb,amsmath,
			   citeautoscript,longbibliography,
			   a4paper
			   ]{revtex4-2}

\usepackage[utf8]{inputenc}
\usepackage[english]{babel}

\usepackage{microtype} 
\usepackage{xspace} 

\usepackage{txfonts}  
\usepackage{txfontsb} 

\usepackage{bm} 

\usepackage{xcolor}
\usepackage[]{graphicx} 
\graphicspath{{figs/}}

\usepackage[]{booktabs}
\usepackage{array}
\usepackage{layouts}
\usepackage{multirow}

\usepackage{enumerate}
\usepackage[inline]{enumitem}

\usepackage{hyperref}
\hypersetup{colorlinks,
    linkcolor=black,
    citecolor=black,
    urlcolor=black
}

\usepackage{pdfpages}
\usepackage{pgffor}
\makeatletter
\AtBeginDocument{\let\LS@rot\@undefined}
\makeatother

\usepackage[capitalize,nameinlink]{cleveref}

\crefname{subequations}{Eqs.}{Eqs.} 
\Crefname{subequations}{Eqs.}{Eqs.}
\crefformat{subequations}{#2Eqs.~(#1)#3}
\Crefformat{subequations}{#2Eqs.~(#1)#3}
\crefname{page}{p.}{p.} 

\usepackage{placeins}

\usepackage{siunitx}
\DeclareSIUnit[number-unit-product = ]\percent{\char`\%} 

\usepackage[centering,hmargin=16mm,tmargin=30mm,bmargin=26mm]{geometry}

\usepackage{soul}

\usepackage{textcomp} 
\usepackage{xifthen}
\usepackage{etoolbox}
\newboolean{togglecomments}
\newboolean{togglechanges} 

\setboolean{togglecomments}{true}
\setboolean{togglechanges}{false}

\newcommand{\textblacksquare}{$\blacksquare$}
\newcommand{\todo}[1]{\ifbool{togglecomments}%
	{\textcolor{green!60!black}{\small\textsf{{}\textsuperscript{\textsc{\textsf{todo}}}}[#1]}} 
	{}}     
\newcommand{\comment}[2]{\ifbool{togglecomments}%
		{\textcolor{blue!70!black}{\small\sf\textsuperscript{\textsc{\textsf{#1}}}[#2]}} 
		{}}     
\newcommand{\swap}[2]{\ifbool{togglechanges}
	{#2}  
	{\textcolor{red!70!black}{[\ignorespaces#1]}\textrightarrow{}\textcolor{green!50!black}{[\ignorespaces#2]}}}
\newcommand{\remove}[1]{\ifbool{togglechanges}
	{}    
	{\textcolor{red!70!black}{\ignorespaces#1}}}
\newcommand{\inset}[1]{\ifbool{togglechanges}
	{#1}  
	{\textcolor{green!50!black}{#1}}}
\newcommand{\citeremind}[1]{%
	[\textcolor{blue!75!black!80!yellow}{\textblacksquare%
		\ifthenelse{\isempty{#1}}{}{\textsuperscript{\tiny\textsf{#1}}}%
	}]\xspace}

\newcommand{\captionbullet}[1]{\textbf{#1}}

\newcommand{\ie}{i.e.\@\xspace} 

\newcommand{\appropto}{\mathrel{\vcenter{
			\offinterlineskip\halign{\hfil$##$\cr
				\propto\cr\noalign{\kern.2pt}\sim\cr\noalign{\kern-2.5pt}}}}}

\newcommand{\suppsec}{Supplementary Section\xspace}

\makeatletter
\newcommand{\raisemath}[1]{\mathpalette{\raisem@th{#1}}}
\newcommand{\raisem@th}[3]{\raisebox{#1}{$#2#3$}}
\makeatother

\renewcommand{\paragraph}[1]{\vskip 1ex\noindent\textbf{#1.}~}

\usepackage[eulergreek]{sansmath}
\makeatletter
\renewcommand\@make@capt@title[2]{%
    \@ifx@empty\float@link{\@firstofone}{\expandafter\href\expandafter{\float@link}}%
    \sisetup{math-sf=\textsf}%
    \sansmath\sffamily\textbf{#1\@caption@fignum@sep}#2 
}%

\makeatother

\newcommand{\mitphysicsaffil}{\footnotesize Department of Physics, Massachusetts Institute of Technology, Cambridge, Massachusetts 02139, USA}
\newcommand{\miteecsaffil}{\footnotesize Department of Electrical Engineering and Computer Science, Massachusetts Institute of Technology, Cambridge, Massachusetts 02139, USA}
\newcommand{\hkustaffil}{\footnotesize Department of Physics, Hong Kong University of Science and Technology, Clear Water Bay, Hong Kong, China}
\newcommand{\fbaffil}{\footnotesize Facebook AI Research, New York, New York 10003, USA}

\begin{document}

\author{Andrew~Ma}
\thanks{A.M. and Y.Z. contributed equally to this work.}
\affiliation{\miteecsaffil}
\author{Yang~Zhang}
\thanks{A.M. and Y.Z. contributed equally to this work.}
\affiliation{\mitphysicsaffil}
\author{Thomas~Christensen}
\affiliation{\mitphysicsaffil}
\author{Hoi~Chun~Po}
\affiliation{\mitphysicsaffil}\affiliation{\hkustaffil}
\author{Li Jing}
\affiliation{\mitphysicsaffil}\affiliation{\fbaffil}
\author{Liang Fu}
\email{liangfu@mit.edu}
\affiliation{\mitphysicsaffil}
\author{Marin~Solja\v{c}i\'{c}}
\email{soljacic@mit.edu}
\affiliation{\mitphysicsaffil}

\title{Topogivity: A Machine-Learned Chemical Rule for Discovering Topological Materials}

\begin{abstract}
    Topological materials present unconventional electronic properties that make them attractive for both basic science and next-generation technological applications.
    The majority of currently known topological materials have been discovered using methods that involve symmetry-based analysis of the quantum wavefunction.
    Here we use machine learning to develop a simple-to-use heuristic chemical rule that diagnoses with a high accuracy whether a material is topological using only its chemical formula.
    This heuristic rule is based on a notion that we term \emph{topogivity}, a machine-learned numerical value for each element that loosely captures its tendency to form topological materials. 
    We next implement a high-throughput procedure for discovering topological materials based on the heuristic topogivity-rule prediction followed by ab initio validation.
    This way, we discover new topological materials that are not diagnosable using symmetry indicators, including several that may be promising for experimental observation.
\end{abstract}

\maketitle

Topological materials, including both topological insulators~\cite{kane2005quantum,bernevig2006quantum,konig2007quantum,fu2007topological3d,hsieh2008topological,fu2011topological,hsieh2012topological_revised,benalcazar2017quantized} and topological semimetals~\cite{liu2014discovery,xu2015discovery,lv2015experimental,burkov2011topological,bradlyn2016beyond}, are unconventional phases of matter characterized by topologically nontrivial electron wavefunctions.  
Since the beginning of the field, an important and enduring question has been how to determine whether a given electronic material is topological. Efforts to answer this question have largely relied on first-principles calculations in synergy with topological band theory~\cite{xiao2021first,bansil2016colloquium}.
In particular, recently developed theories known as symmetry indicators~\cite{po2017symmetry_revised} and topological quantum chemistry~\cite{bradlyn2017topological} allow the diagnosis of a wide range of topological materials using symmetry-based analysis of the wavefunction~\cite{fu2007topologicalinv, slager2017, song2018quantitative_revised}.
These symmetry-based methods require relatively low computational cost and have enabled high-throughput computational searches for topological materials~\cite{tang2019comprehensive,zhang2019catalogue,vergniory2019complete}.
Despite these successes, symmetry indicators have limited diagnostic power for certain forms of band topology or when applied to low-symmetry crystal structures~\cite{po2017symmetry_revised}.
For example, the first experimentally observed Weyl semimetal, tantalum arsenide (TaAs)~\cite{xu2015discovery,lv2015experimental}, is a non-symmetry-diagnosable topological material (i.e., a topological material whose topology is undetectable by symmetry indicators)~\cite{tang2019comprehensive}.
Its topological nature is established by calculating the wavefunction-based topological invariant directly~\cite{weng2015weyl}, which involves significant computational cost. 
From a broad conceptual standpoint, topological materials such as Chern insulators and time-reversal-invariant $Z_2$ topological insulators are robust against any small perturbation breaking all crystal symmetries, which renders symmetry indicators inapplicable even though topology remains intact.
Thus, for both practical and fundamental reasons, it is highly desirable to develop accurate and simple-to-use rules to determine whether {\it any} given material is topological.

\begin{figure*}[!htb]
    \centering
    \includegraphics[width=0.7\textwidth]{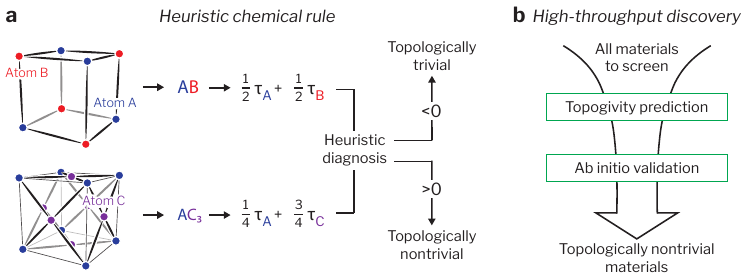}
    \caption{%
    \textbf{Topogivity-based diagnosis and discovery of topological materials.}
    \captionbullet{a}
    Given a stoichiometric material, the topogivity-based heuristic diagnosis is evaluated by simply weighting the material's elements' (A, B, and C) topogivities ($\tau_{\text{A}}$, $\tau_{\text{B}}$, and $\tau_{\text{C}}$) by their relative abundance in the chemical formula (AB and AC\textsubscript{3}).
    The sign indicates the topological classification and the magnitude indicates roughly how confident we are in this classification.
    Each element's topogivity is a machine-learned parameter that loosely captures the element's tendency to form topologically nontrivial materials.
    \captionbullet{b}
    We leverage our framework to perform high-throughput topological materials discovery. First, we use the topogivities to rapidly screen through a suitable collection of materials (i.e., the discovery space) in order to find candidate topological materials. Subsequently, we carry out ab initio validation by performing DFT on these candidates.
    We discovered topological materials that are not diagnosable using the standard symmetry indicators approach~\cite{po2017symmetry_revised}.
    \label{fig_model_and_discovery_strategy}
    }
\end{figure*}

Many aspects of materials can be understood at a heuristic level from a chemistry perspective.
A well-known example is bonding, which can be understood using quantum mechanical approaches such as molecular orbital theory~\cite{hehre1976ab}, as well as using heuristics such as the difference of element electronegativities.
While quantum theory can provide greater detail and accuracy, 
chemical heuristics can often provide valuable insight and a useful guide for materials discovery.
Is there a deep chemical reason why a particular material is topological?
To what extent can topological materials be understood and identified using chemical heuristic approaches? 
While connections between chemistry and electronic band topology (e.g., based on the presence or absence of certain elements) have been explored~\cite{kumar2020topological,schoop2018chemical,gibson2015three_revised,isaeva2020crystal,klemenz2020role,gui2019new}, existing chemical heuristics do not provide a broadly applicable path for finding topological materials.

Here, we use machine learning (ML) to help us search for chemical origins of topological electronic structure in materials.
Recently, ML has become a powerful approach for advancing scientific discovery in materials science~\cite{zhong2020accelerated,deringer2021origins,george2021chemist}.  In the area of topological materials, researchers have begun to apply ML to both toy models~\cite{zhang2017quantum,zhang2018machine,scheurer2020unsupervised,zhang2020interpreting} and ab initio data~\cite{acosta2018analysis_manual,claussen2020detection,andrejevic2020machine,cao2020artificial,liu2020screening_revised_again,schleder2021machine}.
While ML has led to important advances for many applications in engineering and science, most ML models act essentially like black boxes: they are complicated models which often provide correct answers, but because of their complexity, they are difficult to understand, and hence provide little insight and intuition about the systems they are applied to.
Since a key aim of our work is precisely to learn insights about electronic topology, we instead focus our quest onto interpretable ML models, with a goal of finding a broadly applicable heuristic chemical 
rule that diagnoses whether or not a material is topological.

The heuristic rule that our ML approach discovered is based on the notion of a learned parameter for each element that loosely captures the tendency of an element to form topological materials. 
We refer to this as an element's \emph{topogivity}.
This heuristic rule is simple, hand-calculable, and interpretable: a given material is diagnosed with high accuracy
(typically $>80\%$)
as topologically nontrivial (trivial) if the weighted average of its elements' topogivities is positive (negative) (\cref{fig_model_and_discovery_strategy}a).
The heuristic rule does not rely on crystal symmetry, and our approach can be used to make predictions on {\it all} materials. 
We integrate the heuristic rule into a high-throughput procedure to search for non-symmetry-diagnosable topological materials, in which we perform screening using the heuristic rule followed by density functional theory (DFT) validation (\cref{fig_model_and_discovery_strategy}b).
The newly discovered topological materials include several high-quality examples that may be promising for experimental realization.

\section*{Classes of materials}

Conventional textbook chemistry teaches that the electrons of insulators (including semiconductors) are localized to ionic or covalent bonds, while the electrons of metals are delocalized and ``free''.
From a band-theoretic perspective, the former make up part of the class of materials known as atomic insulators and the latter roughly correspond to ``ordinary'' metals.
Topological insulators and topological semimetals do not fit into this conventional dichotomy.
Topological insulators feature a band gap and a nontrivial topological invariant, and as a consequence, their electronic states cannot be reduced to an assembly of localized atomic or molecular orbitals.  
Topological semimetals have band degeneracies protected by symmetry or topology near or at the Fermi level.
Collectively, we refer to topological insulators and topological semimetals as topological materials, and refer to all other materials as trivial materials.

To learn a heuristic chemical rule for diagnosing topological materials, we employ a supervised learning approach.
This requires a labeled dataset in which each material is labeled as ``topological'' or ``trivial''.
The learned heuristic chemical rule is then applied to screen another dataset which we refer to as the discovery space. 
Existing ab initio databases of stoichiometric, non-magnetic, three-dimensional materials~\cite{zhang2019catalogue,vergniory2019complete,vergniory2022all,tang2019comprehensive} offer a convenient source of data.
However, it is important to note that they are imperfect, in part because the symmetry-based high-throughput calculation methods that were used to generate them are inherently incapable of detecting certain topology. 
Taking such limitations into account, we identify the labeled dataset as a subset of the database generated by \citet{tang2019comprehensive} (see \suppsec~S1) (our methodology could also be applied to other databases that contain both trivial materials and topological materials). 
Our labeled dataset consists of 9,026 materials, of which $51\%$ are labeled as trivial and the remaining $49\%$ are labeled as topological. However, due to the aforementioned imperfection, for ML purposes this labeled dataset should effectively be considered as a dataset with noisy labels, e.g., some topological materials are incorrectly labeled as trivial. 
Separately, the discovery space consists of 1,433 materials, whose topology cannot be determined from the symmetry indicators method (see \suppsec~S1). By applying the learned heuristic chemical rule to the discovery space and then performing DFT, we are able to evaluate its ability to predict topological materials beyond those diagnosable by existing standard approaches.  Some of the topological materials that we identify in the discovery space are known elsewhere in the literature and serve primarily as confirmations, whereas others represent instances of truly new materials discovery.

\section*{Learning a heuristic chemical rule}

\begin{figure*}[!htb]
    \centering
    \includegraphics[scale=.85]{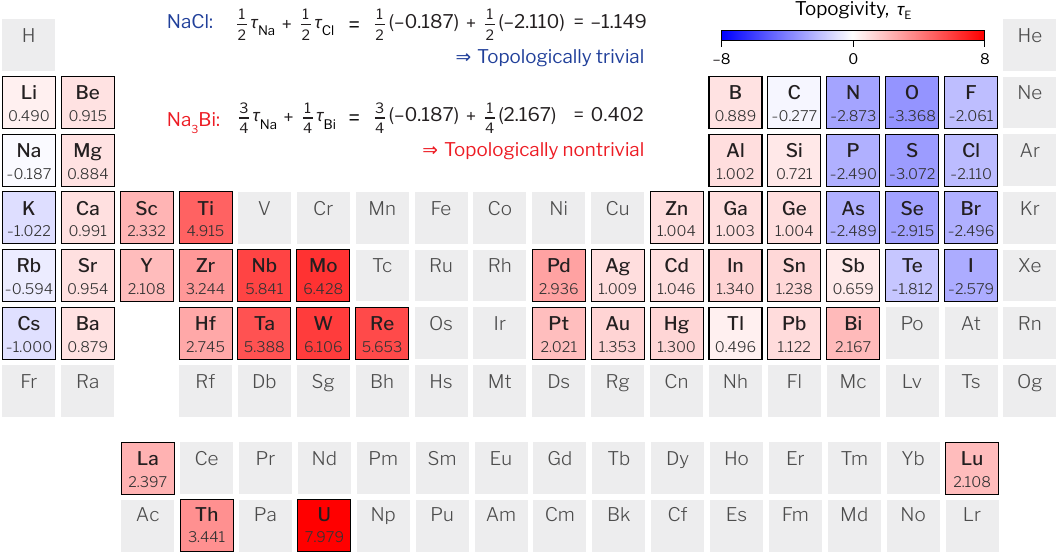}
    \caption{%
    \textbf{Periodic table of topogivities.}
    Machine-learned topogivities $\tau_E$ are shown by color-coding and in values.
    Elements that do not appear in any material in the dataset are shown in gray.
    Example applications of the topogivity-based heuristic chemical rule are shown for two materials, NaCl (trivial) and Na\textsubscript{3}Bi (nontrivial)~\cite{liu2014discovery}.
    \label{fig_periodic_table_visualization_of_topogivities}}
\end{figure*}

Our machine learning model takes the form of a heuristic chemical rule.
Specifically, the model maps each material $M$ to a number $g(M)$ according to the function
\begin{equation*}
    g(M) = \sum_{E} f_E(M) \tau_E,
\end{equation*}
where the summation runs over the elements present in the chemical formula of material $M$, $\tau_E$ is a learned parameter for each element $E$, and $f_E(M)$ is the element fraction for the element $E$ in material $M$ (e.g., for a chemical formula $A_x B_y C_z$, $f_A(M) = \frac{x}{x+y+z}$, $f_B(M) = \frac{y}{x+y+z}$, and $f_C(M) = \frac{z}{x+y+z}$).
Classification decisions are made according to the sign of $g(M)$: classify as topological if positive and classify as trivial if negative.
A greater magnitude of $g(M)$ roughly corresponds to a more confident classification decision.
We refer to the model as a heuristic \emph{chemical} rule in the sense that all the information required for obtaining a diagnosis is contained in the material's chemical formula.
Additional modeling and methodological details are provided in Supplementary Sections S2.A and S2.B; an analysis of information lost from not incorporating spatial information in the classifier is provided in \suppsec~S3.A.

For each element $E$, we refer to the optimized parameter $\tau_E$ as its \emph{topogivity}.
For a given material $M$, $g(M)$ is simply the weighted average of its elements' topogivities, where the weighting is with respect to each element's relative abundance, as identifiable from the material's chemical formula.
Conceptually, an element's topogivity loosely captures its tendency to form topological materials -- greater topogivity is intended to roughly correspond to a greater tendency (see \suppsec~S3.D for details and caveats on the meaning of topogivity).

Before making predictions in the discovery space, we want to first evaluate model performance within the labeled dataset.  To do this, we use a nested cross validation procedure and average the results over multiple test sets.
We find an average of $82.7 \%$ accuracy.
Additionally, we find empirical evidence that as the magnitude of $g(M)$ is increased, the fraction of correctly classified materials first increases and then plateaus, with the plateau beginning around $|g(M)| \approx 1$.
We heuristically set a threshold of 1.0 for a high-confidence topologically nontrivial classification and observe on average that $93.0 \%$ of materials with $g(M) \geq 1.0$ are correctly classified.
Details and extended results are presented in \suppsec~S2.C.
Additionally, we found that the model works better for materials with two or three distinct elements than for materials with one or four distinct elements (see \suppsec~S3.B).
Having completed nested cross validation, we proceed to use the entire labeled dataset to fit the final model (see \suppsec~S2.D), which is what we will use for making predictions in the discovery space.
An additional evaluation of model performance is presented in \suppsec~S2.E.

We visualize the final model's learned topogivities in \cref{fig_periodic_table_visualization_of_topogivities}.
This periodic table of topogivities enables an immediate heuristic diagnosis of any stoichiometric material whose elements are featured in the table.
This is illustrated with examples in \cref{fig_periodic_table_visualization_of_topogivities} for the trivial insulator NaCl and the Dirac semimetal Na$_3$Bi~\cite{liu2014discovery}.
The Weyl semimetal TaAs~\cite{xu2015discovery,lv2015experimental} is also worth highlighting: TaAs is non-symmetry-diagnosable~\cite{tang2019comprehensive} and does not appear in the labeled dataset, but is successfully diagnosed as topological by the topogivity approach: $g(\text{TaAs}) = \tfrac{1}{2}\tau_{\text{Ta}} + \tfrac{1}{2}\tau_{\text{As}} = 1.450$.

\begin{figure*}[!htb]
    \centering
    \includegraphics[scale=1]{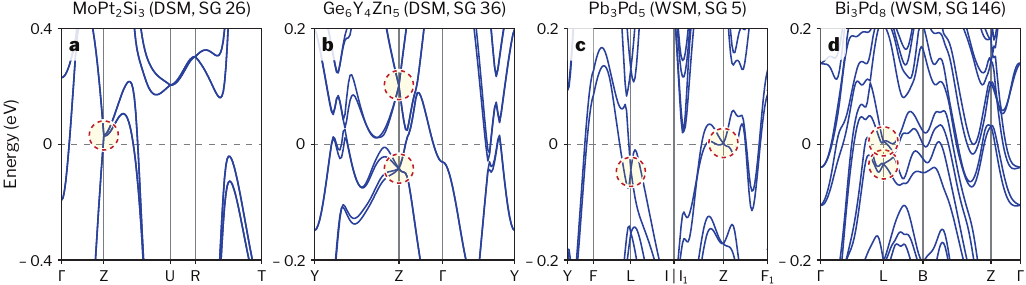}
    \caption{\textbf{Selection of newly discovered topological materials.}
    These materials are not diagnosable using symmetry indicators~\cite{po2017symmetry_revised}, but were successfully discovered using our topogivity-based approach.
    The band structures were computed using DFT.
    MoPt\textsubscript{2}Si\textsubscript{3} (\captionbullet{a}) and Ge\textsubscript{6}Y\textsubscript{4}Zn\textsubscript{5} (\captionbullet{b}) are nonsymmorphic Dirac semimetals.
    Pb\textsubscript{3}Pd\textsubscript{5} (\captionbullet{c}) and Bi\textsubscript{3}Pd\textsubscript{8} (\captionbullet{d}) are Kramers Weyl semimetals.
    For each material, relevant topological degeneracies are highlighted by circles.
    \label{fig_featured_band_structures}}
\end{figure*}

The simplicity of our model enables us to readily extract chemical insights from the periodic table of topogivities.
First, we observe that elements that are near each other in the periodic table tend to have similar topogivities, which is consistent with intuition.
Second, we observe that the elements with negative topogivities are located in two clusters respectively in the top right and bottom left parts of the periodic table.
This is also consistent with intuition, since ionic compounds often have large trivial band gaps and elements from these two clusters tend to form ionic compounds.
Third, considering group 15 (the pnictogens), we observe that while N, P, and As have negative topogivities (and Sb has a small positive topogivity), Bi has a positive topogivity with a relatively large magnitude.
This is consistent with the intuition that Bi often plays a role in topological materials~\cite{isaeva2020crystal}.
Finally, we observe a region of high topogivities in the early transition metals -- future work could attempt to understand the underlying reasons for this (note that there is a chance that typical oxidation states are artificially inflating these topogivities).
Overall, while the element topogivities are parameters whose specific learned values are affected by dataset and modeling limitations (see \suppsec~S3.D for discussion), the fact that we can extract chemical insights that are consistent with intuition is evidence that a topogivity-based approximate picture can provide a meaningful way to study topological materials.

\section*{High-throughput screening and ab initio calculations}

To identify topological materials using the learned topogivities, we compute $g(M)$ for each of the 1,433 materials in the discovery space.
We restrict our attention to the materials that have a $g(M)$ value that corresponds to a topologically nontrivial classification with high-confidence (i.e., $g(M) \geq 1.0$): that leaves 73 materials (after the removal of 2 other materials).
Additionally, since it is difficult to obtain accurate DFT calculations for f-electron materials, we exclude any material that contains a 4f or 5f electron, eliminating 5 materials and thus leaving us with 68 materials for ab initio validation.
Full details of our topogivity-based screening procedure (including filtering criteria) are provided in \suppsec~S4.A.

For each of these 68 materials, we perform DFT within the generalized gradient approximation (GGA)~\cite{perdew1996generalized}.
We include spin-orbital coupling in all of our calculations, which is consistent with the database~\cite{tang2019comprehensive} (note that topological classification can depend on the presence or absence of spin-orbital coupling~\cite{po2017symmetry_revised}).
In our DFT calculations, we checked for many -- but not all -- types of topological materials.
In principle it is possible that some topological materials were not detected by our DFT.
Further methodological details are provided in \suppsec~S4.B.

Of the 68 materials, we find 56 topological materials, corresponding to a success rate of 82.4\%.
We stress that the discovery space and the labeled dataset correspond to different regimes of materials, and so it is quite interesting that a model that was fit on the labeled dataset still works in the discovery space (see \suppsec~S4.C).
We note that there are aspects of our procedure and data analysis that could have introduced some bias into the success rate (see \suppsec~S4.C).

The 56 topological materials that we found consist of 48 Weyl semimetals, 7 Dirac semimetals, and 1 Dirac nodal line semimetal.
The band structures of all 56 of these topological materials are included in \suppsec~S6.
Some of these topological materials have previously been predicted in the literature and a smaller portion have also already been experimentally observed, e.g., TaAs~\cite{xu2015discovery,lv2015experimental}.
More importantly, our DFT calculations also identify multiple new topological materials that to our knowledge have not been previously identified.

We highlight four particularly interesting newly discovered topological materials in \cref{fig_featured_band_structures}.
Each is a topological semimetal with a relatively clean band structure and at least one band crossing within 50 meV of the Fermi level, making it promising for potential experimental investigation. MoPt\textsubscript{2}Si\textsubscript{3} and Ge\textsubscript{6}Y\textsubscript{4}Zn\textsubscript{5} are both nonsymmorphic Dirac semimetals.
At the Z point, the former has a Dirac point in the valence band manifold and the latter has Dirac points in both the valence and conduction band manifolds.
Pb\textsubscript{3}Pd\textsubscript{5} and Bi\textsubscript{3}Pd\textsubscript{8} are both Kramers Weyl semimetals~\cite{chang2018topological}.
The former has Weyl nodes at the L and Z points, and the latter has two Weyl nodes close in energy at the L point.
In particular, we highlight that MoPt\textsubscript{2}Si\textsubscript{3} has a Dirac point close to the Fermi level as well as a relatively clean Fermi surface, and Pb\textsubscript{3}Pd\textsubscript{5} has a Weyl node located at the $Z$ point that is right at the Fermi level.
We emphasize that the reason the band degeneracies in these four materials are non-symmetry-diagnosable is that they are all within the valence band manifold or conduction band manifold.
Such band degeneracies cannot be diagnosed by the symmetry indicators method, which is formulated based on the electron filling and therefore cannot target band degeneracies that are not \emph{between} the valence and conduction bands~\cite{po2017symmetry_revised}. 

Separately, as a preliminary exploration, we performed DFT calculations on a selection of labeled dataset materials that are labeled as trivial but which the model classifies as topological.
Our DFT calculations revealed that in some of these cases, the material is actually topological (\ie, the model is correct and it is actually the label that is wrong).
Further details -- including some selected example materials from this exploration -- are included in \suppsec~S5.

\section*{Discussion and Outlook}

The topogivity approach provides only a coarse-grained topological classification (nontrivial or trivial); it lacks the fine-grained detail of ab initio approaches.
Moreover, it is important to note that topogivity is not an unambiguously defined quantity, as its exact numerical value for each element can depend, for example, on the choice of machine learning algorithm and the use of the weighted average formulation.
This fact is illustrated in \suppsec~S3.C, where we empirically demonstrate the minor impact of making a particular change to the machine learning algorithm.
Further discussion pertaining to this lack of an unambiguous definition is included in \suppsec~S3.D.

Nevertheless, topogivity offers a broadly applicable and simple approach for diagnosing topological materials.
This diagnosis uses only the chemical formula and requires merely a handful of arithmetic operations to evaluate.
An important highlight of the topogivity-based diagnosis approach is that it enables the discovery of non-symmetry-diagnosable topological materials.
Furthermore, the periodic table of topogivities (\cref{fig_periodic_table_visualization_of_topogivities}) provides simple intuition for a complex, exotic phenomenon.

One worthy future direction is to look for a more complete understanding of the underlying reasons for the values of the elements' topogivities, which may in turn shed new light on the fundamental question of why some materials are topological while others are not.
Another promising path forward would be to perform more comprehensive searches for new topological materials using topogivity-based strategies.
Finally, it is intriguing to contemplate whether our interpretable-ML approach, used here to discover topogivity, could perhaps be used for other material properties as well, such as ferroelectricity, ferromagnetism, or maybe even superconductivity.

\FloatBarrier

\section*{Acknowledgements}

We thank Pawan Goyal for assisting preliminary work.
We thank Samuel Kim, Peter Lu, Rumen Dangovski, and Edward Zhang for helpful discussions.
We thank Feng Tang and Xiangang Wan for the sharing of materials data. 
We thank Paola Rebusco for critical reading and editing of the manuscript. 
A.M. acknowledges support from the MIT EECS Alan L. McWhorter Fellowship and from the National Science Foundation Graduate Research Fellowship under Grant No.~1745302.
Research was sponsored in part by the United States Air Force Research Laboratory and the United States Air Force Artificial Intelligence Accelerator and was accomplished under Cooperative Agreement Number FA8750-19-2-1000.
The views and conclusions contained in this document are those of the authors and should not be interpreted as representing the official policies, either expressed or implied, of the United States Air Force or the U.S. Government.
The U.S. Government is authorized to reproduce and distribute reprints for Government purposes notwithstanding any copyright notation herein.
This material is also based upon work supported in part by the Air Force Office of Scientific Research under the awards number FA9550-20-1-0115, FA9550-21-1-0299, and FA9550-21-1-0317, as well as in part by the US Office of Naval Research (ONR) Multidisciplinary University Research Initiative (MURI) grant N00014-20-1-2325 on Robust Photonic Materials with High-Order Topological Protection.
This material is also based upon work supported in part by the U.S. Army Research Office through the Institute for Soldier Nanotechnologies at MIT, under Collaborative Agreement Number W911NF-18-2-0048.
This work is also supported in part by the National Science Foundation under Cooperative Agreement PHY-2019786 (The NSF AI Institute for Artificial Intelligence and Fundamental Interactions, \url{http://iaifi.org/}).
The work of Y.Z. and L.F. was supported by DOE Office of Basic Energy Sciences, Division of Materials Sciences and Engineering under Award DE-SC0018945 (theoretical analysis) and DE-SC0019275 (band structure calculation).
L.F. was partly supported by the Simons Investigator Award from Simons Foundation.
The work of H.C.P. was partly supported by a Pappalardo Fellowship at MIT.

\section*{Code and data availability}
The code and data underlying the machine learning part of this paper is available in our public repository (https://github.com/andrewma8/topogivity).

\def\bibsection{\section*{\refname}}
\bibliographystyle{apsrev4-2-longbib}
\bibliography{refs}

\begin{thebibliography}{47}%
\makeatletter
\providecommand \@ifxundefined [1]{%
 \@ifx{#1\undefined}
}%
\providecommand \@ifnum [1]{%
 \ifnum #1\expandafter \@firstoftwo
 \else \expandafter \@secondoftwo
 \fi
}%
\providecommand \@ifx [1]{%
 \ifx #1\expandafter \@firstoftwo
 \else \expandafter \@secondoftwo
 \fi
}%
\providecommand \natexlab [1]{#1}%
\providecommand \enquote  [1]{``#1''}%
\providecommand \bibnamefont  [1]{#1}%
\providecommand \bibfnamefont [1]{#1}%
\providecommand \citenamefont [1]{#1}%
\providecommand \href@noop [0]{\@secondoftwo}%
\providecommand \href [0]{\begingroup \@sanitize@url \@href}%
\providecommand \@href[1]{\@@startlink{#1}\@@href}%
\providecommand \@@href[1]{\endgroup#1\@@endlink}%
\providecommand \@sanitize@url [0]{\catcode `\\12\catcode `\$12\catcode
  `\&12\catcode `\#12\catcode `\^12\catcode `\_12\catcode `\%12\relax}%
\providecommand \@@startlink[1]{}%
\providecommand \@@endlink[0]{}%
\providecommand \url  [0]{\begingroup\@sanitize@url \@url }%
\providecommand \@url [1]{\endgroup\@href {#1}{\urlprefix }}%
\providecommand \urlprefix  [0]{URL }%
\providecommand \Eprint [0]{\href }%
\providecommand \doibase [0]{https://doi.org/}%
\providecommand \selectlanguage [0]{\@gobble}%
\providecommand \bibinfo  [0]{\@secondoftwo}%
\providecommand \bibfield  [0]{\@secondoftwo}%
\providecommand \translation [1]{[#1]}%
\providecommand \BibitemOpen [0]{}%
\providecommand \bibitemStop [0]{}%
\providecommand \bibitemNoStop [0]{.\EOS\space}%
\providecommand \EOS [0]{\spacefactor3000\relax}%
\providecommand \BibitemShut  [1]{\csname bibitem#1\endcsname}%
\let\auto@bib@innerbib\@empty
\bibitem [{\citenamefont {Kane}\ and\ \citenamefont
  {Mele}(2005)}]{kane2005quantum}%
  \BibitemOpen
  \bibfield  {author} {\bibinfo {author} {\bibfnamefont {C.~L.}\ \bibnamefont
  {Kane}}\ and\ \bibinfo {author} {\bibfnamefont {E.~J.}\ \bibnamefont
  {Mele}},\ }\bibfield  {title} {\bibinfo {title} {Quantum spin {Hall} effect
  in graphene},\ }\href {https://doi.org/10.1103/PhysRevLett.95.226801}
  {\bibfield  {journal} {\bibinfo  {journal} {Phys. Rev. Lett.}\ }\textbf
  {\bibinfo {volume} {95}},\ \bibinfo {pages} {226801} (\bibinfo {year}
  {2005})}\BibitemShut {NoStop}%
\bibitem [{\citenamefont {Bernevig}\ \emph {et~al.}(2006)\citenamefont
  {Bernevig}, \citenamefont {Hughes},\ and\ \citenamefont
  {Zhang}}]{bernevig2006quantum}%
  \BibitemOpen
  \bibfield  {author} {\bibinfo {author} {\bibfnamefont {B.~A.}\ \bibnamefont
  {Bernevig}}, \bibinfo {author} {\bibfnamefont {T.~L.}\ \bibnamefont
  {Hughes}},\ and\ \bibinfo {author} {\bibfnamefont {S.-C.}\ \bibnamefont
  {Zhang}},\ }\bibfield  {title} {\bibinfo {title} {Quantum spin {Hall} effect
  and topological phase transition in {HgTe} quantum wells},\ }\href
  {https://doi.org/10.1126/science.1133734} {\bibfield  {journal} {\bibinfo
  {journal} {Science}\ }\textbf {\bibinfo {volume} {314}},\ \bibinfo {pages}
  {1757} (\bibinfo {year} {2006})}\BibitemShut {NoStop}%
\bibitem [{\citenamefont {Konig}\ \emph {et~al.}(2007)\citenamefont {Konig},
  \citenamefont {Wiedmann}, \citenamefont {Brune}, \citenamefont {Roth},
  \citenamefont {Buhmann}, \citenamefont {Molenkamp}, \citenamefont {Qi},\ and\
  \citenamefont {Zhang}}]{konig2007quantum}%
  \BibitemOpen
  \bibfield  {author} {\bibinfo {author} {\bibfnamefont {M.}~\bibnamefont
  {Konig}}, \bibinfo {author} {\bibfnamefont {S.}~\bibnamefont {Wiedmann}},
  \bibinfo {author} {\bibfnamefont {C.}~\bibnamefont {Brune}}, \bibinfo
  {author} {\bibfnamefont {A.}~\bibnamefont {Roth}}, \bibinfo {author}
  {\bibfnamefont {H.}~\bibnamefont {Buhmann}}, \bibinfo {author} {\bibfnamefont
  {L.~W.}\ \bibnamefont {Molenkamp}}, \bibinfo {author} {\bibfnamefont {X.-L.}\
  \bibnamefont {Qi}},\ and\ \bibinfo {author} {\bibfnamefont {S.-C.}\
  \bibnamefont {Zhang}},\ }\bibfield  {title} {\bibinfo {title} {Quantum spin
  {Hall} insulator state in {HgTe} quantum wells},\ }\href
  {https://doi.org/10.1126/science.1148047} {\bibfield  {journal} {\bibinfo
  {journal} {Science}\ }\textbf {\bibinfo {volume} {318}},\ \bibinfo {pages}
  {766} (\bibinfo {year} {2007})}\BibitemShut {NoStop}%
\bibitem [{\citenamefont {Fu}\ \emph {et~al.}(2007)\citenamefont {Fu},
  \citenamefont {Kane},\ and\ \citenamefont {Mele}}]{fu2007topological3d}%
  \BibitemOpen
  \bibfield  {author} {\bibinfo {author} {\bibfnamefont {L.}~\bibnamefont
  {Fu}}, \bibinfo {author} {\bibfnamefont {C.~L.}\ \bibnamefont {Kane}},\ and\
  \bibinfo {author} {\bibfnamefont {E.~J.}\ \bibnamefont {Mele}},\ }\bibfield
  {title} {\bibinfo {title} {Topological insulators in three dimensions},\
  }\href {https://doi.org/10.1103/PhysRevLett.98.106803} {\bibfield  {journal}
  {\bibinfo  {journal} {Phys. Rev. Lett.}\ }\textbf {\bibinfo {volume} {98}},\
  \bibinfo {pages} {106803} (\bibinfo {year} {2007})}\BibitemShut {NoStop}%
\bibitem [{\citenamefont {Hsieh}\ \emph {et~al.}(2008)\citenamefont {Hsieh},
  \citenamefont {Qian}, \citenamefont {Wray}, \citenamefont {Xia},
  \citenamefont {Hor}, \citenamefont {Cava},\ and\ \citenamefont
  {Hasan}}]{hsieh2008topological}%
  \BibitemOpen
  \bibfield  {author} {\bibinfo {author} {\bibfnamefont {D.}~\bibnamefont
  {Hsieh}}, \bibinfo {author} {\bibfnamefont {D.}~\bibnamefont {Qian}},
  \bibinfo {author} {\bibfnamefont {L.}~\bibnamefont {Wray}}, \bibinfo {author}
  {\bibfnamefont {Y.}~\bibnamefont {Xia}}, \bibinfo {author} {\bibfnamefont
  {Y.~S.}\ \bibnamefont {Hor}}, \bibinfo {author} {\bibfnamefont {R.~J.}\
  \bibnamefont {Cava}},\ and\ \bibinfo {author} {\bibfnamefont {M.~Z.}\
  \bibnamefont {Hasan}},\ }\bibfield  {title} {\bibinfo {title} {A topological
  {Dirac} insulator in a quantum spin {Hall} phase},\ }\href
  {https://doi.org/10.1038/nature06843} {\bibfield  {journal} {\bibinfo
  {journal} {Nature}\ }\textbf {\bibinfo {volume} {452}},\ \bibinfo {pages}
  {970} (\bibinfo {year} {2008})}\BibitemShut {NoStop}%
\bibitem [{\citenamefont {Fu}(2011)}]{fu2011topological}%
  \BibitemOpen
  \bibfield  {author} {\bibinfo {author} {\bibfnamefont {L.}~\bibnamefont
  {Fu}},\ }\bibfield  {title} {\bibinfo {title} {Topological crystalline
  insulators},\ }\href {https://doi.org/10.1103/PhysRevLett.106.106802}
  {\bibfield  {journal} {\bibinfo  {journal} {Phys. Rev. Lett.}\ }\textbf
  {\bibinfo {volume} {106}},\ \bibinfo {pages} {106802} (\bibinfo {year}
  {2011})}\BibitemShut {NoStop}%
\bibitem [{\citenamefont {Hsieh}\ \emph {et~al.}(2012)\citenamefont {Hsieh},
  \citenamefont {Lin}, \citenamefont {Liu}, \citenamefont {Duan}, \citenamefont
  {Bansil},\ and\ \citenamefont {Fu}}]{hsieh2012topological_revised}%
  \BibitemOpen
  \bibfield  {author} {\bibinfo {author} {\bibfnamefont {T.~H.}\ \bibnamefont
  {Hsieh}}, \bibinfo {author} {\bibfnamefont {H.}~\bibnamefont {Lin}}, \bibinfo
  {author} {\bibfnamefont {J.}~\bibnamefont {Liu}}, \bibinfo {author}
  {\bibfnamefont {W.}~\bibnamefont {Duan}}, \bibinfo {author} {\bibfnamefont
  {A.}~\bibnamefont {Bansil}},\ and\ \bibinfo {author} {\bibfnamefont
  {L.}~\bibnamefont {Fu}},\ }\bibfield  {title} {\bibinfo {title} {Topological
  crystalline insulators in the {SnTe} material class},\ }\href
  {https://doi.org/10.1038/ncomms1969} {\bibfield  {journal} {\bibinfo
  {journal} {Nat. Commun.}\ }\textbf {\bibinfo {volume} {3}},\ \bibinfo {pages}
  {982} (\bibinfo {year} {2012})}\BibitemShut {NoStop}%
\bibitem [{\citenamefont {Benalcazar}\ \emph {et~al.}(2017)\citenamefont
  {Benalcazar}, \citenamefont {Bernevig},\ and\ \citenamefont
  {Hughes}}]{benalcazar2017quantized}%
  \BibitemOpen
  \bibfield  {author} {\bibinfo {author} {\bibfnamefont {W.~A.}\ \bibnamefont
  {Benalcazar}}, \bibinfo {author} {\bibfnamefont {B.~A.}\ \bibnamefont
  {Bernevig}},\ and\ \bibinfo {author} {\bibfnamefont {T.~L.}\ \bibnamefont
  {Hughes}},\ }\bibfield  {title} {\bibinfo {title} {Quantized electric
  multipole insulators},\ }\href {https://doi.org/10.1126/science.aah6442}
  {\bibfield  {journal} {\bibinfo  {journal} {Science}\ }\textbf {\bibinfo
  {volume} {357}},\ \bibinfo {pages} {61} (\bibinfo {year} {2017})}\BibitemShut
  {NoStop}%
\bibitem [{\citenamefont {Liu}\ \emph {et~al.}(2014)\citenamefont {Liu},
  \citenamefont {Zhou}, \citenamefont {Zhang}, \citenamefont {Wang},
  \citenamefont {Weng}, \citenamefont {Prabhakaran}, \citenamefont {Mo},
  \citenamefont {Shen}, \citenamefont {Fang}, \citenamefont {Dai} \emph
  {et~al.}}]{liu2014discovery}%
  \BibitemOpen
  \bibfield  {author} {\bibinfo {author} {\bibfnamefont {Z.}~\bibnamefont
  {Liu}}, \bibinfo {author} {\bibfnamefont {B.}~\bibnamefont {Zhou}}, \bibinfo
  {author} {\bibfnamefont {Y.}~\bibnamefont {Zhang}}, \bibinfo {author}
  {\bibfnamefont {Z.}~\bibnamefont {Wang}}, \bibinfo {author} {\bibfnamefont
  {H.}~\bibnamefont {Weng}}, \bibinfo {author} {\bibfnamefont {D.}~\bibnamefont
  {Prabhakaran}}, \bibinfo {author} {\bibfnamefont {S.-K.}\ \bibnamefont {Mo}},
  \bibinfo {author} {\bibfnamefont {Z.}~\bibnamefont {Shen}}, \bibinfo {author}
  {\bibfnamefont {Z.}~\bibnamefont {Fang}}, \bibinfo {author} {\bibfnamefont
  {X.}~\bibnamefont {Dai}}, \emph {et~al.},\ }\bibfield  {title} {\bibinfo
  {title} {Discovery of a three-dimensional topological {Dirac} semimetal,
  {Na\textsubscript{3}Bi}},\ }\href {https://doi.org/10.1126/science.1245085}
  {\bibfield  {journal} {\bibinfo  {journal} {Science}\ }\textbf {\bibinfo
  {volume} {343}},\ \bibinfo {pages} {864} (\bibinfo {year}
  {2014})}\BibitemShut {NoStop}%
\bibitem [{\citenamefont {Xu}\ \emph {et~al.}(2015)\citenamefont {Xu},
  \citenamefont {Belopolski}, \citenamefont {Alidoust}, \citenamefont
  {Neupane}, \citenamefont {Bian}, \citenamefont {Zhang}, \citenamefont
  {Sankar}, \citenamefont {Chang}, \citenamefont {Yuan}, \citenamefont {Lee}
  \emph {et~al.}}]{xu2015discovery}%
  \BibitemOpen
  \bibfield  {author} {\bibinfo {author} {\bibfnamefont {S.-Y.}\ \bibnamefont
  {Xu}}, \bibinfo {author} {\bibfnamefont {I.}~\bibnamefont {Belopolski}},
  \bibinfo {author} {\bibfnamefont {N.}~\bibnamefont {Alidoust}}, \bibinfo
  {author} {\bibfnamefont {M.}~\bibnamefont {Neupane}}, \bibinfo {author}
  {\bibfnamefont {G.}~\bibnamefont {Bian}}, \bibinfo {author} {\bibfnamefont
  {C.}~\bibnamefont {Zhang}}, \bibinfo {author} {\bibfnamefont
  {R.}~\bibnamefont {Sankar}}, \bibinfo {author} {\bibfnamefont
  {G.}~\bibnamefont {Chang}}, \bibinfo {author} {\bibfnamefont
  {Z.}~\bibnamefont {Yuan}}, \bibinfo {author} {\bibfnamefont {C.-C.}\
  \bibnamefont {Lee}}, \emph {et~al.},\ }\bibfield  {title} {\bibinfo {title}
  {Discovery of a {Weyl} fermion semimetal and topological {Fermi} arcs},\
  }\href {https://doi.org/10.1126/science.aaa9297} {\bibfield  {journal}
  {\bibinfo  {journal} {Science}\ }\textbf {\bibinfo {volume} {349}},\ \bibinfo
  {pages} {613} (\bibinfo {year} {2015})}\BibitemShut {NoStop}%
\bibitem [{\citenamefont {Lv}\ \emph {et~al.}(2015)\citenamefont {Lv},
  \citenamefont {Weng}, \citenamefont {Fu}, \citenamefont {Wang}, \citenamefont
  {Miao}, \citenamefont {Ma}, \citenamefont {Richard}, \citenamefont {Huang},
  \citenamefont {Zhao}, \citenamefont {Chen} \emph
  {et~al.}}]{lv2015experimental}%
  \BibitemOpen
  \bibfield  {author} {\bibinfo {author} {\bibfnamefont {B.}~\bibnamefont
  {Lv}}, \bibinfo {author} {\bibfnamefont {H.}~\bibnamefont {Weng}}, \bibinfo
  {author} {\bibfnamefont {B.}~\bibnamefont {Fu}}, \bibinfo {author}
  {\bibfnamefont {X.~P.}\ \bibnamefont {Wang}}, \bibinfo {author}
  {\bibfnamefont {H.}~\bibnamefont {Miao}}, \bibinfo {author} {\bibfnamefont
  {J.}~\bibnamefont {Ma}}, \bibinfo {author} {\bibfnamefont {P.}~\bibnamefont
  {Richard}}, \bibinfo {author} {\bibfnamefont {X.}~\bibnamefont {Huang}},
  \bibinfo {author} {\bibfnamefont {L.}~\bibnamefont {Zhao}}, \bibinfo {author}
  {\bibfnamefont {G.}~\bibnamefont {Chen}}, \emph {et~al.},\ }\bibfield
  {title} {\bibinfo {title} {Experimental discovery of {Weyl} semimetal
  {TaAs}},\ }\href {https://doi.org/10.1103/PhysRevX.5.031013} {\bibfield
  {journal} {\bibinfo  {journal} {Phys. Rev. X}\ }\textbf {\bibinfo {volume}
  {5}},\ \bibinfo {pages} {031013} (\bibinfo {year} {2015})}\BibitemShut
  {NoStop}%
\bibitem [{\citenamefont {Burkov}\ \emph {et~al.}(2011)\citenamefont {Burkov},
  \citenamefont {Hook},\ and\ \citenamefont {Balents}}]{burkov2011topological}%
  \BibitemOpen
  \bibfield  {author} {\bibinfo {author} {\bibfnamefont {A.}~\bibnamefont
  {Burkov}}, \bibinfo {author} {\bibfnamefont {M.}~\bibnamefont {Hook}},\ and\
  \bibinfo {author} {\bibfnamefont {L.}~\bibnamefont {Balents}},\ }\bibfield
  {title} {\bibinfo {title} {Topological nodal semimetals},\ }\href
  {https://doi.org/10.1103/PhysRevB.84.235126} {\bibfield  {journal} {\bibinfo
  {journal} {Phys. Rev. B}\ }\textbf {\bibinfo {volume} {84}},\ \bibinfo
  {pages} {235126} (\bibinfo {year} {2011})}\BibitemShut {NoStop}%
\bibitem [{\citenamefont {Bradlyn}\ \emph {et~al.}(2016)\citenamefont
  {Bradlyn}, \citenamefont {Cano}, \citenamefont {Wang}, \citenamefont
  {Vergniory}, \citenamefont {Felser}, \citenamefont {Cava},\ and\
  \citenamefont {Bernevig}}]{bradlyn2016beyond}%
  \BibitemOpen
  \bibfield  {author} {\bibinfo {author} {\bibfnamefont {B.}~\bibnamefont
  {Bradlyn}}, \bibinfo {author} {\bibfnamefont {J.}~\bibnamefont {Cano}},
  \bibinfo {author} {\bibfnamefont {Z.}~\bibnamefont {Wang}}, \bibinfo {author}
  {\bibfnamefont {M.}~\bibnamefont {Vergniory}}, \bibinfo {author}
  {\bibfnamefont {C.}~\bibnamefont {Felser}}, \bibinfo {author} {\bibfnamefont
  {R.~J.}\ \bibnamefont {Cava}},\ and\ \bibinfo {author} {\bibfnamefont
  {B.~A.}\ \bibnamefont {Bernevig}},\ }\bibfield  {title} {\bibinfo {title}
  {Beyond {Dirac} and {Weyl} fermions: Unconventional quasiparticles in
  conventional crystals},\ }\href {https://doi.org/10.1126/science.aaf5037}
  {\bibfield  {journal} {\bibinfo  {journal} {Science}\ }\textbf {\bibinfo
  {volume} {353}},\ \bibinfo {pages} {558} (\bibinfo {year}
  {2016})}\BibitemShut {NoStop}%
\bibitem [{\citenamefont {Xiao}\ and\ \citenamefont
  {Yan}(2021)}]{xiao2021first}%
  \BibitemOpen
  \bibfield  {author} {\bibinfo {author} {\bibfnamefont {J.}~\bibnamefont
  {Xiao}}\ and\ \bibinfo {author} {\bibfnamefont {B.}~\bibnamefont {Yan}},\
  }\bibfield  {title} {\bibinfo {title} {First-principles calculations for
  topological quantum materials},\ }\href
  {https://doi.org/10.1038/s42254-021-00292-8} {\bibfield  {journal} {\bibinfo
  {journal} {Nat. Rev. Phys.}\ }\textbf {\bibinfo {volume} {3}},\ \bibinfo
  {pages} {283} (\bibinfo {year} {2021})}\BibitemShut {NoStop}%
\bibitem [{\citenamefont {Bansil}\ \emph {et~al.}(2016)\citenamefont {Bansil},
  \citenamefont {Lin},\ and\ \citenamefont {Das}}]{bansil2016colloquium}%
  \BibitemOpen
  \bibfield  {author} {\bibinfo {author} {\bibfnamefont {A.}~\bibnamefont
  {Bansil}}, \bibinfo {author} {\bibfnamefont {H.}~\bibnamefont {Lin}},\ and\
  \bibinfo {author} {\bibfnamefont {T.}~\bibnamefont {Das}},\ }\bibfield
  {title} {\bibinfo {title} {Colloquium: Topological band theory},\ }\href
  {https://doi.org/10.1103/RevModPhys.88.021004} {\bibfield  {journal}
  {\bibinfo  {journal} {Rev. Mod. Phys.}\ }\textbf {\bibinfo {volume} {88}},\
  \bibinfo {pages} {021004} (\bibinfo {year} {2016})}\BibitemShut {NoStop}%
\bibitem [{\citenamefont {Po}\ \emph {et~al.}(2017)\citenamefont {Po},
  \citenamefont {Vishwanath},\ and\ \citenamefont
  {Watanabe}}]{po2017symmetry_revised}%
  \BibitemOpen
  \bibfield  {author} {\bibinfo {author} {\bibfnamefont {H.~C.}\ \bibnamefont
  {Po}}, \bibinfo {author} {\bibfnamefont {A.}~\bibnamefont {Vishwanath}},\
  and\ \bibinfo {author} {\bibfnamefont {H.}~\bibnamefont {Watanabe}},\
  }\bibfield  {title} {\bibinfo {title} {Symmetry-based indicators of band
  topology in the 230 space groups},\ }\href
  {https://doi.org/10.1038/s41467-017-00133-2} {\bibfield  {journal} {\bibinfo
  {journal} {Nat. Commun.}\ }\textbf {\bibinfo {volume} {8}},\ \bibinfo {pages}
  {50} (\bibinfo {year} {2017})}\BibitemShut {NoStop}%
\bibitem [{\citenamefont {Bradlyn}\ \emph {et~al.}(2017)\citenamefont
  {Bradlyn}, \citenamefont {Elcoro}, \citenamefont {Cano}, \citenamefont
  {Vergniory}, \citenamefont {Wang}, \citenamefont {Felser}, \citenamefont
  {Aroyo},\ and\ \citenamefont {Bernevig}}]{bradlyn2017topological}%
  \BibitemOpen
  \bibfield  {author} {\bibinfo {author} {\bibfnamefont {B.}~\bibnamefont
  {Bradlyn}}, \bibinfo {author} {\bibfnamefont {L.}~\bibnamefont {Elcoro}},
  \bibinfo {author} {\bibfnamefont {J.}~\bibnamefont {Cano}}, \bibinfo {author}
  {\bibfnamefont {M.}~\bibnamefont {Vergniory}}, \bibinfo {author}
  {\bibfnamefont {Z.}~\bibnamefont {Wang}}, \bibinfo {author} {\bibfnamefont
  {C.}~\bibnamefont {Felser}}, \bibinfo {author} {\bibfnamefont
  {M.}~\bibnamefont {Aroyo}},\ and\ \bibinfo {author} {\bibfnamefont {B.~A.}\
  \bibnamefont {Bernevig}},\ }\bibfield  {title} {\bibinfo {title} {Topological
  quantum chemistry},\ }\href {https://doi.org/10.1038/nature23268} {\bibfield
  {journal} {\bibinfo  {journal} {Nature}\ }\textbf {\bibinfo {volume} {547}},\
  \bibinfo {pages} {298} (\bibinfo {year} {2017})}\BibitemShut {NoStop}%
\bibitem [{\citenamefont {Fu}\ and\ \citenamefont
  {Kane}(2007)}]{fu2007topologicalinv}%
  \BibitemOpen
  \bibfield  {author} {\bibinfo {author} {\bibfnamefont {L.}~\bibnamefont
  {Fu}}\ and\ \bibinfo {author} {\bibfnamefont {C.~L.}\ \bibnamefont {Kane}},\
  }\bibfield  {title} {\bibinfo {title} {Topological insulators with inversion
  symmetry},\ }\href {https://doi.org/10.1103/PhysRevB.76.045302} {\bibfield
  {journal} {\bibinfo  {journal} {Phys. Rev. B}\ }\textbf {\bibinfo {volume}
  {76}},\ \bibinfo {pages} {045302} (\bibinfo {year} {2007})}\BibitemShut
  {NoStop}%
\bibitem [{\citenamefont {Kruthoff}\ \emph {et~al.}(2017)\citenamefont
  {Kruthoff}, \citenamefont {de~Boer}, \citenamefont {van Wezel}, \citenamefont
  {Kane},\ and\ \citenamefont {Slager}}]{slager2017}%
  \BibitemOpen
  \bibfield  {author} {\bibinfo {author} {\bibfnamefont {J.}~\bibnamefont
  {Kruthoff}}, \bibinfo {author} {\bibfnamefont {J.}~\bibnamefont {de~Boer}},
  \bibinfo {author} {\bibfnamefont {J.}~\bibnamefont {van Wezel}}, \bibinfo
  {author} {\bibfnamefont {C.~L.}\ \bibnamefont {Kane}},\ and\ \bibinfo
  {author} {\bibfnamefont {R.-J.}\ \bibnamefont {Slager}},\ }\bibfield  {title}
  {\bibinfo {title} {Topological classification of crystalline insulators
  through band structure combinatorics},\ }\href
  {https://doi.org/10.1103/PhysRevX.7.041069} {\bibfield  {journal} {\bibinfo
  {journal} {Phys. Rev. X}\ }\textbf {\bibinfo {volume} {7}},\ \bibinfo {pages}
  {041069} (\bibinfo {year} {2017})}\BibitemShut {NoStop}%
\bibitem [{\citenamefont {Song}\ \emph {et~al.}(2018)\citenamefont {Song},
  \citenamefont {Zhang}, \citenamefont {Fang},\ and\ \citenamefont
  {Fang}}]{song2018quantitative_revised}%
  \BibitemOpen
  \bibfield  {author} {\bibinfo {author} {\bibfnamefont {Z.}~\bibnamefont
  {Song}}, \bibinfo {author} {\bibfnamefont {T.}~\bibnamefont {Zhang}},
  \bibinfo {author} {\bibfnamefont {Z.}~\bibnamefont {Fang}},\ and\ \bibinfo
  {author} {\bibfnamefont {C.}~\bibnamefont {Fang}},\ }\bibfield  {title}
  {\bibinfo {title} {Quantitative mappings between symmetry and topology in
  solids},\ }\href {https://doi.org/10.1038/s41467-018-06010-w} {\bibfield
  {journal} {\bibinfo  {journal} {Nat. Commun.}\ }\textbf {\bibinfo {volume}
  {9}},\ \bibinfo {pages} {3530} (\bibinfo {year} {2018})}\BibitemShut
  {NoStop}%
\bibitem [{\citenamefont {Tang}\ \emph {et~al.}(2019)\citenamefont {Tang},
  \citenamefont {Po}, \citenamefont {Vishwanath},\ and\ \citenamefont
  {Wan}}]{tang2019comprehensive}%
  \BibitemOpen
  \bibfield  {author} {\bibinfo {author} {\bibfnamefont {F.}~\bibnamefont
  {Tang}}, \bibinfo {author} {\bibfnamefont {H.~C.}\ \bibnamefont {Po}},
  \bibinfo {author} {\bibfnamefont {A.}~\bibnamefont {Vishwanath}},\ and\
  \bibinfo {author} {\bibfnamefont {X.}~\bibnamefont {Wan}},\ }\bibfield
  {title} {\bibinfo {title} {Comprehensive search for topological materials
  using symmetry indicators},\ }\href
  {https://doi.org/10.1038/s41586-019-0937-5} {\bibfield  {journal} {\bibinfo
  {journal} {Nature}\ }\textbf {\bibinfo {volume} {566}},\ \bibinfo {pages}
  {486} (\bibinfo {year} {2019})}\BibitemShut {NoStop}%
\bibitem [{\citenamefont {Zhang}\ \emph {et~al.}(2019)\citenamefont {Zhang},
  \citenamefont {Jiang}, \citenamefont {Song}, \citenamefont {Huang},
  \citenamefont {He}, \citenamefont {Fang}, \citenamefont {Weng},\ and\
  \citenamefont {Fang}}]{zhang2019catalogue}%
  \BibitemOpen
  \bibfield  {author} {\bibinfo {author} {\bibfnamefont {T.}~\bibnamefont
  {Zhang}}, \bibinfo {author} {\bibfnamefont {Y.}~\bibnamefont {Jiang}},
  \bibinfo {author} {\bibfnamefont {Z.}~\bibnamefont {Song}}, \bibinfo {author}
  {\bibfnamefont {H.}~\bibnamefont {Huang}}, \bibinfo {author} {\bibfnamefont
  {Y.}~\bibnamefont {He}}, \bibinfo {author} {\bibfnamefont {Z.}~\bibnamefont
  {Fang}}, \bibinfo {author} {\bibfnamefont {H.}~\bibnamefont {Weng}},\ and\
  \bibinfo {author} {\bibfnamefont {C.}~\bibnamefont {Fang}},\ }\bibfield
  {title} {\bibinfo {title} {Catalogue of topological electronic materials},\
  }\href {https://doi.org/10.1038/s41586-019-0944-6} {\bibfield  {journal}
  {\bibinfo  {journal} {Nature}\ }\textbf {\bibinfo {volume} {566}},\ \bibinfo
  {pages} {475} (\bibinfo {year} {2019})}\BibitemShut {NoStop}%
\bibitem [{\citenamefont {Vergniory}\ \emph {et~al.}(2019)\citenamefont
  {Vergniory}, \citenamefont {Elcoro}, \citenamefont {Felser}, \citenamefont
  {Regnault}, \citenamefont {Bernevig},\ and\ \citenamefont
  {Wang}}]{vergniory2019complete}%
  \BibitemOpen
  \bibfield  {author} {\bibinfo {author} {\bibfnamefont {M.}~\bibnamefont
  {Vergniory}}, \bibinfo {author} {\bibfnamefont {L.}~\bibnamefont {Elcoro}},
  \bibinfo {author} {\bibfnamefont {C.}~\bibnamefont {Felser}}, \bibinfo
  {author} {\bibfnamefont {N.}~\bibnamefont {Regnault}}, \bibinfo {author}
  {\bibfnamefont {B.~A.}\ \bibnamefont {Bernevig}},\ and\ \bibinfo {author}
  {\bibfnamefont {Z.}~\bibnamefont {Wang}},\ }\bibfield  {title} {\bibinfo
  {title} {A complete catalogue of high-quality topological materials},\ }\href
  {https://doi.org/10.1038/s41586-019-0954-4} {\bibfield  {journal} {\bibinfo
  {journal} {Nature}\ }\textbf {\bibinfo {volume} {566}},\ \bibinfo {pages}
  {480} (\bibinfo {year} {2019})}\BibitemShut {NoStop}%
\bibitem [{\citenamefont {Weng}\ \emph {et~al.}(2015)\citenamefont {Weng},
  \citenamefont {Fang}, \citenamefont {Fang}, \citenamefont {Bernevig},\ and\
  \citenamefont {Dai}}]{weng2015weyl}%
  \BibitemOpen
  \bibfield  {author} {\bibinfo {author} {\bibfnamefont {H.}~\bibnamefont
  {Weng}}, \bibinfo {author} {\bibfnamefont {C.}~\bibnamefont {Fang}}, \bibinfo
  {author} {\bibfnamefont {Z.}~\bibnamefont {Fang}}, \bibinfo {author}
  {\bibfnamefont {B.~A.}\ \bibnamefont {Bernevig}},\ and\ \bibinfo {author}
  {\bibfnamefont {X.}~\bibnamefont {Dai}},\ }\bibfield  {title} {\bibinfo
  {title} {{Weyl} semimetal phase in noncentrosymmetric transition-metal
  monophosphides},\ }\href {https://doi.org/10.1103/PhysRevX.5.011029}
  {\bibfield  {journal} {\bibinfo  {journal} {Phys. Rev. X}\ }\textbf {\bibinfo
  {volume} {5}},\ \bibinfo {pages} {011029} (\bibinfo {year}
  {2015})}\BibitemShut {NoStop}%
\bibitem [{\citenamefont {Hehre}(1976)}]{hehre1976ab}%
  \BibitemOpen
  \bibfield  {author} {\bibinfo {author} {\bibfnamefont {W.~J.}\ \bibnamefont
  {Hehre}},\ }\bibfield  {title} {\bibinfo {title} {Ab initio molecular orbital
  theory},\ }\href {https://doi.org/10.1021/ar50107a003} {\bibfield  {journal}
  {\bibinfo  {journal} {Acc. Chem. Res.}\ }\textbf {\bibinfo {volume} {9}},\
  \bibinfo {pages} {399} (\bibinfo {year} {1976})}\BibitemShut {NoStop}%
\bibitem [{\citenamefont {Kumar}\ \emph {et~al.}(2021)\citenamefont {Kumar},
  \citenamefont {Guin}, \citenamefont {Manna}, \citenamefont {Shekhar},\ and\
  \citenamefont {Felser}}]{kumar2020topological}%
  \BibitemOpen
  \bibfield  {author} {\bibinfo {author} {\bibfnamefont {N.}~\bibnamefont
  {Kumar}}, \bibinfo {author} {\bibfnamefont {S.~N.}\ \bibnamefont {Guin}},
  \bibinfo {author} {\bibfnamefont {K.}~\bibnamefont {Manna}}, \bibinfo
  {author} {\bibfnamefont {C.}~\bibnamefont {Shekhar}},\ and\ \bibinfo {author}
  {\bibfnamefont {C.}~\bibnamefont {Felser}},\ }\bibfield  {title} {\bibinfo
  {title} {Topological quantum materials from the viewpoint of chemistry},\
  }\href {https://doi.org/10.1021/acs.chemrev.0c00732} {\bibfield  {journal}
  {\bibinfo  {journal} {Chem. Rev.}\ }\textbf {\bibinfo {volume} {121}},\
  \bibinfo {pages} {2780} (\bibinfo {year} {2021})}\BibitemShut {NoStop}%
\bibitem [{\citenamefont {Schoop}\ \emph {et~al.}(2018)\citenamefont {Schoop},
  \citenamefont {Pielnhofer},\ and\ \citenamefont
  {Lotsch}}]{schoop2018chemical}%
  \BibitemOpen
  \bibfield  {author} {\bibinfo {author} {\bibfnamefont {L.~M.}\ \bibnamefont
  {Schoop}}, \bibinfo {author} {\bibfnamefont {F.}~\bibnamefont {Pielnhofer}},\
  and\ \bibinfo {author} {\bibfnamefont {B.~V.}\ \bibnamefont {Lotsch}},\
  }\bibfield  {title} {\bibinfo {title} {Chemical principles of topological
  semimetals},\ }\href {https://doi.org/10.1021/acs.chemmater.7b05133}
  {\bibfield  {journal} {\bibinfo  {journal} {Chem. Mater.}\ }\textbf {\bibinfo
  {volume} {30}},\ \bibinfo {pages} {3155} (\bibinfo {year}
  {2018})}\BibitemShut {NoStop}%
\bibitem [{\citenamefont {Gibson}\ \emph {et~al.}(2015)\citenamefont {Gibson},
  \citenamefont {Schoop}, \citenamefont {Muechler}, \citenamefont {Xie},
  \citenamefont {Hirschberger}, \citenamefont {Ong}, \citenamefont {Car},\ and\
  \citenamefont {Cava}}]{gibson2015three_revised}%
  \BibitemOpen
  \bibfield  {author} {\bibinfo {author} {\bibfnamefont {Q.}~\bibnamefont
  {Gibson}}, \bibinfo {author} {\bibfnamefont {L.~M.}\ \bibnamefont {Schoop}},
  \bibinfo {author} {\bibfnamefont {L.}~\bibnamefont {Muechler}}, \bibinfo
  {author} {\bibfnamefont {L.}~\bibnamefont {Xie}}, \bibinfo {author}
  {\bibfnamefont {M.}~\bibnamefont {Hirschberger}}, \bibinfo {author}
  {\bibfnamefont {N.~P.}\ \bibnamefont {Ong}}, \bibinfo {author} {\bibfnamefont
  {R.}~\bibnamefont {Car}},\ and\ \bibinfo {author} {\bibfnamefont {R.~J.}\
  \bibnamefont {Cava}},\ }\bibfield  {title} {\bibinfo {title}
  {Three-dimensional {Dirac} semimetals: Design principles and predictions of
  new materials},\ }\href {https://doi.org/10.1103/PhysRevB.91.205128}
  {\bibfield  {journal} {\bibinfo  {journal} {Phys. Rev. B}\ }\textbf {\bibinfo
  {volume} {91}},\ \bibinfo {pages} {205128} (\bibinfo {year}
  {2015})}\BibitemShut {NoStop}%
\bibitem [{\citenamefont {Isaeva}\ and\ \citenamefont
  {Ruck}(2020)}]{isaeva2020crystal}%
  \BibitemOpen
  \bibfield  {author} {\bibinfo {author} {\bibfnamefont {A.}~\bibnamefont
  {Isaeva}}\ and\ \bibinfo {author} {\bibfnamefont {M.}~\bibnamefont {Ruck}},\
  }\bibfield  {title} {\bibinfo {title} {Crystal chemistry and bonding patterns
  of bismuth-based topological insulators},\ }\href
  {https://doi.org/10.1021/acs.inorgchem.9b03461} {\bibfield  {journal}
  {\bibinfo  {journal} {Inorg. Chem.}\ }\textbf {\bibinfo {volume} {59}},\
  \bibinfo {pages} {3437} (\bibinfo {year} {2020})}\BibitemShut {NoStop}%
\bibitem [{\citenamefont {Klemenz}\ \emph {et~al.}(2020)\citenamefont
  {Klemenz}, \citenamefont {Hay}, \citenamefont {Teicher}, \citenamefont
  {Topp}, \citenamefont {Cano},\ and\ \citenamefont
  {Schoop}}]{klemenz2020role}%
  \BibitemOpen
  \bibfield  {author} {\bibinfo {author} {\bibfnamefont {S.}~\bibnamefont
  {Klemenz}}, \bibinfo {author} {\bibfnamefont {A.~K.}\ \bibnamefont {Hay}},
  \bibinfo {author} {\bibfnamefont {S.~M.}\ \bibnamefont {Teicher}}, \bibinfo
  {author} {\bibfnamefont {A.}~\bibnamefont {Topp}}, \bibinfo {author}
  {\bibfnamefont {J.}~\bibnamefont {Cano}},\ and\ \bibinfo {author}
  {\bibfnamefont {L.~M.}\ \bibnamefont {Schoop}},\ }\bibfield  {title}
  {\bibinfo {title} {The role of delocalized chemical bonding in
  square-net-based topological semimetals},\ }\href
  {https://doi.org/10.1021/jacs.0c01227} {\bibfield  {journal} {\bibinfo
  {journal} {J. Am. Chem. Soc.}\ }\textbf {\bibinfo {volume} {142}},\ \bibinfo
  {pages} {6350} (\bibinfo {year} {2020})}\BibitemShut {NoStop}%
\bibitem [{\citenamefont {Gui}\ \emph {et~al.}(2019)\citenamefont {Gui},
  \citenamefont {Pletikosic}, \citenamefont {Cao}, \citenamefont {Tien},
  \citenamefont {Xu}, \citenamefont {Zhong}, \citenamefont {Wang},
  \citenamefont {Chang}, \citenamefont {Jia}, \citenamefont {Valla} \emph
  {et~al.}}]{gui2019new}%
  \BibitemOpen
  \bibfield  {author} {\bibinfo {author} {\bibfnamefont {X.}~\bibnamefont
  {Gui}}, \bibinfo {author} {\bibfnamefont {I.}~\bibnamefont {Pletikosic}},
  \bibinfo {author} {\bibfnamefont {H.}~\bibnamefont {Cao}}, \bibinfo {author}
  {\bibfnamefont {H.-J.}\ \bibnamefont {Tien}}, \bibinfo {author}
  {\bibfnamefont {X.}~\bibnamefont {Xu}}, \bibinfo {author} {\bibfnamefont
  {R.}~\bibnamefont {Zhong}}, \bibinfo {author} {\bibfnamefont
  {G.}~\bibnamefont {Wang}}, \bibinfo {author} {\bibfnamefont {T.-R.}\
  \bibnamefont {Chang}}, \bibinfo {author} {\bibfnamefont {S.}~\bibnamefont
  {Jia}}, \bibinfo {author} {\bibfnamefont {T.}~\bibnamefont {Valla}}, \emph
  {et~al.},\ }\bibfield  {title} {\bibinfo {title} {A new magnetic topological
  quantum material candidate by design},\ }\href
  {https://doi.org/10.1021/acscentsci.9b00202} {\bibfield  {journal} {\bibinfo
  {journal} {ACS Cent. Sci.}\ }\textbf {\bibinfo {volume} {5}},\ \bibinfo
  {pages} {900} (\bibinfo {year} {2019})}\BibitemShut {NoStop}%
\bibitem [{\citenamefont {Zhong}\ \emph {et~al.}(2020)\citenamefont {Zhong},
  \citenamefont {Tran}, \citenamefont {Min}, \citenamefont {Wang},
  \citenamefont {Wang}, \citenamefont {Dinh}, \citenamefont {De~Luna},
  \citenamefont {Yu}, \citenamefont {Rasouli}, \citenamefont {Brodersen} \emph
  {et~al.}}]{zhong2020accelerated}%
  \BibitemOpen
  \bibfield  {author} {\bibinfo {author} {\bibfnamefont {M.}~\bibnamefont
  {Zhong}}, \bibinfo {author} {\bibfnamefont {K.}~\bibnamefont {Tran}},
  \bibinfo {author} {\bibfnamefont {Y.}~\bibnamefont {Min}}, \bibinfo {author}
  {\bibfnamefont {C.}~\bibnamefont {Wang}}, \bibinfo {author} {\bibfnamefont
  {Z.}~\bibnamefont {Wang}}, \bibinfo {author} {\bibfnamefont {C.-T.}\
  \bibnamefont {Dinh}}, \bibinfo {author} {\bibfnamefont {P.}~\bibnamefont
  {De~Luna}}, \bibinfo {author} {\bibfnamefont {Z.}~\bibnamefont {Yu}},
  \bibinfo {author} {\bibfnamefont {A.~S.}\ \bibnamefont {Rasouli}}, \bibinfo
  {author} {\bibfnamefont {P.}~\bibnamefont {Brodersen}}, \emph {et~al.},\
  }\bibfield  {title} {\bibinfo {title} {Accelerated discovery of
  {CO}\textsubscript{2} electrocatalysts using active machine learning},\
  }\href {https://doi.org/10.1038/s41586-020-2242-8} {\bibfield  {journal}
  {\bibinfo  {journal} {Nature}\ }\textbf {\bibinfo {volume} {581}},\ \bibinfo
  {pages} {178} (\bibinfo {year} {2020})}\BibitemShut {NoStop}%
\bibitem [{\citenamefont {Deringer}\ \emph {et~al.}(2021)\citenamefont
  {Deringer}, \citenamefont {Bernstein}, \citenamefont {Cs{\'a}nyi},
  \citenamefont {Mahmoud}, \citenamefont {Ceriotti}, \citenamefont {Wilson},
  \citenamefont {Drabold},\ and\ \citenamefont
  {Elliott}}]{deringer2021origins}%
  \BibitemOpen
  \bibfield  {author} {\bibinfo {author} {\bibfnamefont {V.~L.}\ \bibnamefont
  {Deringer}}, \bibinfo {author} {\bibfnamefont {N.}~\bibnamefont {Bernstein}},
  \bibinfo {author} {\bibfnamefont {G.}~\bibnamefont {Cs{\'a}nyi}}, \bibinfo
  {author} {\bibfnamefont {C.~B.}\ \bibnamefont {Mahmoud}}, \bibinfo {author}
  {\bibfnamefont {M.}~\bibnamefont {Ceriotti}}, \bibinfo {author}
  {\bibfnamefont {M.}~\bibnamefont {Wilson}}, \bibinfo {author} {\bibfnamefont
  {D.~A.}\ \bibnamefont {Drabold}},\ and\ \bibinfo {author} {\bibfnamefont
  {S.~R.}\ \bibnamefont {Elliott}},\ }\bibfield  {title} {\bibinfo {title}
  {Origins of structural and electronic transitions in disordered silicon},\
  }\href {https://doi.org/10.1038/s41586-020-03072-z} {\bibfield  {journal}
  {\bibinfo  {journal} {Nature}\ }\textbf {\bibinfo {volume} {589}},\ \bibinfo
  {pages} {59} (\bibinfo {year} {2021})}\BibitemShut {NoStop}%
\bibitem [{\citenamefont {George}\ and\ \citenamefont
  {Hautier}(2021)}]{george2021chemist}%
  \BibitemOpen
  \bibfield  {author} {\bibinfo {author} {\bibfnamefont {J.}~\bibnamefont
  {George}}\ and\ \bibinfo {author} {\bibfnamefont {G.}~\bibnamefont
  {Hautier}},\ }\bibfield  {title} {\bibinfo {title} {Chemist versus machine:
  Traditional knowledge versus machine learning techniques},\ }\href
  {https://doi.org/10.1016/j.trechm.2020.10.007} {\bibfield  {journal}
  {\bibinfo  {journal} {Trends Chem.}\ }\textbf {\bibinfo {volume} {3}},\
  \bibinfo {pages} {86} (\bibinfo {year} {2021})}\BibitemShut {NoStop}%
\bibitem [{\citenamefont {Zhang}\ and\ \citenamefont
  {Kim}(2017)}]{zhang2017quantum}%
  \BibitemOpen
  \bibfield  {author} {\bibinfo {author} {\bibfnamefont {Y.}~\bibnamefont
  {Zhang}}\ and\ \bibinfo {author} {\bibfnamefont {E.-A.}\ \bibnamefont
  {Kim}},\ }\bibfield  {title} {\bibinfo {title} {Quantum loop topography for
  machine learning},\ }\href {https://doi.org/10.1103/PhysRevLett.118.216401}
  {\bibfield  {journal} {\bibinfo  {journal} {Phys. Rev. Lett.}\ }\textbf
  {\bibinfo {volume} {118}},\ \bibinfo {pages} {216401} (\bibinfo {year}
  {2017})}\BibitemShut {NoStop}%
\bibitem [{\citenamefont {Zhang}\ \emph {et~al.}(2018)\citenamefont {Zhang},
  \citenamefont {Shen},\ and\ \citenamefont {Zhai}}]{zhang2018machine}%
  \BibitemOpen
  \bibfield  {author} {\bibinfo {author} {\bibfnamefont {P.}~\bibnamefont
  {Zhang}}, \bibinfo {author} {\bibfnamefont {H.}~\bibnamefont {Shen}},\ and\
  \bibinfo {author} {\bibfnamefont {H.}~\bibnamefont {Zhai}},\ }\bibfield
  {title} {\bibinfo {title} {Machine learning topological invariants with
  neural networks},\ }\href {https://doi.org/10.1103/PhysRevLett.120.066401}
  {\bibfield  {journal} {\bibinfo  {journal} {Phys. Rev. Lett.}\ }\textbf
  {\bibinfo {volume} {120}},\ \bibinfo {pages} {066401} (\bibinfo {year}
  {2018})}\BibitemShut {NoStop}%
\bibitem [{\citenamefont {Scheurer}\ and\ \citenamefont
  {Slager}(2020)}]{scheurer2020unsupervised}%
  \BibitemOpen
  \bibfield  {author} {\bibinfo {author} {\bibfnamefont {M.~S.}\ \bibnamefont
  {Scheurer}}\ and\ \bibinfo {author} {\bibfnamefont {R.-J.}\ \bibnamefont
  {Slager}},\ }\bibfield  {title} {\bibinfo {title} {Unsupervised machine
  learning and band topology},\ }\href
  {https://doi.org/10.1103/PhysRevLett.124.226401} {\bibfield  {journal}
  {\bibinfo  {journal} {Phys. Rev. Lett.}\ }\textbf {\bibinfo {volume} {124}},\
  \bibinfo {pages} {226401} (\bibinfo {year} {2020})}\BibitemShut {NoStop}%
\bibitem [{\citenamefont {Zhang}\ \emph {et~al.}(2020)\citenamefont {Zhang},
  \citenamefont {Ginsparg},\ and\ \citenamefont {Kim}}]{zhang2020interpreting}%
  \BibitemOpen
  \bibfield  {author} {\bibinfo {author} {\bibfnamefont {Y.}~\bibnamefont
  {Zhang}}, \bibinfo {author} {\bibfnamefont {P.}~\bibnamefont {Ginsparg}},\
  and\ \bibinfo {author} {\bibfnamefont {E.-A.}\ \bibnamefont {Kim}},\
  }\bibfield  {title} {\bibinfo {title} {Interpreting machine learning of
  topological quantum phase transitions},\ }\href
  {https://doi.org/10.1103/PhysRevResearch.2.023283} {\bibfield  {journal}
  {\bibinfo  {journal} {Phys. Rev. Research}\ }\textbf {\bibinfo {volume}
  {2}},\ \bibinfo {pages} {023283} (\bibinfo {year} {2020})}\BibitemShut
  {NoStop}%
\bibitem [{\citenamefont {Acosta}\ \emph {et~al.}(2021)\citenamefont {Acosta},
  \citenamefont {Ouyang}, \citenamefont {Fazzio}, \citenamefont {Scheffler},
  \citenamefont {Ghiringhelli},\ and\ \citenamefont
  {Carbogno}}]{acosta2018analysis_manual}%
  \BibitemOpen
  \bibfield  {author} {\bibinfo {author} {\bibfnamefont {C.~M.}\ \bibnamefont
  {Acosta}}, \bibinfo {author} {\bibfnamefont {R.}~\bibnamefont {Ouyang}},
  \bibinfo {author} {\bibfnamefont {A.}~\bibnamefont {Fazzio}}, \bibinfo
  {author} {\bibfnamefont {M.}~\bibnamefont {Scheffler}}, \bibinfo {author}
  {\bibfnamefont {L.~M.}\ \bibnamefont {Ghiringhelli}},\ and\ \bibinfo {author}
  {\bibfnamefont {C.}~\bibnamefont {Carbogno}},\ }\bibfield  {title} {\bibinfo
  {title} {Analysis of topological transitions in two-dimensional materials by
  compressed sensing},\ }\href {https://arxiv.org/abs/1805.10950} {\bibfield
  {journal} {\bibinfo  {journal} {2018, 1805.10950. arXiv;
  https://arxiv.org/abs/1805.10950}\ } (\bibinfo {year} {accessed Aug. 5,
  2021})}\BibitemShut {NoStop}%
\bibitem [{\citenamefont {Claussen}\ \emph {et~al.}(2020)\citenamefont
  {Claussen}, \citenamefont {Bernevig},\ and\ \citenamefont
  {Regnault}}]{claussen2020detection}%
  \BibitemOpen
  \bibfield  {author} {\bibinfo {author} {\bibfnamefont {N.}~\bibnamefont
  {Claussen}}, \bibinfo {author} {\bibfnamefont {B.~A.}\ \bibnamefont
  {Bernevig}},\ and\ \bibinfo {author} {\bibfnamefont {N.}~\bibnamefont
  {Regnault}},\ }\bibfield  {title} {\bibinfo {title} {Detection of topological
  materials with machine learning},\ }\href
  {https://doi.org/10.1103/PhysRevB.101.245117} {\bibfield  {journal} {\bibinfo
   {journal} {Phys. Rev. B}\ }\textbf {\bibinfo {volume} {101}},\ \bibinfo
  {pages} {245117} (\bibinfo {year} {2020})}\BibitemShut {NoStop}%
\bibitem [{\citenamefont {Andrejevic}\ \emph {et~al.}(2022)\citenamefont
  {Andrejevic}, \citenamefont {Andrejevic}, \citenamefont {Bernevig},
  \citenamefont {Regnault}, \citenamefont {Han}, \citenamefont {Fabbris},
  \citenamefont {Nguyen}, \citenamefont {Drucker}, \citenamefont {Rycroft},\
  and\ \citenamefont {Li}}]{andrejevic2020machine}%
  \BibitemOpen
  \bibfield  {author} {\bibinfo {author} {\bibfnamefont {N.}~\bibnamefont
  {Andrejevic}}, \bibinfo {author} {\bibfnamefont {J.}~\bibnamefont
  {Andrejevic}}, \bibinfo {author} {\bibfnamefont {B.~A.}\ \bibnamefont
  {Bernevig}}, \bibinfo {author} {\bibfnamefont {N.}~\bibnamefont {Regnault}},
  \bibinfo {author} {\bibfnamefont {F.}~\bibnamefont {Han}}, \bibinfo {author}
  {\bibfnamefont {G.}~\bibnamefont {Fabbris}}, \bibinfo {author} {\bibfnamefont
  {T.}~\bibnamefont {Nguyen}}, \bibinfo {author} {\bibfnamefont {N.~C.}\
  \bibnamefont {Drucker}}, \bibinfo {author} {\bibfnamefont {C.~H.}\
  \bibnamefont {Rycroft}},\ and\ \bibinfo {author} {\bibfnamefont
  {M.}~\bibnamefont {Li}},\ }\bibfield  {title} {\bibinfo {title}
  {Machine-learning spectral indicators of topology},\ }\href
  {https://doi.org/10.1002/adma.202204113} {\bibfield  {journal} {\bibinfo
  {journal} {Adv. Mater.}\ }\textbf {\bibinfo {volume} {\kern-.5ex}},\ \bibinfo
  {pages} {2204113} (\bibinfo {year} {2022})}\BibitemShut {NoStop}%
\bibitem [{\citenamefont {Cao}\ \emph {et~al.}(2020)\citenamefont {Cao},
  \citenamefont {Ouyang}, \citenamefont {Ghiringhelli}, \citenamefont
  {Scheffler}, \citenamefont {Liu}, \citenamefont {Carbogno},\ and\
  \citenamefont {Zhang}}]{cao2020artificial}%
  \BibitemOpen
  \bibfield  {author} {\bibinfo {author} {\bibfnamefont {G.}~\bibnamefont
  {Cao}}, \bibinfo {author} {\bibfnamefont {R.}~\bibnamefont {Ouyang}},
  \bibinfo {author} {\bibfnamefont {L.~M.}\ \bibnamefont {Ghiringhelli}},
  \bibinfo {author} {\bibfnamefont {M.}~\bibnamefont {Scheffler}}, \bibinfo
  {author} {\bibfnamefont {H.}~\bibnamefont {Liu}}, \bibinfo {author}
  {\bibfnamefont {C.}~\bibnamefont {Carbogno}},\ and\ \bibinfo {author}
  {\bibfnamefont {Z.}~\bibnamefont {Zhang}},\ }\bibfield  {title} {\bibinfo
  {title} {Artificial intelligence for high-throughput discovery of topological
  insulators: The example of alloyed tetradymites},\ }\href
  {https://doi.org/10.1103/PhysRevMaterials.4.034204} {\bibfield  {journal}
  {\bibinfo  {journal} {Phys. Rev. Mater.}\ }\textbf {\bibinfo {volume} {4}},\
  \bibinfo {pages} {034204} (\bibinfo {year} {2020})}\BibitemShut {NoStop}%
\bibitem [{\citenamefont {Liu}\ \emph {et~al.}(2021)\citenamefont {Liu},
  \citenamefont {Cao}, \citenamefont {Zhou},\ and\ \citenamefont
  {Liu}}]{liu2020screening_revised_again}%
  \BibitemOpen
  \bibfield  {author} {\bibinfo {author} {\bibfnamefont {H.}~\bibnamefont
  {Liu}}, \bibinfo {author} {\bibfnamefont {G.}~\bibnamefont {Cao}}, \bibinfo
  {author} {\bibfnamefont {Z.}~\bibnamefont {Zhou}},\ and\ \bibinfo {author}
  {\bibfnamefont {J.}~\bibnamefont {Liu}},\ }\bibfield  {title} {\bibinfo
  {title} {Screening potential topological insulators in half-{Heusler}
  compounds via compressed-sensing},\ }\href
  {https://doi.org/10.1088/1361-648X/abba8d} {\bibfield  {journal} {\bibinfo
  {journal} {J. Phys.: Condens. Matter}\ }\textbf {\bibinfo {volume} {33}},\
  \bibinfo {pages} {325501} (\bibinfo {year} {2021})}\BibitemShut {NoStop}%
\bibitem [{\citenamefont {Schleder}\ \emph {et~al.}(2021)\citenamefont
  {Schleder}, \citenamefont {Focassio},\ and\ \citenamefont
  {Fazzio}}]{schleder2021machine}%
  \BibitemOpen
  \bibfield  {author} {\bibinfo {author} {\bibfnamefont {G.~R.}\ \bibnamefont
  {Schleder}}, \bibinfo {author} {\bibfnamefont {B.}~\bibnamefont {Focassio}},\
  and\ \bibinfo {author} {\bibfnamefont {A.}~\bibnamefont {Fazzio}},\
  }\bibfield  {title} {\bibinfo {title} {Machine learning for materials
  discovery: Two-dimensional topological insulators},\ }\href
  {https://doi.org/10.1063/5.0055035} {\bibfield  {journal} {\bibinfo
  {journal} {Appl. Phys. Rev.}\ }\textbf {\bibinfo {volume} {8}},\ \bibinfo
  {pages} {031409} (\bibinfo {year} {2021})}\BibitemShut {NoStop}%
\bibitem [{\citenamefont {Vergniory}\ \emph {et~al.}(2022)\citenamefont
  {Vergniory}, \citenamefont {Wieder}, \citenamefont {Elcoro}, \citenamefont
  {Parkin}, \citenamefont {Felser}, \citenamefont {Bernevig},\ and\
  \citenamefont {Regnault}}]{vergniory2022all}%
  \BibitemOpen
  \bibfield  {author} {\bibinfo {author} {\bibfnamefont {M.~G.}\ \bibnamefont
  {Vergniory}}, \bibinfo {author} {\bibfnamefont {B.~J.}\ \bibnamefont
  {Wieder}}, \bibinfo {author} {\bibfnamefont {L.}~\bibnamefont {Elcoro}},
  \bibinfo {author} {\bibfnamefont {S.~S.}\ \bibnamefont {Parkin}}, \bibinfo
  {author} {\bibfnamefont {C.}~\bibnamefont {Felser}}, \bibinfo {author}
  {\bibfnamefont {B.~A.}\ \bibnamefont {Bernevig}},\ and\ \bibinfo {author}
  {\bibfnamefont {N.}~\bibnamefont {Regnault}},\ }\bibfield  {title} {\bibinfo
  {title} {All topological bands of all nonmagnetic stoichiometric materials},\
  }\href {https://doi.org/10.1126/science.abg9094} {\bibfield  {journal}
  {\bibinfo  {journal} {Science}\ }\textbf {\bibinfo {volume} {376}},\ \bibinfo
  {pages} {eabg9094} (\bibinfo {year} {2022})}\BibitemShut {NoStop}%
\bibitem [{\citenamefont {Perdew}\ \emph {et~al.}(1996)\citenamefont {Perdew},
  \citenamefont {Burke},\ and\ \citenamefont
  {Ernzerhof}}]{perdew1996generalized}%
  \BibitemOpen
  \bibfield  {author} {\bibinfo {author} {\bibfnamefont {J.~P.}\ \bibnamefont
  {Perdew}}, \bibinfo {author} {\bibfnamefont {K.}~\bibnamefont {Burke}},\ and\
  \bibinfo {author} {\bibfnamefont {M.}~\bibnamefont {Ernzerhof}},\ }\bibfield
  {title} {\bibinfo {title} {Generalized gradient approximation made simple},\
  }\href {https://doi.org/10.1103/PhysRevLett.77.3865} {\bibfield  {journal}
  {\bibinfo  {journal} {Phys. Rev. Lett.}\ }\textbf {\bibinfo {volume} {77}},\
  \bibinfo {pages} {3865} (\bibinfo {year} {1996})}\BibitemShut {NoStop}%
\bibitem [{\citenamefont {Chang}\ \emph {et~al.}(2018)\citenamefont {Chang},
  \citenamefont {Wieder}, \citenamefont {Schindler}, \citenamefont {Sanchez},
  \citenamefont {Belopolski}, \citenamefont {Huang}, \citenamefont {Singh},
  \citenamefont {Wu}, \citenamefont {Chang}, \citenamefont {Neupert} \emph
  {et~al.}}]{chang2018topological}%
  \BibitemOpen
  \bibfield  {author} {\bibinfo {author} {\bibfnamefont {G.}~\bibnamefont
  {Chang}}, \bibinfo {author} {\bibfnamefont {B.~J.}\ \bibnamefont {Wieder}},
  \bibinfo {author} {\bibfnamefont {F.}~\bibnamefont {Schindler}}, \bibinfo
  {author} {\bibfnamefont {D.~S.}\ \bibnamefont {Sanchez}}, \bibinfo {author}
  {\bibfnamefont {I.}~\bibnamefont {Belopolski}}, \bibinfo {author}
  {\bibfnamefont {S.-M.}\ \bibnamefont {Huang}}, \bibinfo {author}
  {\bibfnamefont {B.}~\bibnamefont {Singh}}, \bibinfo {author} {\bibfnamefont
  {D.}~\bibnamefont {Wu}}, \bibinfo {author} {\bibfnamefont {T.-R.}\
  \bibnamefont {Chang}}, \bibinfo {author} {\bibfnamefont {T.}~\bibnamefont
  {Neupert}}, \emph {et~al.},\ }\bibfield  {title} {\bibinfo {title}
  {Topological quantum properties of chiral crystals},\ }\href
  {https://doi.org/10.1038/s41563-018-0169-3} {\bibfield  {journal} {\bibinfo
  {journal} {Nat. Mater.}\ }\textbf {\bibinfo {volume} {17}},\ \bibinfo {pages}
  {978} (\bibinfo {year} {2018})}\BibitemShut {NoStop}%
\end{thebibliography}%

\clearpage


\section*{Supplementary material}

# SUPPLEMENTARY INFORMATION

# Topogivity: A Machine-Learned Chemical Rule for Discovering Topological Materials

Andrew Ma,[1, *] Yang Zhang,[2, *] Thomas Christensen,[2] Hoi Chun Po,[2, 3] Li Jing,[2, 4] Liang Fu,[2, †] and Marin Soljačić[2, ‡]

[1] Department of Electrical Engineering and Computer Science, Massachusetts Institute of Technology, Cambridge, Massachusetts 02139, USA

[2] Department of Physics, Massachusetts Institute of Technology, Cambridge, Massachusetts 02139, USA

[3] Department of Physics, Hong Kong University of Science and Technology, Clear Water Bay, Kowloon, Hong Kong

[4] Facebook AI Research, New York, New York 10003, USA

## CONTENTS

## S1. DESCRIPTION OF DATASETS

### A. Intended Purposes

There are two primary purposes for which we make use of data in this project:

1. As data for performing supervised learning.

2. As input into the high-throughput screening and ab initio validation process.

We require a separate dataset for each purpose.

We refer to the dataset used for the first purpose as the labeled dataset. This first purpose includes (i) training, validation, and testing (more specifically, in our case we will use a nested cross validation procedure) and (ii) fitting a final model. For this purpose, we need a dataset for supervised learning where each data point consists of a material that is labeled as either "topological" or "trivial".

We refer to the dataset used for the second purpose as the discovery space. The materials in the discovery space do not have labels. In the first step of our high-throughput process, the final model is used to screen through all of the materials in the discovery space in order to identify candidate topological materials. Subsequently, we perform density functional theory (DFT) calculations on candidates to determine which are actually topological (as a minor detail, a small number of candidates are filtered out prior to DFT). By choosing a suitable discovery space, this process of screening followed by ab initio calculations can serve two aims. First, it allows us to evaluate our model's performance in a particularly interesting regime, which contains a different type of materials from those in the labeled dataset. Second, any topological material identified in this manner that has not previously been identified in the literature represents a newly discovered topological material.

## B. Partitioning Materials Based on Symmetry Indicator Categorization and Space Group

Theories known as symmetry indicators [S1] and topological quantum chemistry [S2] have enabled the generation of ab initio databases consisting of electronic materials and their symmetry-based categorizations [S3; S4; S5; S6]. These databases provide a convenient source of readily available data, and in this work we will make use of the database generated in Tang *et al.* [S3]. We will partition this database into subsets by making use of the symmetry indicator categorizations [S1] as well as the space groups of the materials. Each subset will be used for a different role within our approach.

The Tang *et al.* [S3] database consists of stoichiometric, non-magnetic, three-dimensional materials treated with spin-orbital coupling. Consequently, our modeling approach and results will apply to this setting. All of our discussions of materials and of the symmetry indicators framework [S1] will be specialized to this setting as well.

In the symmetry indicators framework, each material is categorized as one of three broad categories using symmetry-based analysis [S1]. Note that the symmetry indicators framework can actually give more fine-grained information than just which of these three categories the material is in, but we will not be making use of that fine-grained information here. The three categories are:

(A) Compatibility relationships are violated, so we must have a band degeneracy.

(B) Compatibility relationships are not violated, yet at the same time the material is incompatible with being an atomic insulator. Hence, the material can either be (i) a material with a continuous gap that has a nontrivial topological invariant or (ii) a material with a band degeneracy that is not detectable from checking the compatibility relationships.

(C) Everything else. This means that insofar as symmetry indicators can tell, there is no way to distinguish the material from being an atomic insulator. However, being in this category *does not guarantee* that a material is an atomic insulator.

In the terminology of [S7], bullet (A) above is referred to as "Case 3", bullet (B) above is referred to as "Case 2", and bullet (C) above is referred to as "Case 1". In our work, we lump together materials that correspond to the bullets (A) and (B) and refer to them as NAI (Not an Atomic Insulator). We refer to materials that correspond to bullet (C) as USI (Undiagnosable by Symmetry Indicators). We use the phrase "non-symmetry-diagnosable topological material" to refer to a material that is topologically nontrivial but categorized as USI by the symmetry indicators framework.

In the theory of symmetry indicators, each space group has a symmetry indicator group $X_{BS}$ [S1]. Here, we will use the term NT-IG (Nontrivial Indicator Group) to refer to the set of materials whose space group has a nontrivial symmetry indicator group (i.e., $X_{BS} \neq \mathbb{Z}_1$). We will use the term T-IG (Trivial Indicator Group) to refer to the set of materials whose space group has a trivial symmetry indicator group (i.e., $X_{BS} = \mathbb{Z}_1$). The diagnostic power of symmetry indicators (by which we loosely mean the ability of the method to detect topological materials) is greater in NT-IG than in T-IG in the following two senses. First, it is possible for the symmetry indicators method to yield a categorization corresponding to bullet (B) above (i.e., "Case 2" in the terminology of [S7]) for a material in NT-IG, whereas it is not possible for the method to yield such a categorization for a material in T-IG. Second, the fraction of NT-IG materials that are categorized as NAI is much greater than the fraction of T-IG materials that are categorized as NAI (as illustrated by the pie charts in Fig. S1).

We can partition our set of materials into the following four subsets:

1. Materials that are categorized as USI and in NT-IG

2. Materials that are categorized as USI and in T-IG

3. Materials that are categorized as NAI and in NT-IG

4. Materials that are categorized as NAI and in T-IG

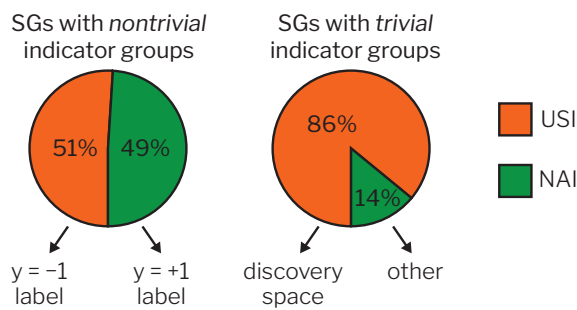

**Supplementary Figure S1. Materials dataset.** The symmetry-indicator-generated ab initio dataset from Tang *et al.* [S3] can be partitioned into two sets of materials based on whether the material's space group has a nontrivial symmetry indicator group (NT-IG) or a trivial symmetry indicator group (T-IG) [S1]. The two sets have substantially different ratios of NAI (Not an Atomic Insulator) to USI (Undiagnosable by Symmetry Indicators). The percentages shown were calculated for the dataset after preprocessing was completed. The NAI portion of the NT-IG materials is used as labeled data with topological labels ($y = +1$) and the USI portion of the NT-IG materials is used as labeled data with trivial labels ($y = -1$). The USI portion of the T-IG materials is the discovery space.

We use the first subset as labeled data with trivial labels and use the second subset as the discovery space. These two choices can be justified by the greater diagnostic power of symmetry indicators in NT-IG than in T-IG, which suggests that a USI material in NT-IG is more likely to be trivial than a USI material in T-IG. We use the third subset as labeled data with topological labels. The fourth subset is used for an additional evaluation of the model's performance that will be discussed in Supplementary Section S2.E. These four subsets and the roles that they play are depicted in Fig. S1. We note that another reasonable choice might have been to also include the fourth subset as part of the labeled data with topological labels (in that case, the labeled data with topological labels would just be all of the NAI materials).

Our choices of which materials to use as labeled data with topological labels and labeled data with trivial labels effectively entail that our labeled dataset consists of data points with noisy labels. Some of the primary sources of noise are as follows. First, there are some materials that are topologically nontrivial but indistinguishable from an atomic insulator insofar as the symmetry indicators framework is able to tell, and such topological materials are thus categorized as USI [S1]. Since this can occur even for materials in NT-IG, this means that some truly topological materials are incorrectly labeled as trivial in the labeled dataset. Second, whether it is actually the case that an NAI categorization implies that a material is topological depends on what one considers as a suitable definition of topological material. For example, given that the symmetry indicator categorization ignores energetic aspects of the electronic bands [S1], a material could still be categorized as NAI even if all of its band degeneracies are actually far from the Fermi level. When this occurs for a material in NT-IG, then if one is using a definition for topological semimetal that includes a criterion for how close a band degeneracy is to the Fermi level (which is a criterion that we do employ in our own DFT calculations), then it would correspond to a material that is incorrectly labeled as topological in the labeled dataset. Third, there are also incorrect categorizations that arise from the DFT calculations themselves that were used to generate the ab initio data in the Tang *et al.* [S3] database, since DFT results are not always accurate [S8]. This source of noise can cause both incorrect topological labels and incorrect trivial labels. Supplementary Section S5 includes some of our own ab initio calculation results that illustrate particular examples where a topological material is indeed incorrectly labeled as trivial. Note that the fact that the labels are noisy means that a hypothetical model that actually achieved perfect accuracy on the labeled dataset would be fitting in part to noise, and so such a model may not actually be the most desirable one (even if one only cared about predictive performance and did not care about interpretability).

A useful implication of our choice of discovery space is that if a material in the discovery space does turn out to be topological, then it would be a non-symmetry-diagnosable topological material (assuming that there was no error in the categorization of the material as USI). Thus, our choice of discovery space enables us to demonstrate and utilize our approach in a regime where it is presently difficult to discover new topological materials – the current first-principles approaches for diagnosing topological materials

in this regime typically involve significant computational cost (e.g., using Wilson loops [S9]).

We note that there are some materials in the discovery space that have previously been identified as topologically nontrivial in the literature by using methods other than symmetry indicators. Such materials are still included as part of our process of high-throughput screening and ab initio validation. They are still relevant for evaluating how successful our approach is at diagnosing non-symmetry-diagnosable topological materials, but they are not relevant for discovering truly new topological materials (in the sense of identifying topological materials that are not yet known in the literature).

### C. Data Source and Preprocessing

As previously mentioned, we make use of the database generated by Tang *et al.* [S3]. This database is available at the following url: ccmp.nju.edu.cn

The materials in this database were present in the Inorganic Crystal Structure Database (ICSD) [S10] (they were at least present at the time that Tang *et al.* [S3] generated this database, but since ICSD sometimes gets updated they are not guaranteed to all still be present in ICSD). The ICSD number, which we will refer to, is an identifier of materials in ICSD.

The four relevant categories on the database website are "Topological insulators", "Topological crystalline insulators", "Topological (semi-)metals", and "Materials with band crossings in case 1". These categories on the database website do not represent complete characterizations of the materials and should actually be understood in terms of symmetry indicator categorizations. Specifically, using the terminology of [S7], the "Topological (semi-)metals" category corresponds to "Case 3" (i.e., bullet (A) in Supplementary Section S1.B), the "Topological insulators" and "Topological crystalline insulators" categories together correspond to "Case 2" (i.e., bullet (B) in Supplementary Section S1.B), and the "Materials with band crossings in case 1" category corresponds to a subset of "Case 1" (i.e., a subset of bullet (C) in Supplementary Section S1.B). So, for example, a material in the "Topological insulators" category on the database website might not actually be insulating since the symmetry indicator method ignores energetic aspects. For each of these four relevant categories, we scrape the database website to get the reduced formula and space group of each entry in the category, using the following criterion: if two or more entries in a given category on the website have the same reduced formula *and* same space group, then we collapse them into a single entry. By reduced formula, we mean the formula that one obtains by dividing out the greatest common divisor of the subscripts of a chemical formula (e.g., $O_5Ti_5$ would be expressed as $O_1Ti_1$ or equivalently OTi). Although the combination of reduced formula and space group does not uniquely identify a material (e.g., multiple different ICSD numbers can correspond to the same combination of reduced formula and space group), we use this combination in our approach for simplicity, as two materials with the same combination are typically fairly similar. Additionally, our machine learning model uses only the element fractions to make predictions (see Supplementary Section S2.A for details), and hence an overall scale factor in the chemical formula is not utilized by the model (so we do not need to keep track of this overall scale factor for the purpose of doing machine learning). The raw data that we scraped from the database website is included in our public repository (https://github.com/andrewma8/topogivity).

We next process the scraped data as follows. First, to form the set of NAI materials, we merge the following three sets of scraped data: the "Topological (semi-)metals" set, the "Topological insulators" set, and the "Topological crystalline insulators" set. We merge by taking the union of the three sets (so if the same combination of reduced formula *and* space group appears as an entry in more than one of the three original sets, it will appear as only a single entry in the NAI set). The set of USI materials is simply the "Materials with band crossings in case 1" set that we scraped from the database website. Next, any entry (i.e., combination of reduced formula and space group) that is present in both the NAI set and the USI set is removed from both sets. Partitioning the USI set into two sets based on space group type gives us the set of materials that are both USI and NT-IG (labeled data with trivial labels) as well as the set of materials that are both USI and T-IG (discovery space). Partitioning the NAI set into two sets based on space group type gives us the set of materials that are both NAI and NT-IG (labeled data with topological labels) as well as the set of materials that are both NAI and T-IG (used for an additional evaluation of

model performance). Additionally, we consider any element that occurs in a low number of the NT-IG materials to be a rare element. To improve reliability of the modeling, any material in any of the sets that contains at least one rare element is removed (although rare element is defined based on NT-IG materials, T-IG materials containing at least one rare element must also be removed since the learned model will not be able to be applied to elements that were never seen in the labeled dataset). There are 54 elements that occur among the materials that remain. Nominally, we used "occurring less than 25 times in the NT-IG materials" (prior to removal of rares) as the criterion for being a rare element, but we note that all 54 of the elements that were not rare occur in the remaining NT-IG materials at least 143 times (after removal of rares). This marks the end of data preprocessing, and the materials that still remain are the ones that we end up actually using. We used the pymatgen library [S11] to assist in preprocessing of this data.

Note that since two materials with the same reduced formula but different space groups are distinct entries, there are some reduced formulas that appear multiple times in the data even after preprocessing. Two particular instances of this are relevant to point out. First, there can be a data point with a topological label and a data point with a trivial label that have the same reduced formula. This limits the maximum accuracy that can be obtained using our machine learning model – in fact it limits the maximum accuracy of any model that only takes into account the reduced formula. This issue is analyzed in more depth in Supplementary Section S3.A. Second, there can be a data point with a topological label and an entry in the discovery space that have the same reduced formula. This issue is relevant to part of the discussion in Supplementary Section S4.C.

## D. Size and Composition of the Datasets

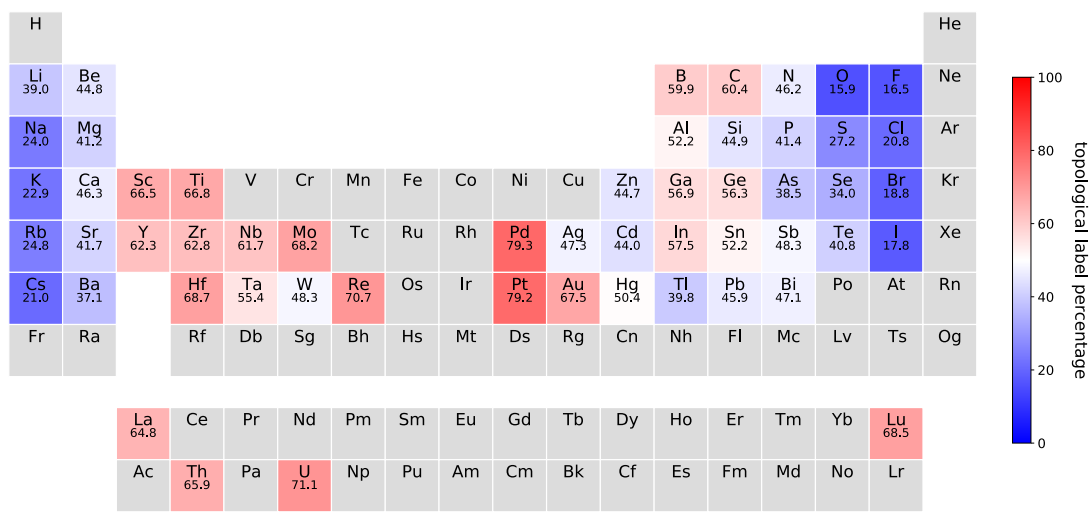

**Supplementary Figure S2. Topological label percentage for each element.** Percentages are shown by color-coding and in values. These are calculated using just the labeled dataset (after data preprocessing was completed). For each element, the value shown represents the percentage of materials with the topological label among the materials that contain that given element. The elements shown in gray are not present in the labeled dataset (after data preprocessing). Note that the labels are noisy.

After preprocessing, we are left with 9,026 materials in the labeled dataset, of which 4,604 are labeled as trivial and 4,422 are labeled as topological (recall that from our previous discussion, these should be regarded as noisy labels). As such, 49% of the materials among the labeled dataset have a topological label (see Fig. S1). Additionally, we have 1,433 materials in the discovery space and 238 materials for the additional evaluation of model performance.

For each element, we calculate the percentage of materials that are labeled as topological among the labeled dataset materials that contain that element. We visualize these percentages on the periodic table in Fig. S2. Note that Claussen *et al.* [S12] and Andrejevic *et al.* [S13] also include periodic table visu-

alizations for their datasets that illustrate the prevalence of topology for each element (the visualization in the former is for just topological insulators). The way that the visualization in Fig. S2 of our paper was generated differs from the ways that the visualizations in Claussen *et al.* [S12] and Andrejevic *et al.* [S13] were generated (e.g., different datasets), but we can observe some similarities in the trends.

### S2. DESCRIPTION, IMPLEMENTATION, AND EVALUATION OF THE MACHINE-LEARNED TOPOGIVITY MODEL

### A. Modeling Using the Topogivity Framework

Here we provide some more detail on our machine learning model. As stated in the main text, our model maps each material to a number using the function

$$g(M) = \sum_E f_E(M)\tau_E. \tag{S1}$$

Here $M$ denotes a material and $E$ denotes an element. $f_E(M)$ is the element fraction for element $E$ in material $M$ as determinable from the chemical formula (e.g., for the material $Na_3Bi$, $f_{Na}(Na_3Bi) = \frac{3}{4}$ and $f_{Bi}(Na_3Bi) = \frac{1}{4}$). $\tau_E$ is a parameter for each element that is learned from data, which we term an element's *topogivity*. The summation can be viewed as running over the elements that are present in material $M$. Equivalently, taking the natural definition that $f_E(M) = 0$ for all elements that are not present in $M$, then we can simply consider this summation to run over all elements $E$ that are present in the dataset. The classification decision for a given material $M$ is given by

$$\hat{y}(M) = \text{sign}[g(M)], \tag{S2}$$

where $\hat{y}(M) = 1$ corresponds to a classification as topological and $\hat{y}(M) = -1$ corresponds to a classification as trivial. Although $g(M)$ clearly does not represent a probability (it takes values outside of $[0, 1]$), it can provide us with some information about how confident we are in a classification decision. Specifically, in Supplementary Section S2.C we will show empirical evidence that as the magnitude of g(M) increases, the fraction of of correctly classified samples first increases and then plateaus. Empirically, it appears that up to a certain point, increasing magnitude $|g(M)|$ corresponds to increasing confidence in the classification decision. Beyond this point, the confidence does not noticeably appear to increase further with further increase in magnitude $|g(M)|$.

In practice, we will not learn a topogivity for every single element, since we will learn topogivities only for those elements that appear in our dataset. As such, if a material contains an element that does not appear in the dataset, then our actual learned model cannot make a prediction on it. However, this is a limitation of the dataset rather than the modeling approach itself, which could be extended to materials containing other elements if given a suitable dataset.

In order to make a diagnosis using our model, the only information that needs to be inputted are the element fractions (which can be determined from the chemical formula). *No spatial information is explicitly used.* However, this does not necessarily mean that spatial information does play an implicit role, since there are relationships between chemical composition and crystal structure [S14].

In the area of machine learning for materials science, there has been previous work on neural network methods that use only a material's chemical composition to make predictions [S15; S16]. More generally, there are a variety of approaches to machine learning for materials science [S17], some of which have been applied to the study of topological materials [S12; S13; S18; S19; S20; S21; S22; S23; S24; S25; S26; S27].

### B. Method for Learning Topogivities from Data

In this subsection, we will describe the method that we employed for learning the topogivities (i.e., the values of the parameters $\{\tau_E\}$ in Eq. (S1)) from data. Note, however, that there are also other possible

methods for learning the topogivities (i.e., other methods that are also compatible with the form of the model specified by Eqs. (S1) and (S2)). At the end of this subsection, we comment on a potential change to the method that might make it even more straightforward. Separately, in Supplementary Section S3.C, we empirically demonstrate the effect of making a particular change to the machine learning algorithm (including an analysis of how this change affects the learned values of the topogivities).

We define the element fraction vector $\mathbf{f}(M)$ for a material $M$ to be a vector that contains the element fraction $f_E(M)$ for each element. This includes element fractions for elements that are not present in the material $M$, which as we stated in Supplementary Section S2.A are defined to simply be zero. However, we don't include the element fractions for elements that are not present in the dataset at all (since those entries would just be zero for every single $M$ that appears in our dataset). Thus, for the case of our labeled dataset, $\mathbf{f}(M)$ is a vector with 54 entries, given explicitly as

$$\mathbf{f}(M) = (f_{\text{Li}}(M), f_{\text{Be}}(M), f_{\text{B}}(M), f_{\text{C}}(M), f_{\text{N}}(M), f_{\text{O}}(M), f_{\text{F}}(M), ..., f_{\text{U}}(M))^T \tag{S3}$$

where we have ordered the entries based on ascending atomic number solely for convenience. Next, we define $\tilde{\mathbf{f}}(M)$ as the vector that one obtains by taking $\mathbf{f}(M)$ and deleting one entry (i.e., it is a vector that contains all of the element fractions except for one). For our case, that means $\tilde{\mathbf{f}}(M)$ is a vector with 53 entries. $\tilde{\mathbf{f}}(M)$ still contains all of the information that was contained in $\mathbf{f}(M)$, since the entries of $\mathbf{f}(M)$ sum up to one and so the value of the entry that is missing in $\tilde{\mathbf{f}}(M)$ can be recovered. Heuristically, we choose to drop the entry that contains the element fraction for the element that occurs in the greatest number of materials in our labeled dataset, which is oxygen. Thus, in our case we have

$$\tilde{\mathbf{f}}(M) = (f_{\text{Li}}(M), f_{\text{Be}}(M), f_{\text{B}}(M), f_{\text{C}}(M), f_{\text{N}}(M), f_{\text{F}}(M), ..., f_{\text{U}}(M))^T. \tag{S4}$$

We will denote the $i$-th entry of $\tilde{\mathbf{f}}(M)$ as $\tilde{f}_i(M)$ (e.g., $\tilde{f}_6(M) = f_F(M)$).

Let $n_E$ be the number of elements present in the dataset (54 in the case of our labeled dataset). Now, define a function $h(M)$ parameterized by a vector $\mathbf{w} = (w_1, w_2, ..., w_{n_E-1})^T$ and scalar $b$:

$$h(M) = \mathbf{w}^T \tilde{\mathbf{f}}(M) + b \tag{S5}$$

For any given values of $\mathbf{w}$ and $b$, one can map to corresponding values of $\{\tau_E\}$ such that $g(\cdot)$ in Eq. S1 represents the *same* function as $h(\cdot)$ in Eq. S5. Moreover, there is a unique mapping such that this is the case. The mapping is given by:

$$\tau_E = \begin{cases} b, & \text{if } E = \tilde{E} \\ w_{\iota(E)} + b, & \text{otherwise} \end{cases} \tag{S6}$$

Here $\tilde{E}$ is the element whose element fraction appears in $\mathbf{f}(M)$ but does not appear in $\tilde{\mathbf{f}}(M)$. So for our specific implementation, $\tilde{E}$ is oxygen. $\iota(E)$ denotes the index of the entry of $\tilde{\mathbf{f}}(M)$ that contains the element fraction for element $E$. E.g., for our particular implementation, $\iota(\text{Li}) = 1$, $\iota(\text{N}) = 5$, $\iota(\text{F}) = 6$, and $\iota(\text{U}) = 53$.

The fact that Eq. (S6) is indeed a mapping such that $g(\cdot)$ and $h(\cdot)$ are the same function can be seen as follows. For clarity, we will use $\Omega$ to explicitly denote the set of all elements that are present in the dataset. The functions $h(\cdot)$ and $g(\cdot)$ have the same domain and the same codomain. For every $M$ in the

domain, we have:

$$
\begin{aligned}
g(M) &= \sum_{E \in \Omega} f_E(M) \tau_E \\
&= f_{\tilde{E}}(M) b + \sum_{E \in \Omega \backslash \{\tilde{E}\}} f_E(M)(w_{\iota(E)} + b) \\
&= b \sum_{E \in \Omega} f_E(M) + \sum_{E \in \Omega \backslash \{\tilde{E}\}} f_E(M) w_{\iota(E)} \\
&= b + \sum_{E \in \Omega \backslash \{\tilde{E}\}} \tilde{f}_{\iota(E)}(M) w_{\iota(E)} \\
&= b + \sum_{i=1}^{n_E - 1} \tilde{f}_i(M) w_i \\
&= b + \mathbf{w}^T \tilde{\mathbf{f}}(M) \\
&= h(M)
\end{aligned}
\tag{S7}
$$

So $g(\cdot)$ and $h(\cdot)$ are the same function.

The fact that the mapping in Eq. (S6) is the unique mapping such that $g(\cdot)$ and $h(\cdot)$ are the same function can be seen as follows. Suppose it were not unique, then there would be two different collections $\{\tau_E\}$ and $\{\tau'_E\}$ that correspond to the same function $h(\cdot) = g(\cdot)$, which is impossible since in general there cannot be two different collections corresponding to the same function $g(\cdot)$. (To see that in general there cannot be two different collections corresponding to the same function $g(\cdot)$, simply consider all of the pure element materials – specifying the function $g(\cdot)$ specifies the value of $g(M)$ for each pure element material, which specifies the value of $\tau_E$ for each element.)

As such, we can (i) first learn values of $\mathbf{w}$ and $b$ in the $h(\cdot)$ formulation and then (ii) map to the corresponding values of $\{\tau_E\}$ in the $g(\cdot)$ formulation using Eq. (S6). Additionally, note that since the functions $h(\cdot)$ and $g(\cdot)$ are mathematically equivalent when one maps according to Eq. (S6), it is not necessary to explicitly perform this mapping when simply trying to compute $g(M)$ for a given material $M$ (one can simply just compute it within the $h(M)$ formulation and get the same result). As such, in our code we only explicitly perform this mapping when we want to examine the values of the topogivities. Otherwise, in our code, we just calculate things within the $h(M)$ formulation.

The function $h(\cdot)$ can be thought of as follows: first map the material $M$ to its associated vector $\tilde{\mathbf{f}}(M)$, and then linearly map $\tilde{\mathbf{f}}(M)$ to a scalar $h(M)$ using $\mathbf{w}$ and $b$. Also, recall that we have a binary classification problem (in which the topological label and trivial label are represented using $+1$ and $-1$ respectively). Thus, we can learn the function $h(\cdot)$ (i.e., learn the values of the parameters $\mathbf{w}$ and $b$) by representing each material as its associated vector $\tilde{\mathbf{f}}(M)$, and then using a machine learning algorithm for learning a linear binary classifier.

There are many machine learning algorithms for fitting linear binary classifiers that could be compatible with our framework. In our work, we choose the soft-margin linear support vector machine (SVM), which we provide a brief overview of here. For more details on SVM, see e.g., [S28]. Soft-margin linear SVM is a supervised learning approach that can fit a linear model (i.e., optimize the values of the parameters $\mathbf{w}$ and $b$) given a set of $N$ data points (e.g., a training set) of the form $\{(\mathbf{x}^{(1)}, y^{(1)}), ..., (\mathbf{x}^{(N)}, y^{(N)})\}$, where $\mathbf{x}^{(i)} \in \mathbb{R}^p$ and $y^{(i)} \in \{-1, 1\}$. Mathematically, it can be formulated as the following optimization problem:

$$
\begin{aligned}
&\min_{\mathbf{w}, b, \xi^{(1)}, ..., \xi^{(N)}} \left( \frac{1}{N} \sum_{i=1}^{N} \xi^{(i)} + \gamma \|w\|^2 \right) \\
&\text{subject to } \begin{cases} y^{(i)}(\mathbf{w}^T \mathbf{x}^{(i)} + b) \geq 1 - \xi^{(i)} \\ \xi^{(i)} \geq 0 \end{cases}, \ i = 1, ..., N
\end{aligned}
\tag{S8}
$$

$\{\xi^{(1)}, ..., \xi^{(N)}\}$ are slack variables. $\gamma$ is the only hyperparameter. The above formulation is referred to as the primal problem, which also has a corresponding dual problem formulation. Mathematically, it can also

be reformulated as the following regularized empirical risk minimization problem for finding $\mathbf{w}$ and $b$:

$$\min_{\mathbf{w},b} L(\mathbf{w}, b), \tag{S9}$$

where

$$L(\mathbf{w}, b) = \frac{1}{N} \sum_{i=1}^{N} \max\left[0, 1 - y^{(i)}(\mathbf{w}^T \mathbf{x}^{(i)} + b)\right] + \gamma \|\mathbf{w}\|^2. \tag{S10}$$

From this regularized empirical risk minimization formulation, one can see that the hyperparameter $\gamma$ controls the regularization strength (greater $\gamma$ means greater regularization), and that the regularization penalizes $\mathbf{w}$ but does not penalize $b$. In our context, for a given data point corresponding to a material $M$ and label of either topological or trivial, $\mathbf{x}^{(i)} = \tilde{\mathbf{f}}(M)$ and $y^{(i)} = 1$ if topological and $y^{(i)} = -1$ if trivial.

Note that regularization could introduce some small artifacts into the learned topogivities. Regarding this point, in our context, a drawback of our particular approach to learning topogivities is that the way the topogivity of element $\tilde{E}$ (oxygen in our case) is treated by the regularization is different from the way the rest of the topogivities are treated by the regularization. We can see this as follows. Observe from Eq. (S6) that (i) for $E \neq \tilde{E}$ we have $w_{\iota(E)} = \tau_E - b$, and (ii) $b = \tau_{\tilde{E}}$. This means that for $E \neq \tilde{E}$, $w_{\iota(E)}$ represents the difference between the topogivity of element $E$ and the topogivity of element $\tilde{E}$. What is penalized by the regularization in Eq. (S10) is the square of $w_{\iota(E)}$ for each $E \neq \tilde{E}$. As such, the topogivity of element $\tilde{E}$ is treated differently by this regularization than the rest of the topogivities.

We implemented the soft-margin linear SVM using the scikit-learn library [S29] (specifically, we used the sklearn.svm.SVC class). We used the pymatgen library [S11] to assist with various tasks, such as creating the vectors $\tilde{\mathbf{f}}(M)$ and visualizing the topogivities $\tau_E$.

Note that Eq. (S1) can also be re-written as $g(M) = \boldsymbol{\tau}^T \mathbf{f}(M)$, where $\boldsymbol{\tau}$ is a vector that contains all of the topogivities. As such, in hindsight, it might also have been reasonable to learn the topogivities using an approach in which each material $M$ is represented as $\mathbf{f}(M)$, and then directly optimizing $\boldsymbol{\tau}$ using an algorithm for learning a linear binary classifier that does not have an intercept. In the approach that we actually implemented, one should think of the $h(\cdot)$ formulation merely as an intermediary that enables us to determine the topogivities in the $g(\cdot)$ formulation.

## C. Nested Cross Validation

Before we apply the model to the discovery space, we want to first evaluate how well it performs within the labeled dataset. By using a nested cross validation procedure, we are able to both (i) tune the hyperparamter $\gamma$ and (ii) evaluate our approach's performance on test sets within the labeled dataset.

In addition to the accuracy (the number of correctly classified samples divided by the total number of samples), we will also consider several other metrics. Note that our convention is that positive label means topological label ($y = +1$) and the negative label means trivial label ($y = -1$). We use TP to denote the number of true positives (i.e., number of samples with $\hat{y}(M) = 1$ and $y = 1$), FP to denote the number of false positives (i.e., number of samples with $\hat{y}(M) = 1$ and $y = -1$), TN to denote the number of true negatives (i.e., number of samples with $\hat{y}(M) = -1$ and $y = -1$), and FN to denote the number of false negatives (i.e., number of samples with $\hat{y}(M) = -1$ and $y = 1$). Then we have:

$$\text{recall} = \frac{\text{TP}}{\text{TP} + \text{FN}} \tag{S11}$$

$$\text{precision} = \frac{\text{TP}}{\text{TP} + \text{FP}} \tag{S12}$$

$$\text{F}_1 \text{ score} = 2 \times \frac{\text{recall} \times \text{precision}}{\text{recall} + \text{precision}} \tag{S13}$$

Additionally, we expect that the numerical value of $g(M)$ provides information beyond simply the classification decision (which is indicated by just the sign of $g(M)$), which we empirically characterize as

follows. For each $g(M)$ bin $B$ (i.e., interval of values for $g(M)$), we consider the topological fraction $\sigma(B)$, defined as

$$\sigma(B) = \frac{\text{number of samples with } g(M) \in B \text{ and } y = 1}{\text{number of samples with } g(M) \in B}, \tag{S14}$$

It indicates the fraction of samples that have the topological label ($y = 1$) among the samples that have a $g(M)$ value in $B$. The $g(M)$ bins that we will consider are $B = (-\infty, -2)$, $B = [-2, -1.5)$, $B = [-1.5, -1)$, $B = [-1, -0.5)$, $B = [-0.5, 0)$, $B = [0, 0.5)$, $B = [0.5, 1)$, $B = [1, 1.5)$, $B = [1.5, 2)$, and $B = [2, \infty)$. For the first five of these 10 bins, if a sample has a $g(M)$ value in $B$, then that sample is classified by the model as trivial. Thus, for these first five bins, $\sigma(B)$ indicates the fraction of *incorrectly* classified samples among the samples with $g(M) \in B$. For the other five of these 10 bins, if a sample has a $g(M)$ value in $B$, then that sample is classified by the model as topological (note that we implement the convention that sign[0] = +1, which is an arbitrary choice but in practice it is very unlikely that any material actually has $g(M) = 0$). Thus, for these other five bins, $\sigma(B)$ indicates the fraction of *correctly* classified samples among the samples with $g(M) \in B$.

We use the following $k \times (k - 1)$-fold nested cross validation procedure, with $k = 11$. We partition the labeled dataset into $k$ subsets of approximately equal size. We do this partitioning in a stratified way, so that each of the $k$ subsets has approximately the same ratio of topological labels to trivial labels. There are $k$ iterations in total in the outer loop, which we index as $i = 1, 2, ..., k$. In iteration $i$, we do the following:

1. Choose the $i$-th subset as the test set.

2. Perform inner cross validation on the remaining $(k - 1)$ subsets in order to select the value of the hyperparameter $\gamma$. Specifically, we consider a set of 75 values of the hyperparameter $\gamma$ that are evenly spaced on a log scale, ranging from $10^{-6}$ to $10^{-4}$. For each of these values of $\gamma$, we determine the corresponding mean validation $F_1$ score using $(k - 1)$ rounds of training and validation. In each round of training and validation, we do the the following:

   (a) Choose one of the remaining $(k - 1)$ subsets as the choice of validation set (this choice is different in each of the $(k - 1)$ rounds).

   (b) Merge the other $(k - 2)$ subsets into the training set. Fit the model's parameters on this training set.

   (c) Apply the fitted model from step (b) to the samples in the validation set and compute the validation $F_1$ score.

   Then, take the average of the validation $F_1$ scores obtained from these $(k - 1)$ rounds of training and validation to obtain the mean validation $F_1$ score for the given value of $\gamma$. The value of $\gamma$ with the greatest mean validation $F_1$ score is the selected hyperparameter value for this iteration of the outer loop.

3. Merge together the $(k - 1)$ subsets excluding subset $i$ into a single set (i.e., this merged set consists of all the samples that are not in the test set of this iteration of the outer loop). Using the value of the hyperparameter that was selected in step 2, fit the model's parameters on this merged set. We will refer to this step as "retrain" (e.g., retrain accuracy means the accuracy of the model fitted during this step evaluated on the samples that were used to fit it).

4. Apply the model that was fitted in step 3 to the test set, and compute all metrics of interest on this test set.

For each retrain metric of interest and test metric of interest, $k$ values were obtained in this nested cross validation process (one for each iteration of the outer loop). We calculate the mean and standard deviation of these $k$ values. The standard deviation for $k$ numbers $x_1, ..., x_k$ is calculated using the sample standard deviation given by $\sqrt{\frac{1}{k-1} \sum_{i=1}^{k} (x_i - \bar{x})^2}$ (where $\bar{x}$ is the sample mean).

The retrain and test results for accuracy, recall, precision, and $F_1$ score are shown in Table S1. It is worth pointing out that the mean test value of each metric in this table is only slightly lower than the mean retrain value of the metric.

| | Accuracy | Recall | Precision | $F_1$ score |
|---|---|---|---|---|
| **Retrain** | $83.1 \pm 0.1\%$ | $78.3 \pm 0.3\%$ | $85.9 \pm 0.2\%$ | $81.9 \pm 0.2\%$ |
| **Test** | $82.7 \pm 1.0\%$ | $78.0 \pm 1.8\%$ | $85.6 \pm 1.9\%$ | $81.6 \pm 1.1\%$ |

**Supplementary Table S1.** Nested cross validation results. The first row shows the mean $\pm$ standard deviation of each retrain metric. The second row shows the mean $\pm$ standard deviation of each test metric.

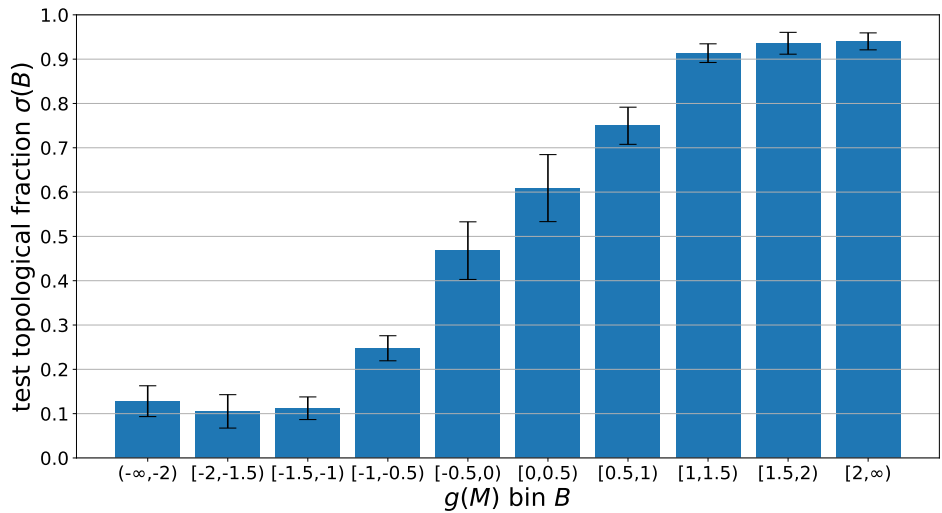

**Supplementary Figure S3. Topological fraction vs. $g(M)$ bin.** For each bin $B$ of $g(M)$ values, the height of the corresponding bar indicates the mean topological fraction $\sigma(B)$, and the error bars represent $\pm$ one standard deviation. Each mean and standard deviation was calculated from the $k = 11$ corresponding test set values obtained in the nested cross validation process.

In Fig. S3, we show the mean and standard deviation of the test results for the topological fraction $\sigma(B)$ corresponding to each $g(M)$ bin $B$. We observe evidence that in a certain range (roughly $-1$ to 1), increasing the value of $g(M)$ increases the fraction of materials that are topological. For $g(M)$ values below $-1$ (roughly) this fraction is low and appears to be relatively flat, and for $g(M)$ values above $+1$ (roughly) this fraction is high and also appears to be relatively flat. Additionally, we note that since we used soft-margin linear SVM to fit the model, $g(M) = 1$ and $g(M) = -1$ actually have some significance in the mathematical formulation (this can be seen e.g., from the loss function in the regularized empirical risk minimization formulation; see Eq. (S10)). Also note that if one were to change the machine learning algorithm, one of the effects could roughly be a re-scaling of the values of $g(M)$.

Another way to look at these topological fraction results is as follows. For the five bins on the left side of Fig. S3, $\sigma(B)$ represents the fraction of samples that are incorrectly classified among samples with $g(M) \in B$, and so the fraction of samples that are correctly classified among samples with $g(M) \in B$ is given by $1 - \sigma(B)$. And for the five bins on the right side of Fig. S3, $\sigma(B)$ directly represents the fraction of samples that are correctly classified among samples with $g(M) \in B$. Thus, we see evidence that as the *magnitude* $|g(M)|$ is increased, the fraction of correctly classified samples first increases and then plateaus. The plateau appears to begin around $|g(M)| \approx 1$ (granted this is a rough estimate, seeing as all bins except the first and last ones have a width of 0.5). Thus, the magnitude of $|g(M)|$ can provide us with useful information about how confident we are in the classification decision.

When the model says that a material $M$ has a value of $g(M) \geq 1.0$, we call it a high-confidence topologically nontrivial classification. Our heuristic choice of 1.0 as the threshold here is intended to make it so that high-confidence topologically nontrivial classifications have a high chance of being correct. In our nested cross validation process, we found that among samples with a high-confidence topologically nontrivial classification, the percentage that were correctly classified was $93.0 \pm 1.2\%$ (mean $\pm$ standard

deviation computed over the $k = 11$ test set results). This threshold will be used in our high-throughput screening process described in Supplementary Section S4.A (implications of this choice are discussed in Supplementary Section S4.C).

The values of the selected hyperparameter $\gamma$ in the nested cross validation procedure were (in ascending order of $\gamma$): $1.00 \times 10^{-6}$ (3 times), $1.13 \times 10^{-6}$ (1 time), $1.28 \times 10^{-6}$ (2 times), $1.45 \times 10^{-6}$ (1 time), $2.25 \times 10^{-6}$ (1 time), $3.26 \times 10^{-6}$ (1 time), $4.74 \times 10^{-6}$ (1 time), and $6.88 \times 10^{-6}$ (1 time).

Supplementary Section S3.B includes additional results from our nested cross validation experiment, which analyze the model performance versus the number of distinct elements in the material.

### D. Final Model

We now proceed to fit the final model. In our nested cross validation process (described in Supplementary Section S2.C), a value of the hyperparameter $\gamma$ was selected in each iteration of the outer loop, so there are a total of $k = 11$ selected values (not all of which are distinct). For fitting our final model, we chose the median of these 11 selected values, which is $\gamma = 1.28 \times 10^{-6}$. Using this value of $\gamma$, we fit the parameters of the final model on the entire labeled dataset. This final model obtained 82.9% accuracy, 78.2% recall, 85.8% precision, and 81.8% $F_1$ score when evaluated on the samples that were used for fitting it.

The values of the learned topogivities of this final model are what are shown in the periodic table of topogivities in Fig. 2 of the main text, which is what we analyzed in order to extract chemical insights. This final model is also what we will use in the high-throughput screening process described in Supplementary Section S4.A.

### E. Additional Evaluation of Model Performance

Here, we present an additional evaluation that characterizes our model in a different setting. We will make use of the set of 238 materials that are both T-IG and NAI (i.e., the materials that are marked "other" in Fig. S1.)

In this work, we chose to use only the NT-IG portion of the NAI materials as labeled data with topological labels. As we mentioned in Supplementary Section S1.B, it might also have been reasonable to just use all of the NAI materials as labeled data with topological labels. Since in this work we did not use the T-IG portion of the NAI materials in the labeled dataset, we can instead use this portion for an additional evaluation of performance by applying the model that was fitted on our labeled dataset to the materials in this portion. Specifically, if the model is working well, it would probably be reasonable to expect that it typically classifies these materials as topological.

We use the final model (i.e., the one that was fitted as described in Supplementary Section S2.D) to compute $\hat{y}(M)$ for each member of the set of materials that are both NAI and T-IG. We find that the model classifies 64.3% of the materials in this set as topological. Since here we are essentially treating this set as materials that are likely to be actually topological and then asking what fraction of this set is classified as topological by the model, it is interesting to compare this 64.3% number to the recall, which is the fraction of samples classified as topological among all the samples that have a label of topological ($y = 1$). In the nested cross validation process described in Supplementary Section S2.C, we found a test recall of $78.0 \pm 1.8\%$ (mean $\pm$ standard deviation). Thus, we observe empirical evidence of some deterioration in model performance when applying the model to the set of materials that are both NAI and T-IG. One possible source of this deterioration may be the fact that we are applying a model that was fitted on materials from one set of space groups to materials from another set of space groups. Although the spatial structure is not explicitly used as an input to the model, it is possible that our model does not have uniform performance across all spatial structure settings (e.g., since spatial structure is related to chemical composition [S14] and chemical composition actually is explicitly used).

## S3. ANALYSIS, PROPERTIES, AND LIMITATIONS OF THE TOPOGIVITY APPROACH

### A. Information Lost Due to Omission of Spatial Information

While we have emphasized the benefits of having such a simple model, it is worth considering the information lost due to these simplifications. One of the key simplifications is that we do not explicitly use spatial information. We analyze one of the implications of this omission in the present subsection.

After the data has been preprocessed as described in Supplementary Section S1.C, each data point is represented by a combination of reduced formula and space group, and no two data points have the same combination of reduced formula and space group. This is true when we consider the entirety of all the data (i.e., considering together the labeled dataset, the discovery space, and the T-IG NAI data); and so, in particular, it is also true for just the labeled dataset itself. Therefore, a hypothetical machine learning model that is allowed to use both the reduced formula and space group could in principle achieve 100% accuracy on the labeled dataset. However, if the machine learning model is not allowed to use the space group, then there are cases where two or more data points are guaranteed to have an identical representation for the machine learning model. This has an interesting implication in the situation where the data points with identical reduced formulas but different space groups do not all have the same label. In that case, the machine learning model is guaranteed to make misclassifications (even if one were considering training accuracy and the model were allowed to simply memorize data), because it must predict the same class for all of these data points, but not all of these data points have the same label.

Consequently, since the space group is excluded from the representation accessible to ML in our setting, the best that any hypothetical model can do in terms of accuracy on the labeled dataset is the following. For each distinct reduced formula that appears in the labeled dataset:

- If this reduced formula is associated with an equal number of topological labels and trivial labels, then either choice for what the model should predict for this reduced formula leads to the same amount of errors incurred (i.e., in terms of counting errors we can just make an arbitrary choice).

- If this reduced formula is associated with an unequal number of topological labels and trivial labels, then for this reduced formula the model should predict the label that is associated with it more times. (Note that this encompasses the case where only one of the labels occurs for that reduced formula, since in that case the other label occurs zero times.)

Therefore, we can count the total number of errors that must be incurred by determining $\min(n, m)$ for each distinct reduced formula and then computing their sum, where $n$ is the number of times the reduced formula occurs with a trivial label and $m$ is the number of times the reduced formula occurs with a topological label. The result is that we find that 203 errors must be incurred in the entire labeled dataset due to these considerations alone. A model that incurs *only* these errors on the entire labeled dataset would have an accuracy of 97.75% when used to classify the entire labeled dataset.

This analysis establishes an upper bound on the accuracy of any model that is not allowed to use space group information for the classification of the entire labeled dataset. This analysis illustrates one aspect of the information lost from not including spatial information. Of course, the omission of spatial information could worsen model accuracy in ways beyond those accounted for in this analysis. Further, for the particular case of the topogivity-based model that we use in this paper, a significant restriction on what kinds of classification decisions are possible comes from its linear form. Finally, even if a model does not use the space group information explicitly, it may still incorporate some aspects of spatial structure implicitly due to the fact that there are relationships between a material's chemical composition and its structure [S14].

### B. Decomposed Model Performance Based on Number of Distinct Elements in the Material

In Supplementary Section S2.C, we showed results for how well the machine learning model performs overall (i.e., its performance across *all* materials in the test sets). On the other hand, we can also consider

the more fine-grained question of how well the machine learning model performs on *particular kinds* of materials within the test sets, which can give us a better sense of when we should trust the model more and when we should trust it less. Here we explore this question by investigating how the model performance varies depending on the number of distinct elements present in the material.

| Subset of labeled dataset | 1 element | 2 elements | 3 elements | 4 elements | 5 elements | 6 elements |
|:---:|:---:|:---:|:---:|:---:|:---:|:---:|
| Number of materials | 213 | 3317 | 4483 | 944 | 67 | 2 |
| Topological percentage | 76.5% | 69.1% | 40.6% | 14.7% | 13.4% | 0.0% |

**Supplementary Table S2.** Size and makeup of labeled dataset subsets consisting of materials that have a given number of distinct elements. For each subset, we show the total number of materials it contains as well as the percentage of its materials that have a topological label.

Before showing the results for machine learning model performance, we first show some basic information about the composition of the labeled dataset when it is broken down into subsets in Table S2. Each subset contains all of the labeled dataset materials with a given number of distinct elements (e.g., the "2 elements" column is for all binary compounds in the labeled dataset). Shown are the number of materials in each subset as well as the percentage of materials with a topological label in each subset. We observe an interesting trend that as the number of distinct elements is increased, the percentage of materials with a topological label decreases. One caveat to this trend is that there is fairly little data for materials with five distinct elements and almost no data for materials with six distinct elements, so the listed percentages may be heavily affected by noise.

One obvious metric to evaluate is the accuracy for a given subset of the test materials (i.e., the fraction correctly classified when considering only the materials in that subset). However, since the individual subsets we will consider are heavily imbalanced in terms of the ratio of topological labels to trivial labels, the accuracy for the subset can be a fairly misleading metric (in contrast, when considering the entire labeled dataset, the ratio of topological labels to trivial labels happened to be close to 1:1). For instance, a naive model that just classifies every material as trivial would do quite well on the subset of test materials containing four distinct elements. An alternative metric that is designed to help address this type of issue is the balanced accuracy, which is given by

$$\text{balanced accuracy} = \frac{1}{2}\left(\frac{\text{TP}}{\text{TP} + \text{FN}} + \frac{\text{TN}}{\text{TN} + \text{FP}}\right). \tag{S15}$$

Note that in the special case of a set where there is an equal number of positive labels and negative labels, the balanced accuracy simply equals the accuracy. To get the balanced accuracy for a given subset of the test materials, TP (number of true positives), FP (number of false positives), TN (number of true negatives), and FN (number of false negatives) are evaluated using only the materials in that subset.

Note the counter-intuitive fact that it can be possible to partition a set of data points into subsets such that the balanced accuracy for each subset is lower than the balanced accuracy for the entire set. As a simple example of this, consider labels $\mathbf{y} = [1, 1, 1, -1, -1, -1, -1, 1]^T$ and predictions $\hat{\mathbf{y}} = [1, 1, 1, 1, -1, -1, -1, -1]^T$. The balanced accuracy for the entire set is 0.75. However, "balanced accuracy for the subset consisting of the first half of the data points" and "balanced accuracy for the subset consisting of the second half of the data points" are both equal to 0.5.

| Subset of test data | 1 element | 2 elements | 3 elements | 4 elements |
|:---:|:---:|:---:|:---:|:---:|
| Accuracy | 66.3 ± 8.4% | 81.5 ± 2.2% | 83.6 ± 1.1% | 86.4 ± 3.3% |
| Balanced accuracy | 55.6 ± 10.3% | 80.4 ± 2.9% | 82.5 ± 1.2% | 59.3 ± 4.9% |

**Supplementary Table S3.** Results for accuracy and balanced accuracy computed on subsets of test data consisting of materials with one, two, three, or four distinct elements. Shown are mean and standard deviations of test results from nested cross validation.

The results reported in this subsection are from the same nested cross validation experiment that we reported in Supplementary Section S2.C; they were simply not described in that part of the paper. The

accuracy and balanced accuracy for materials with a given number of distinct elements are determined by evaluation on the relevant subset of each test set during nested cross validation. (Note that only the test set evaluation is being decomposed based on number of distinct elements; the training, validation, and retraining of the model are not.) This gives $k = 11$ test values of accuracy and $k = 11$ test values of balanced accuracy for each given number of distinct elements. Shown in Table S3 are the means and standard deviations of test accuracy and test balanced accuracy for each given number of distinct elements. Due to the very limited data for materials with five or six distinct elements, we only reported results for the cases of materials with one, two, three, or four distinct elements. Specifically, there are only two total materials in the entire labeled dataset with six distinct elements. Even for materials with five distinct elements, there are only nine topological labels in the entire labeled dataset, and so there are not even enough topological labels for there to be at least one topological label in each test set (and so the test balanced accuracy is not even defined for every fold of nested cross validation).

We can see from the balanced accuracy row in Table S3 that the model performs well for materials with two distinct elements or three distinct elements, but does not perform that well for materials with one distinct element or four distinct elements. This provides information about the kinds of materials for which the model can be trusted more. Additionally, it suggests that at least in terms of predictive performance, there may be more room for improvement on the topogivity approach for materials with one or four distinct elements. Note that for four element materials, the accuracy is actually quite high (despite the balanced accuracy being low), but as we discussed earlier in this section the accuracy can be misleading when the ratio of labels is highly imbalanced (which is indeed the case for four element materials).

It is worth pointing out the percentages of materials that have two or three distinct elements for the data that we used in this work. In the labeled dataset, 36.7% of materials have two distinct elements and 49.7% of materials have three distinct elements. In the discovery space, 22.6% of materials have two distinct elements and 49.9% of materials have three distinct elements. Among the T-IG NAI materials (i.e., the materials that we used for an additional evaluation of model performance in Supplementary Section S2.E), 39.1% of materials have two distinct elements and 52.9% of materials have three distinct elements.

### C. Impact of Changing the Machine Learning Algorithm

In the method for learning topogivities described in Supplementary Section S2.B (which is the method used for the main results in this paper), one of the steps was to use soft-margin linear SVM to determine the values of $\mathbf{w}$ and $b$. There are also other machine learning algorithms capable of fitting a linear binary classifier other than soft-margin linear SVM. One could switch out soft-margin linear SVM for one of these other machine learning algorithms while still retaining the rest of the method described in Supplementary Section S2.B. In this subsection, we explore what happens when we switch out soft-margin linear SVM for logistic regression.

Specifically, we use logistic regression without regularization. We implement this with the scikit-learn library [S29] (using the sklearn.linear_model.LogisticRegression class).

| | Accuracy | Recall | Precision | $F_1$ score |
|---|---|---|---|---|
| **Train** | $83.5 \pm 0.2\%$ | $80.1 \pm 0.2\%$ | $85.3 \pm 0.2\%$ | $82.7 \pm 0.2\%$ |
| **Test** | $83.4 \pm 1.4\%$ | $80.0 \pm 2.3\%$ | $85.2 \pm 1.8\%$ | $82.5 \pm 1.5\%$ |

**Supplementary Table S4.** Cross validation results that are achieved when using logistic regression to fit the parameters. Results are shown in the format of mean $\pm$ standard deviation.

We are not tuning any hyperparameters in this subsection. (When we were using soft-margin linear SVM, the only hyperparameter we had to tune was the regularization strength $\gamma$, but here we are not using regularization.) As such, here we just do a $k$-fold cross validation procedure, rather than the nested cross validation procedure described in Supplementary Section S2.C. We use the exact same subsets that we used when doing nested cross validation for soft-margin linear SVM (i.e., the ones that were obtained

by partitioning the labeled dataset into 11 subsets, stratified so that the ratio of topological to trivial labels is approximately the same in each subset). There are $k = 11$ iterations in the cross validation procedure. In iteration $i$, the $i$-th subset is selected as test set, and the other $k - 1$ subsets are merged into the training set. The logistic regression model is fit on the training set and then applied to make classification decisions on the test set. For each metric, we end up with $k$ train values and $k$ test values. For both train and test for each metric, the mean and standard deviations computed over the associated $k$ values are shown in Table S4.

Comparing the test results for logistic regression in Table S4 with the test results for soft-margin linear SVM in Table S1, we can see that logistic regression yielded greater values for accuracy, recall, and $F_1$ score, while soft-margin linear SVM yielded a greater value for precision. Overall, it looks like logistic regression did slightly better than soft-margin linear SVM in terms of predictive performance, but the difference is minor and it is possible that it is just due to noise. Note that although the test results in this subsection provide some evidence that logistic regression is at least as good as soft-margin linear SVM, we stick to just showing the performance and visualization obtained from soft-margin linear SVM in the main text, since soft-margin linear SVM is what was used to fit the final model that was actually employed to screen for new topological materials in the discovery space.

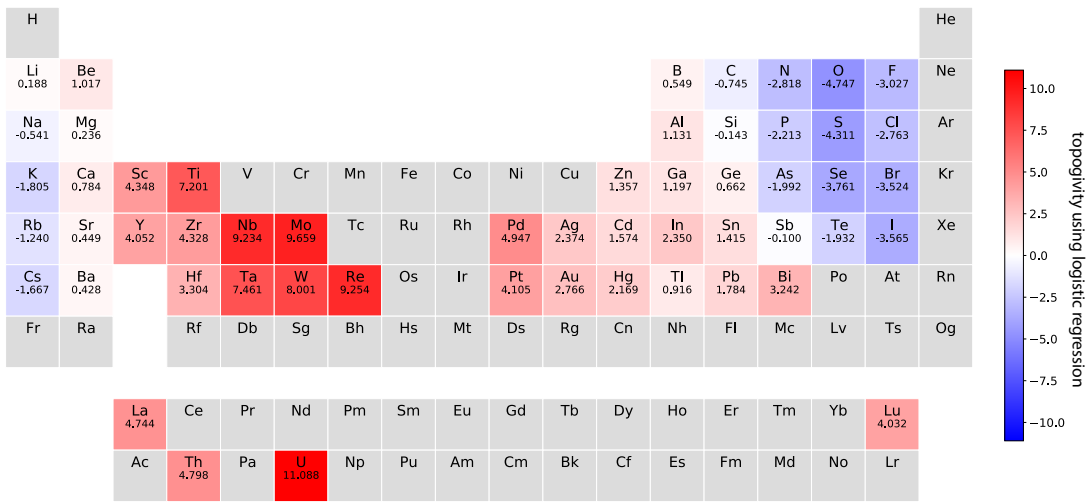

**Supplementary Figure S4. Topogivities learned using logistic regression.** Shown are the values of $\tau_E$ for each element $E$ that are obtained when using logistic regression instead of soft-margin linear SVM. The results are indicated both numerically and by color coding. Elements not present in the labeled dataset are shown in gray.

We then fit a final model on the entire labeled dataset using logistic regression. When evaluated on the same samples that were used to fit it, this model achieves 83.6% accuracy, 80.1% recall, 85.4% precision, and 82.7% $F_1$ score. In Fig. S4, we visualize the topogivities $\{\tau_E\}$ corresponding to this final model on the periodic table. Comparing Fig. S4 to Fig. 2 in the main text, we observe that the topogivities learned using logistic regression exhibit similar trends as the topogivities learned using soft-margin linear SVM.

For a given collection of learned topogivities, rescaling every topogivity by the same positive constant leaves the model's classification decisions unchanged. We can see this as follows. Substituting Eq. (S1) into Eq. (S2), we have

$$\hat{y}(M) = \text{sign}\left[\sum_E f_E(M)\tau_E\right].$$ (S16)

From Eq. (S16), we can see that for any $\alpha > 0$, setting $\tau_E \to \alpha\tau_E \ \forall E$ leaves $\hat{y}(M)$ unchanged.

As such, in order to more fairly compare the information learned by using logistic regression with the information learned by using soft-margin linear SVM, we normalize the values of each collection of learned topogivities. Specifically, let $\tau_E^{\text{logistic}}$ denote the final model topogivity of element $E$ learned

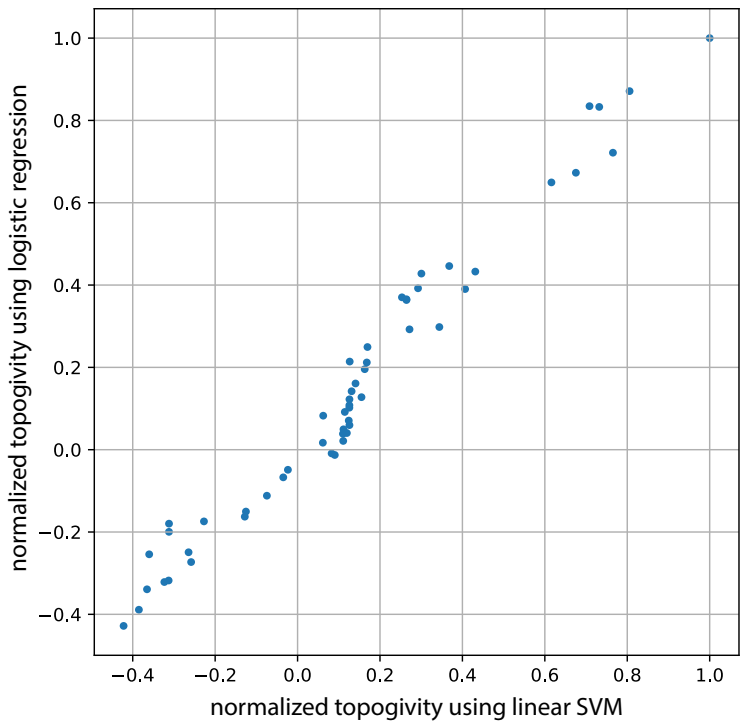

**Supplementary Figure S5. Comparison of topogivities learned using two different machine learning algorithms.** Plotted is the normalized topogivities obtained when using logistic regression ($\tau_E^{\text{logistic}}/|\tau^{\text{logistic}}|_{\text{max}}$) versus the normalized topogivities obtained when using soft-margin linear SVM ($\tau_E^{\text{SVM}}/|\tau^{\text{SVM}}|_{\text{max}}$).

by using logistic regression, and let $\tau_E^{\text{SVM}}$ denote the final model topogivity of element $E$ learned by using soft-margin linear SVM. Let $|\tau^{\text{logistic}}|_{\text{max}}$ denote the maximum magnitude of $\tau_E^{\text{logistic}}$ over all elements, and let $|\tau^{\text{SVM}}|_{\text{max}}$ denote the maximum magnitude of $\tau_E^{\text{SVM}}$ over all elements. In Fig. S5, we plot $\tau_E^{\text{logistic}}/|\tau^{\text{logistic}}|_{\text{max}}$ versus $\tau_E^{\text{SVM}}/|\tau^{\text{SVM}}|_{\text{max}}$. We see that although there are some differences between $\tau_E^{\text{logistic}}/|\tau^{\text{logistic}}|_{\text{max}}$ and $\tau_E^{\text{SVM}}/|\tau^{\text{SVM}}|_{\text{max}}$, the points on the scatter plot lie close to the $y = x$ line. This indicates that fairly similar information is learned by using soft-margin linear SVM and by using logistic regression.

See Supplementary Section S3.D for further discussion on how the results in this subsection shed light on the meaningfulness of the notion of topogivity.

## D. Discussion of the Topogivity-Based Picture

Conceptually, greater topogivity is intended to roughly correspond to a greater tendency to form topological materials. For example, the topogivity of bismuth ($\tau_{\text{Bi}} = 2.167$) suggests that bismuth likely has a relatively high tendency to form topological materials, whereas the topogivity of fluorine ($\tau_{\text{F}} = -2.061$) suggests that fluorine likely has a relatively low tendency to form topological materials (i.e., a relatively high tendency to form trivial materials). The sense in which topogivity captures a tendency to form topological materials is reflected empirically in our nested cross validation results. Specifically, we observed evidence that the fraction of topological materials increases as $g(M)$ is increased within a certain range (roughly from $-1$ to 1), and that this fraction is low for $g(M)$ values below $-1$ and high for $g(M)$ values above 1 (see Fig. S3). Since for a given material $M$, $g(M)$ is a weighted average of its elements' topogivities, these results provide empirical support for the the interpretation that greater topogivity loosely corresponds to a greater tendency to form topological materials.

However, it is also important to recognize that there may be certain limitations on the extent to which the learned values can be meaningfully interpreted. One reason is that the presence of correlations between occurrences and/or element fractions of different elements may influence the learned values of the topogivities. For example, if two elements often occur together in materials, then our modeling approach may have difficulty capturing whether both elements contribute to topology or just one of them (and if so, which one). Additionally, it is possible that the weighted average form of our model (as opposed to e.g., an unweighted average) could influence the values of learned topogivities. For instance, if one element typically occurs with a smaller element fraction than another element but both elements tend to form topological materials, then our modeling might favor a relatively greater topogivity for the first element in order to compensate for its smaller element fraction. Note that the typical oxidation states of an element could potentially be related to its typical element fractions. The learned values of topogivities that were found in our work should thus be understood while taking into account the context of our particular heuristic chemical rule. Moreover, given that our model is a *heuristic* rule that does not correctly diagnose every material, chemical insights extracted from the table of topogivities also should be viewed as providing a heuristic picture that cannot capture everything. One particular thing to keep in mind is that the model's predictive performance was much better for materials with two or three distinct elements than for materials with one or four distinct elements (see Supplementary Section S3.B), and so it is reasonable to think that the chemical insights extracted are more applicable for materials with two or three distinct elements.

We used a labeled dataset with noisy labels to fit our model (the noise arises from multiple sources, as discussed in Supplementary Section S1.B). Insofar as the distinction between topological labels and trivial labels is not completely the same as the distinction between true topological materials and true trivial materials, the learned topogivities likely better reflect the former distinction than the latter distinction. For example, hypothetically, if an element tends to form materials that are not atomic insulators but does not tend to form topological materials (e.g., assuming that one is using a definition for topological material that includes energetic considerations), then it would be quite possible that the element would get an undeserved positive topogivity in our fitted model. However, it is important to emphasize that these issues arising from the imperfections of the existing data are not inherent to the topogivity modeling approach itself. If provided a better labeled dataset, one could easily fit the topogivity parameters in Eq. (S1) on that labeled dataset.

The topogivity of an element is not actually an unambiguously defined quantity, since there are many ways to define parameters such that each parameter loosely captures the tendency of its corresponding element to form topological materials. It is worth pointing out that the lack of an unambiguous definition is not unique to the concept of topogivity. For example, there are multiple definitions of electronegativity, which lead to different numerical values for elements' electronegativities. In the case of topogivity, changing the form of the model and/or the machine learning algorithm can in some sense be thought of as changing the definition of topogivity. Regarding the form of the model, one could imagine changing it so that while one still has a parameter for each element, instead of the sign of a weighted average one gets the classification decision in another way (e.g., by the sign of an unweighted average). If one were to fit a model with a changed form on the labeled dataset, the learned values of the parameters would be different from those in our model, yet each parameter might still reasonably be interpreted as capturing something about the tendency of an element to form topological materials. Even for a fixed form of the model, one still has many choices for how to determine the values of the parameters within the model. Supplementary Section S3.C showed what happened when we kept the form of the model fixed but switched the machine learning algorithm from soft-margin linear SVM to logistic regression. We saw that the values of the normalized topogivities did indeed change. Moreover, we saw that the logistic regression model and the soft-margin linear SVM model had similar predictive performance on test sets, and so there is not a strong reason to prefer one or the other. As such, this indicates that one should use some caution and not be overly confident in interpreting the precise value of any individual element's topogivity. Nevertheless, the normalized topogivities from logistic regression and soft-margin linear SVM were still quite similar (see Fig. S5), which provides empirical support for the idea that the overall qualitative trends in the periodic table of topogivities can be meaningfully interpreted.

## S4. DETAILS ON THE HIGH-THROUGHPUT SCREENING AND AB INITIO VALIDATION PROCESS

### A. Procedure for Topogivity-Based Screening

Using the final model (i.e., the model that was fitted as described in Supplementary Section S2.D), we calculate $g(M)$ for each of the 1,433 entries in the discovery space. Of these 1,433 entries, 140 are classified as topological (i.e., $g(M) \geq 0$). Of the 140 entries that are classified as topological, 75 of them are high-confidence topologically nontrivial classifications (i.e., $g(M) \geq 1$).

From this list of 75 high-confidence topologically nontrivial classifications, a small number of entries are further removed for the following reasons:

- One entry is removed because it is not present in the ICSD (AgPb$_4$Pd$_6$, SG 152). Note that the ICSD is updated over time, so this material may have been present in the past.

- One entry is removed because it is not actually distinct from another entry that is also in the list of 75. Specifically, the two entries have the same reduced formula, and although they are listed as having different space groups, the actual structures corresponding to the entries are in fact identical. We removed the entry with the incorrectly listed space group (AsNb, SG 80).

- Five entries are removed because they contain a 4f or 5f electron (obtaining accurate DFT calculations for f-electron materials is difficult). The five were [PdSnU, SG 186], [GePtU, SG 44], [GeLuPd, SG 44], [CBTh, SG 91], and [Ge$_6$Lu$_4$Zn$_5$, SG 36].

The 68 entries that still remain constitute the list of materials for ab initio validation.

### B. Methods for Ab Initio Calculations

For topological material candidates, we first perform density functional theory calculation with Full-Potential Local-Orbital program (FPLO) [S30]. The band structure calculation is done with fine **k**-mesh including up to 50 points per line to exclude the possible anti-crossings. We then construct the symmetry adapted Wannier tight-binding models by projecting the Bloch wavefunction to localized Wannier functions. To identify topological semimetals such as Weyl semimetals and Dirac semimetals [S31; S32], we scan the entire Brillouin zone for degenerate points and calculate the Chern number over the sphere enclosing them [S33]. The energy criterion for topological crossings is set to be 200 meV around the Fermi level ($E_f - 200$ meV to $E_f + 200$ meV). When we identify both Dirac and Weyl nodes within 200 meV of the Fermi level in the same material, then the material is classified as a Dirac semimetal in our paper. For possible topological insulators without inversion symmetry, we perform Wilson loop calculation to obtain the $\mathbb{Z}_2$ index, and Berry curvature integral over wavefunction manifold indexed by mirror eigenvalue to obtain mirror Chern number [S9; S34; S35]. Note that we did not end up finding any topological insulators in the calculations that we performed on materials in the discovery space.

### C. Analysis and Discussion of the Strategy and its Performance

For our strategy, the success rate (i.e., the percentage of materials that were found to be topological among all of the materials that we performed DFT on) was 82.4%. The essence of our topogivity-based screening procedure was to compute $g(M)$ for each material in the discovery space and then restrict to those with $g(M) \geq 1$ (as a minor detail, there were also a small number of materials that were removed by other filters prior to DFT, as we described in Supplementary Section S4.A). As such, it is useful to compare this 82.4% success rate with our nested cross validation result that the percentage of samples with $g(M) \geq 1$ that were correctly classified was $93.0 \pm 1.2\%$ (mean ± standard deviation over test set results; see Supplementary Section S2.C). As such, we see that there was some deterioration in performance between how well the model did in the labeled dataset and how well it did in the discovery space. It

should be emphasized, however, that the materials in the discovery space and the materials in the labeled dataset correspond to *different regimes*, and so *a priori* it was unclear whether the model would succeed at all in the discovery space. In particular, the topological labels in the labeled dataset correspond to NAI diagnoses by symmetry indicators, whereas the topological materials in the discovery space are non-symmetry-diagnosable topological materials.

Our choice to screen for topological material candidates by restricting to materials with $g(M) \geq 1$ is not the only reasonable choice. In particular, one could choose some threshold $t$ and then screen using a criterion of restricting to the materials with $g(M) \geq t$. The choice we made in our strategy corresponds to choosing $t = 1$ as the threshold, which means that we are prioritizing precision over recall. Different choices might be suitable for different goals. For example, if the goal were instead to get a more comprehensive catalog of topological materials, then it would probably make sense to give some more consideration to recall by lowering the threshold $t$ (which would likely lead to a greater number of identified topological materials, but a lower success rate).

As mentioned in Supplementary Section S1.C, there are instances where a material that appears in the discovery space has the the same reduced formula as a material that appeared as labeled data with a topological label. This can occur because materials with the same reduced formula but different space group are distinct entries (which can have different symmetry indicator categorizations). These instances were included in our high-throughput screening and ab initio validation process, but excluding them would not have resulted in a lower success rate actually. Specifically, if these instances had instead been excluded, we would have had 43 successes out of 52 materials, which would correspond to a success rate of 82.7%.

As is often the case in data science, there are potential sources that may have introduced some bias into the results, and the actual amount of bias is difficult to account for quantitatively. For our success rate of 82.4%, there are possibilities of both upwards or downwards bias. Two sources that are worth noting are as follows. First, the amount of bias would probably be better reduced if all goals, procedures, and data analysis approaches were decided at the start of the project (i.e., prior to performing machine learning, or at least prior to seeing the model's predictions in the discovery space and portions of the DFT results). However, due to the nature of this project, many of these things were decided as we went along. For example, (i) the criteria for which materials to run in DFT calculations and (ii) how the success rate should be calculated (including, e.g., what counts as a topological semimetal) were not pre-planned in this work. Second, we did not exhaustively check for all possible nontrivial topology in our DFT calculations. As such, in principle it is possible that some of the candidate materials that were not explicitly identified as topologically nontrivial by our DFT could in fact be topologically nontrivial.

As we have previously discussed, the imperfections of the labeled dataset that was used to fit the model are not inherent to the topogivity-based modeling approach itself. With a better source of labeled data, the model would likely be better and there is a good chance that the success rate of the high-throughput screening and ab initio validation process would then be greater.

### S5. AB INITIO CALCULATIONS ON POTENTIALLY MISLABELED MATERIALS

As discussed in Supplementary Section S1.B, the labels in the labeled dataset should be considered "noisy" (i.e., some materials with trivial labels are actually topological and vice versa). Therefore, when the model disagrees with the label on a material, it is possible that it is actually the case that the model is correct and the material is mislabeled. This has implications for understanding the nature of the model that we developed. Moreover, the case where a material labeled as trivial is potentially topological is also interesting for the reason that this is potentially a source of candidate topological materials.

As a preliminary exploration, we consider cases in the labeled dataset where the label is trivial but the final model (the one fit on the entire labeled dataset as described in Supplementary Section S2.D) gives a high-confidence topologically nontrivial classification (i.e., $g(M) \geq 1.0$). We perform DFT calculations on a selection of these cases, and find that it indeed turns out that a portion of these materials are actually topological. We feature some selected examples of topological materials identified in this manner in

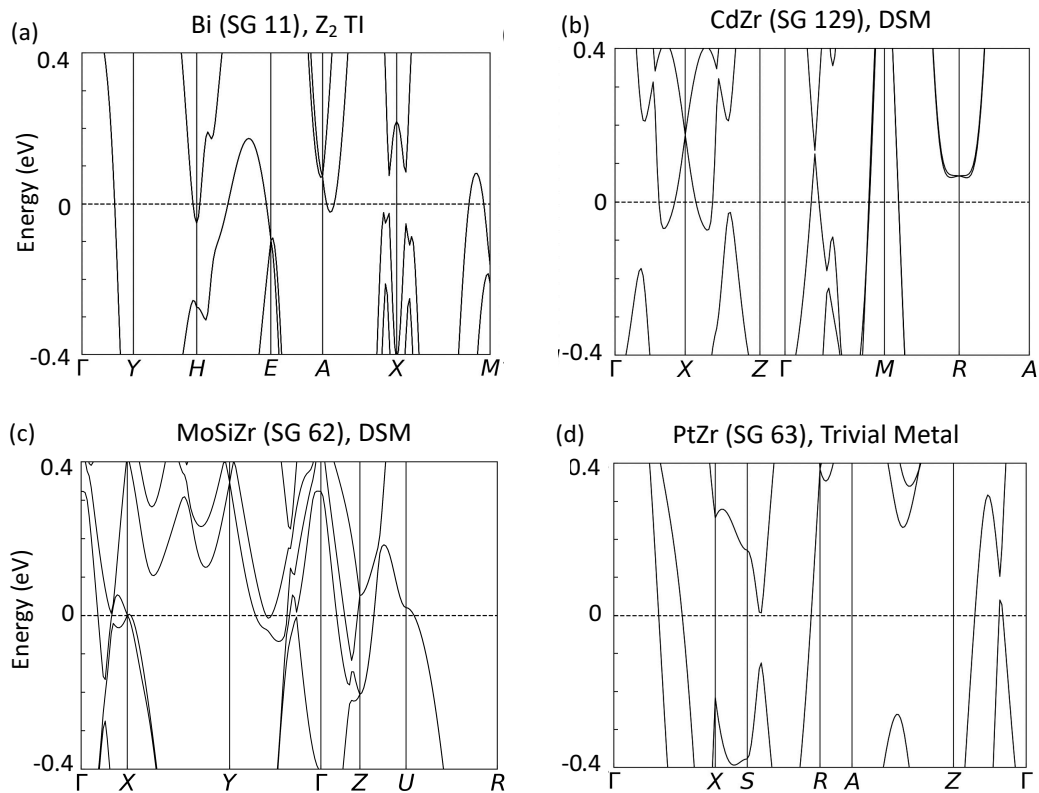

**Supplementary Figure S6. Selected results from ab initio calculations on potentially mislabeled materials.** Band structures computed using DFT are shown. Bi in SG 11 (**a**), CdZr in SG 129 (**b**), and MoSiZr in SG 62 (**c**) are examples of materials that were found to actually be topological (i.e., the trivial label was indeed actually wrong). PtZr in SG 63 is an example of a material that was found to be truly trivial (i.e., the trivial label was correct).

Fig. S6(a)-(c). Specifically, Bi in SG 11 is a $Z_2$ topological insulator (TI), CdZr in SG 129 is a Dirac semimetal, and MoSiZr in SG 62 is also a Dirac semimetal. To illustrate the other possibility, in Fig. S6(d), we show one example where the material turned out to be truly trivial in our DFT calculations – PtZr in SG 63 is a trivial metal.

The selected results from the preliminary exploration presented here are intended just to illustrate for the case of trivial labels that there are indeed cases where an apparent model misclassification actually corresponds to the situation where the material is mislabeled. We did not perform a systematic study that allows us to draw conclusions on the success rate of our approach, and in any case there may be more efficient ways to identify such cases than the crude approach we took here. Developing strategies for identifying topological materials among the materials with trivial labels in the labeled dataset could be an interesting direction for future research in topological materials discovery.

## S6. CATALOG OF TOPOLOGICAL MATERIALS IDENTIFIED BY THE HIGH-THROUGHPUT PROCESS

56 materials were identified as topological by our high-throughput screening and ab initio validation process. We list all of these materials in Table S5. The computed band structures of these materials are shown in Figs. S7-S62. Figs. S7-S54 correspond to Weyl semimetals (WSMs), Figs. S55-S61 correspond to Dirac semimetals (DSMs), and Fig. S62 corresponds to a Dirac nodal line semimetal (DNLSM). Note that some of these materials have already previously been identified as topological by other works in the literature.

**Weyl semimetals**

| Formula | Space group | ICSD ID |
|---|---|---|
| AlGeLa | 109 | 105149 |
| $Al_{22}Mo_5$ | 43 | 400888 |
| AsNb | 109 | 16585, 44027 |
| $AsPd_5$ | 5 | 44042 |
| AsTa | 109 | 44068, 611457 |
| AuCaSn | 44 | 195943 |
| $Au_2Ga$ | 36 | 58459 |
| $BLaPt_2$ | 180 | 98425 |
| $BaGe_3Pt$ | 107 | 174262, 409867 |
| $BaPdSi_3$ | 107 | 174266 |
| $BaPdSn_3$ | 107 | 58673 |
| $BaPtSi_3$ | 107 | 174267 |
| $BaPtSn_3$ | 107 | 58677 |
| $BiPbPd_2$ | 36 | 56278 |
| BiPd | 4 | 54976 |
| $Bi_2Pt$ | 157 | 195775, 428088 |
| $Bi_3Pd_8$ | 146 | 616947 |
| $CaGe_2Pt_2$ | 4 | 619327 |
| $CaPtSi_3$ | 107 | 181448 |
| $Ca_2GePd_2$ | 43 | 251662 |
| $Ca_3Cd_2$ | 102 | 30082 |
| $GePd_5$ | 5 | 637537 |
| $Ge_2Nb$ | 180 | 637208 |
| $Ge_3PdSr$ | 107 | 168862 |
| $Ge_3PtSr$ | 107 | 168863 |
| $Ge_4Zr_5$ | 92 | 638154, 638164 |
| $Hf_5Si_4$ | 92 | 197030, 53043, 638914, 638928 |
| HgPd | 198 | 40321 |
| InSr | 43 | 414234 |
| $La_5Si_4$ | 92 | 247813 |
| MgPt | 198 | 109237, 642775 |
| NbP | 109 | 645167, 645171, 81493 |
| NbReSi | 46 | 600059 |
| $OTi_3$ | 149 | 36055 |
| PTa | 109 | 196965, 648185 |
| $Pb_3Pd_5$ | 5 | 648361 |
| $PdSn_3Sr$ | 107 | 105692 |
| $Pd_2Sb$ | 36 | 77889 |
| $Pd_8Sb_3$ | 146 | 41748, 655017 |
| $Pd_8Sb_3$ | 161 | 648777, 77891 |
| ReSiTa | 46 | 600060 |
| $Re_2ScSi_3$ | 38 | 41742 |
| $Re_2Sc_3Si_3$ | 5 | 77997 |
| $Re_8Sc_5Si_{12}$ | 38 | 5412, 5414, 62203 |
| $Sc_9Te_2$ | 36 | 421464 |
| SiTi | 25 | 166580, 20375, 652436 |
| $Si_4Zr_5$ | 92 | 20357, 43214 |
| $Ta_{21}Te_{13}$ | 183 | 91811 |

**Dirac semimetals**

| Formula | Space group | ICSD ID |
|---|---|---|
| BiPd | 36 | 56279 |
| $CAl_2Mo_3$ | 213 | 42917 |
| $Ge_6La_4Mg_5$ | 36 | 262219 |
| $Ge_8Y_4Zn_5$ | 36 | 425800 |
| $MoPt_2Si_3$ | 26 | 174199 |
| $OTi_6$ | 159 | 17009 |
| $Pd_5Sb_2$ | 185 | 648767, 648776 |

**Dirac nodal line semimetals**

| Formula | Space group | ICSD ID |
|---|---|---|
| $Al_2Y_3$ | 102 | 609643 |

**Supplementary Table S5.** Enumeration of the 56 topological materials that were identified by our high-throughput screening and ab initio validation process. The topological materials consist of 48 Weyl semimetals, 7 Dirac semimetals, and 1 Dirac nodal line semimetal.

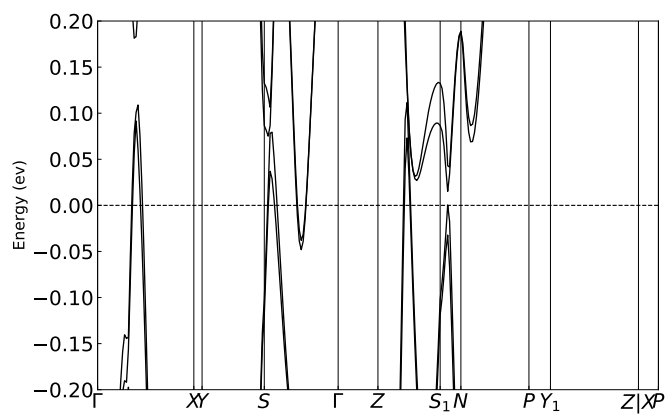

**Supplementary Figure S7.** AlGeLa, SG 109, WSM

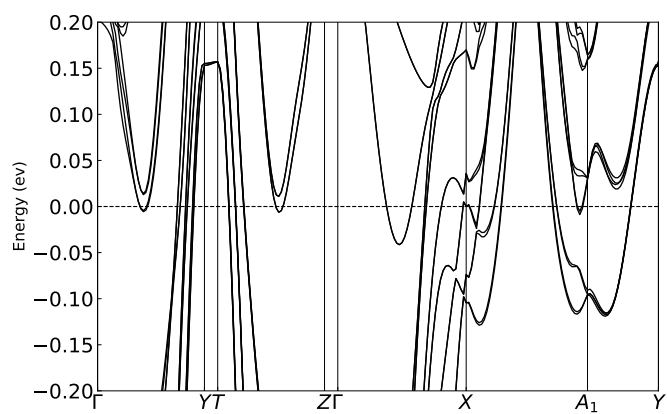

**Supplementary Figure S8.** Al$_{22}$Mo$_5$, SG 43, WSM

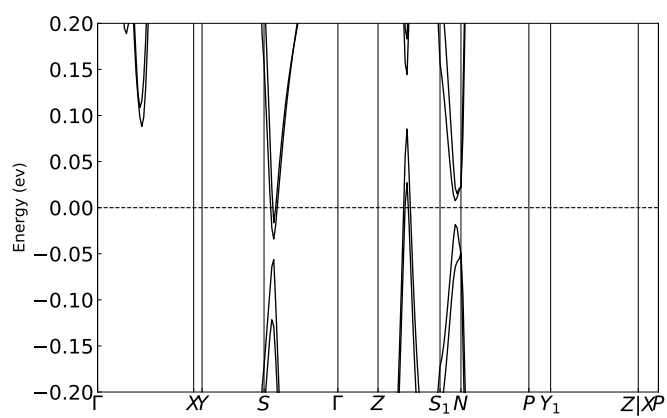

**Supplementary Figure S9.** AsNb, SG 109, WSM

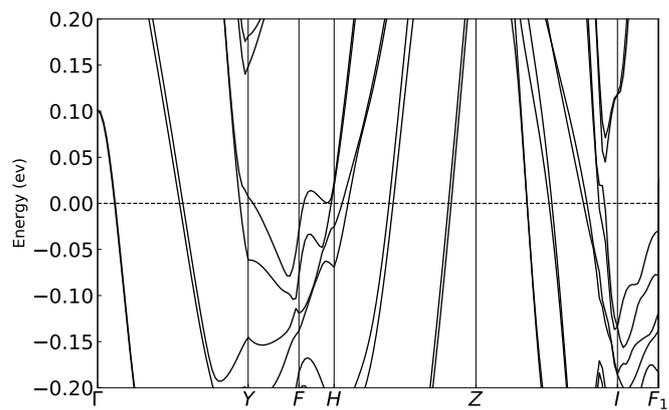

**Supplementary Figure S10.** AsPd$_5$, SG 5, WSM

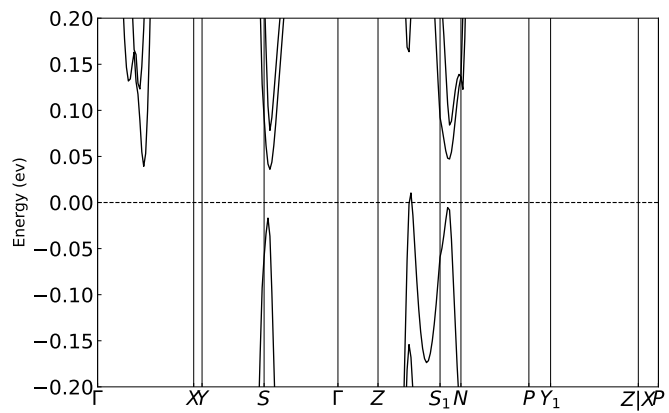

**Supplementary Figure S11.** AsTa, SG 109, WSM

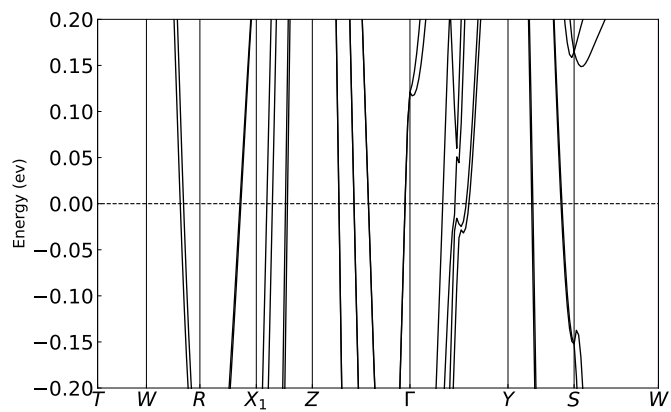

**Supplementary Figure S12.** AuCaSn, SG 44, WSM

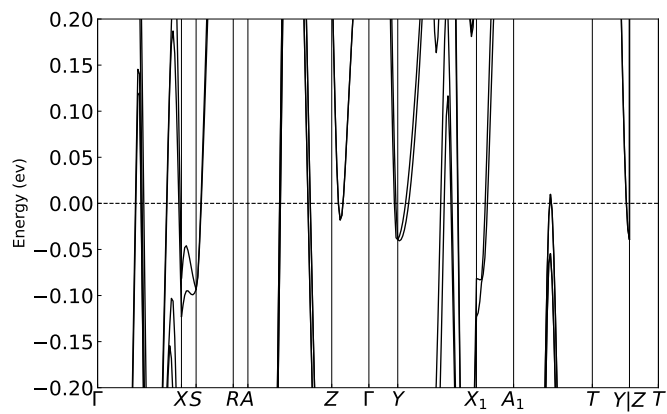

**Supplementary Figure S13.** Au$_2$Ga, SG 36, WSM

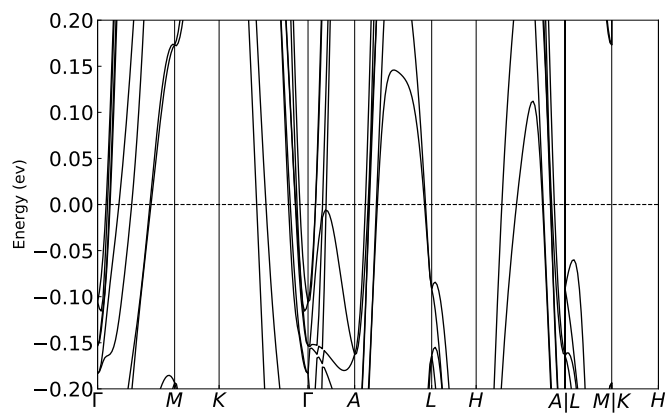

**Supplementary Figure S14.** BLaPt$_2$, SG 180, WSM

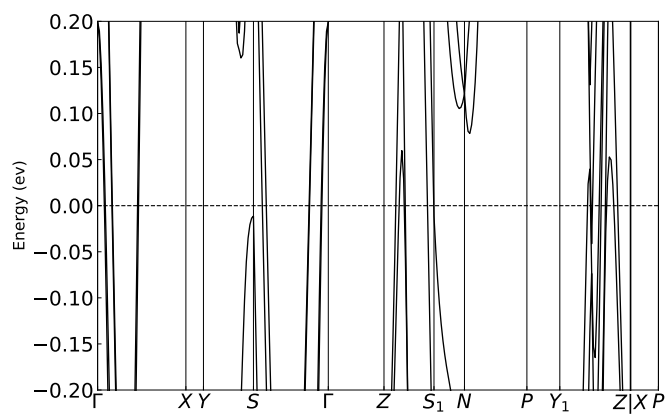

**Supplementary Figure S15.** BaGe$_3$Pt, SG 107, WSM

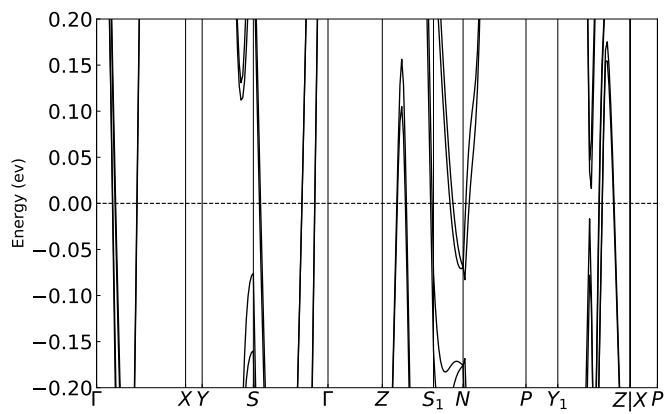

**Supplementary Figure S16.** BaPdSi₃, SG 107, WSM

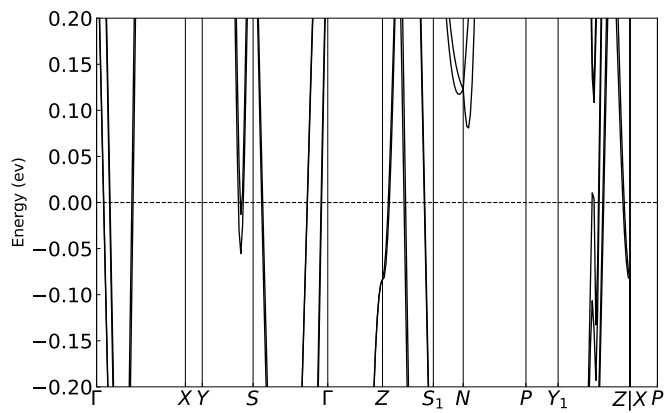

**Supplementary Figure S17.** BaPdSn₃, SG 107, WSM

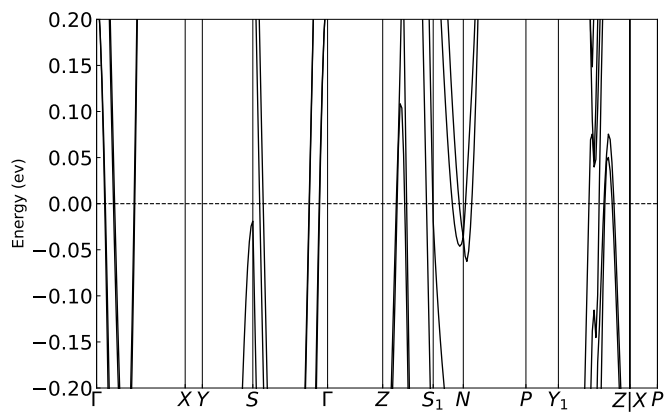

**Supplementary Figure S18.** BaPtSi₃, SG 107, WSM

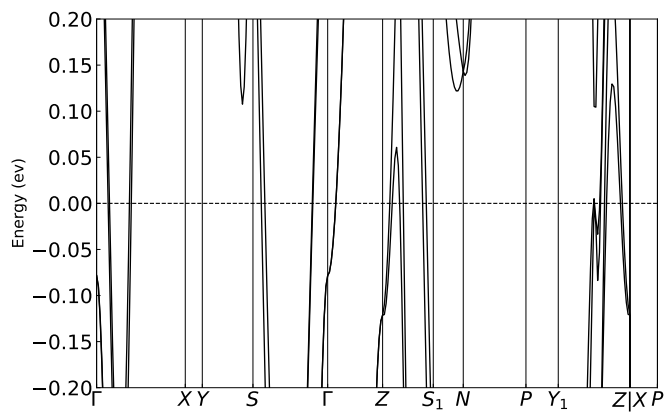

**Supplementary Figure S19.** BaPtSn$_3$, SG 107, WSM

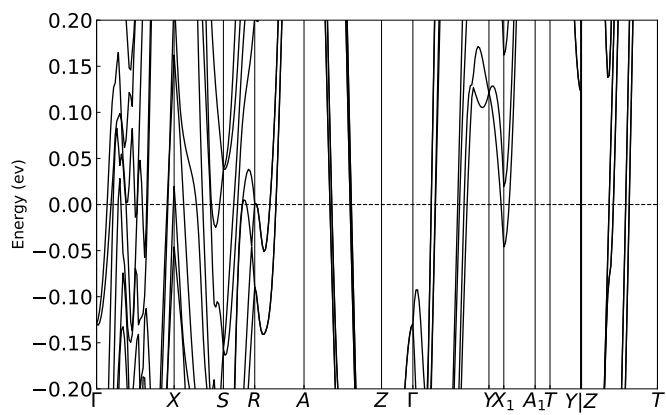

**Supplementary Figure S20.** BiPbPd$_2$, SG 36, WSM

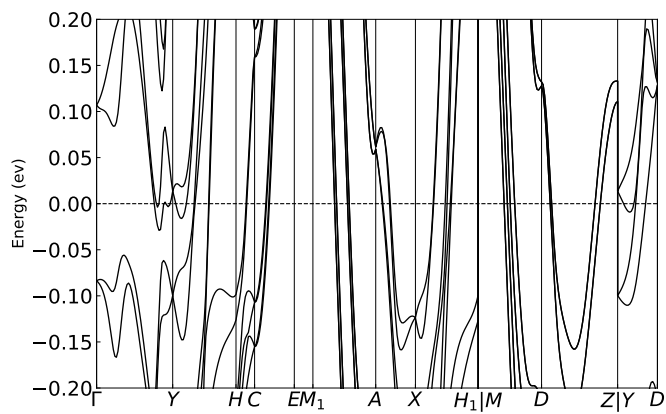

**Supplementary Figure S21.** BiPd, SG 4, WSM

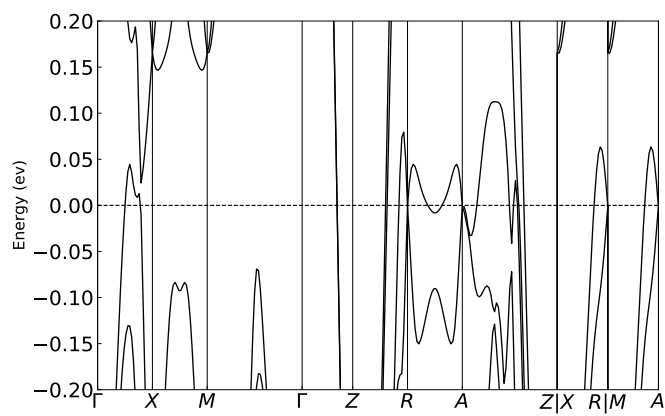

**Supplementary Figure S22.** Bi$_2$Pt, SG 157, WSM

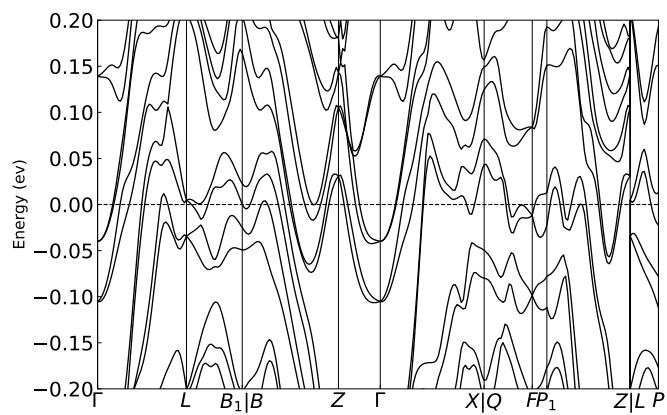

**Supplementary Figure S23.** Bi$_3$Pd$_8$, SG 146, WSM

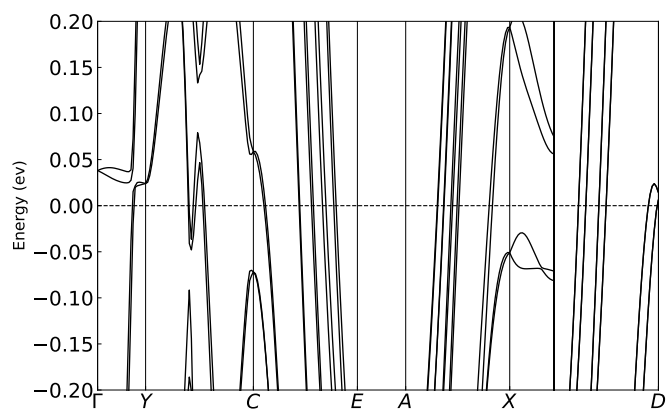

**Supplementary Figure S24.** CaGe$_2$Pt$_2$, SG 4, WSM

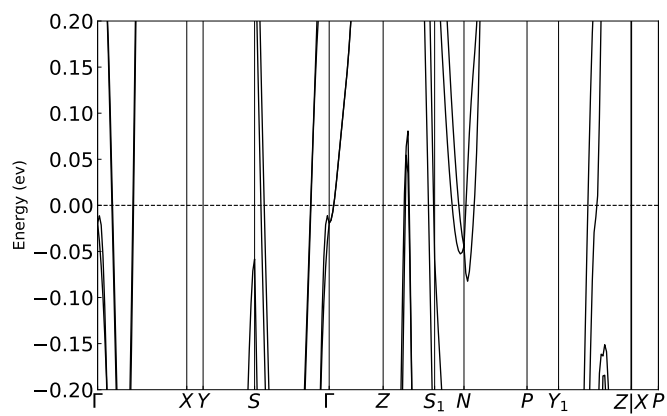

**Supplementary Figure S25.** CaPtSi$_3$, SG 107, WSM

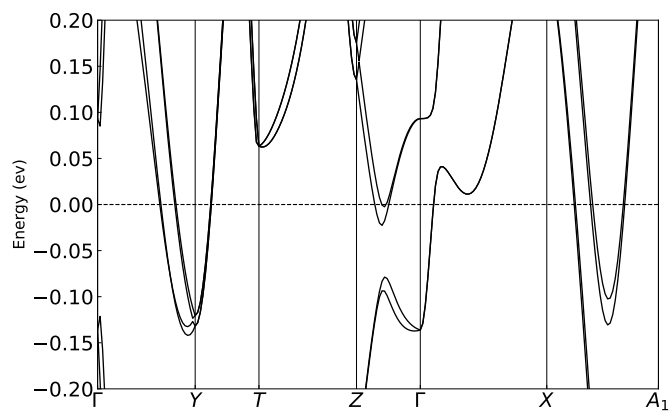

**Supplementary Figure S26.** Ca$_2$GePd$_2$, SG 43, WSM

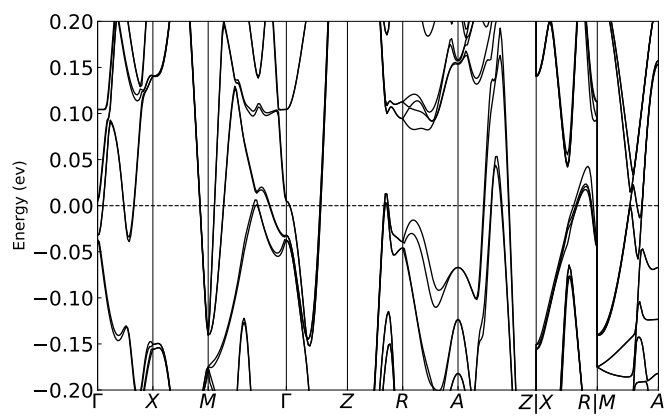

**Supplementary Figure S27.** Ca$_3$Cd$_2$, SG 102, WSM

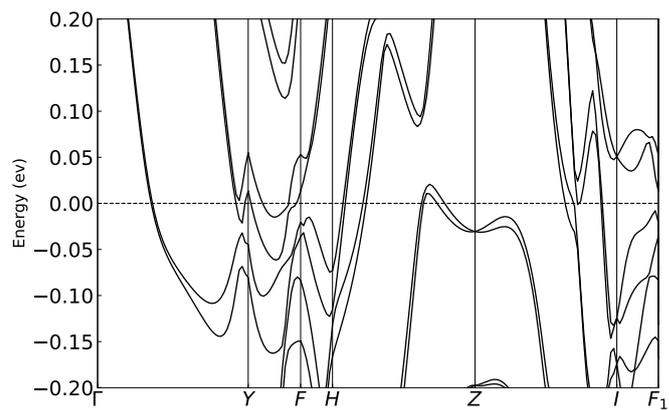

**Supplementary Figure S28.** GePd$_5$, SG 5, WSM

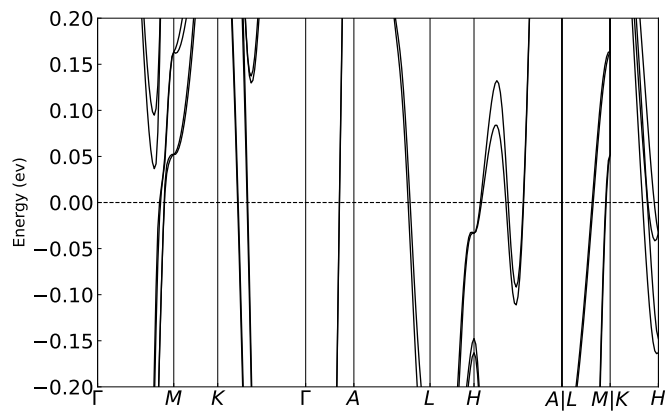

**Supplementary Figure S29.** Ge$_2$Nb, SG 180, WSM

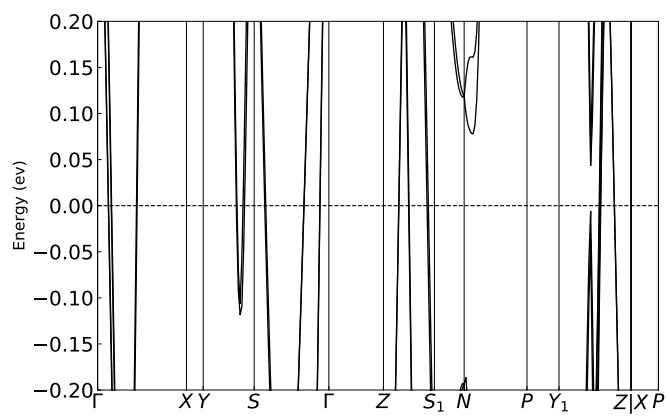

**Supplementary Figure S30.** Ge$_3$PdSr, SG 107, WSM

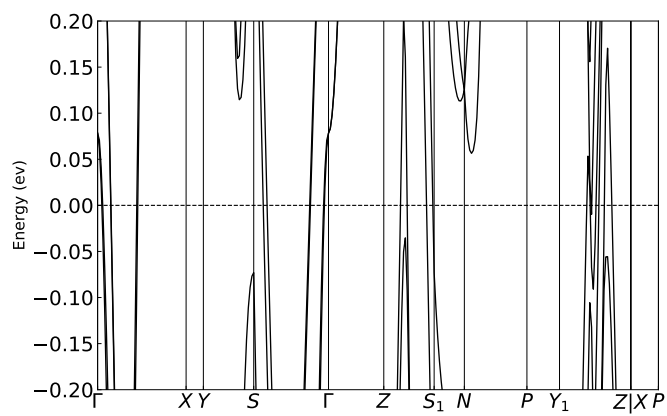

**Supplementary Figure S31.** Ge$_3$PtSr, SG 107, WSM

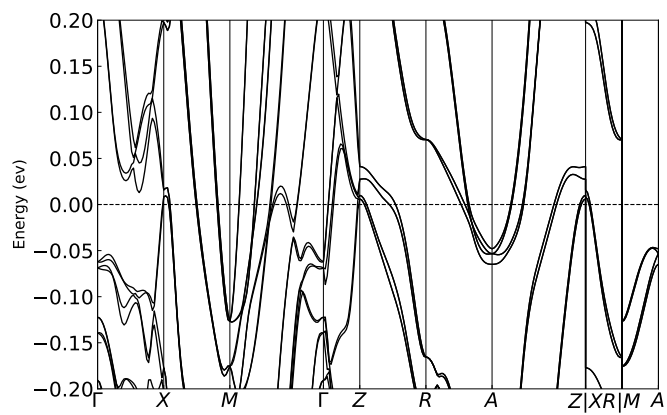

**Supplementary Figure S32.** Ge$_4$Zr$_5$, SG 92, WSM

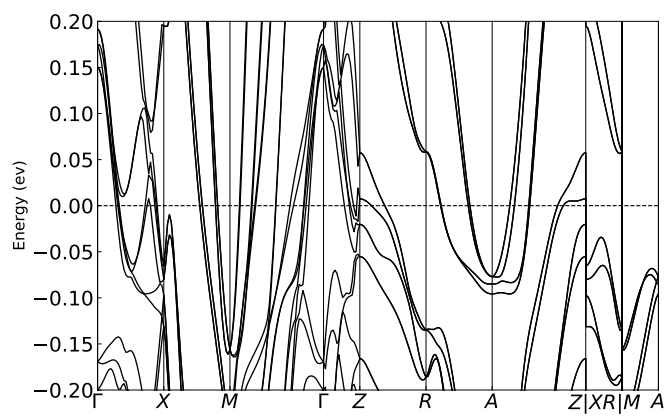

**Supplementary Figure S33.** Hf$_5$Si$_4$, SG 92, WSM

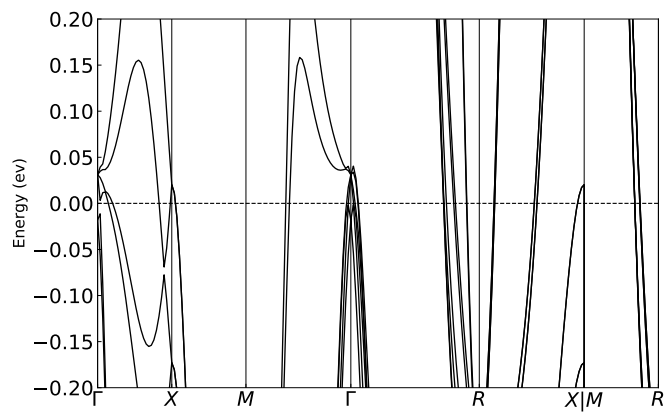

**Supplementary Figure S34.** HgPd, SG 198, WSM

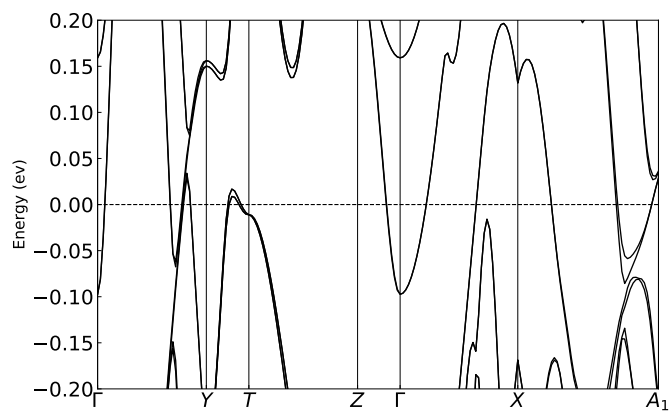

**Supplementary Figure S35.** InSr, SG 43, WSM

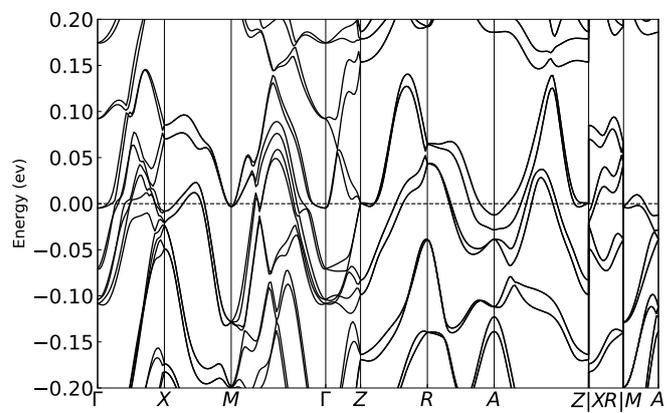

**Supplementary Figure S36.** La$_5$Si$_4$, SG 92, WSM

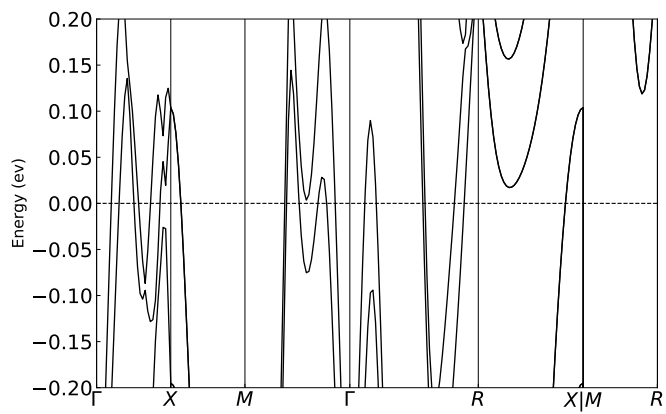

**Supplementary Figure S37.** MgPt, SG 198, WSM

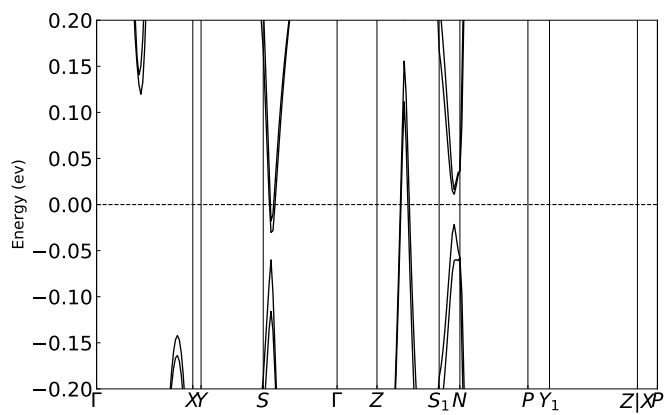

**Supplementary Figure S38.** NbP, SG 109, WSM

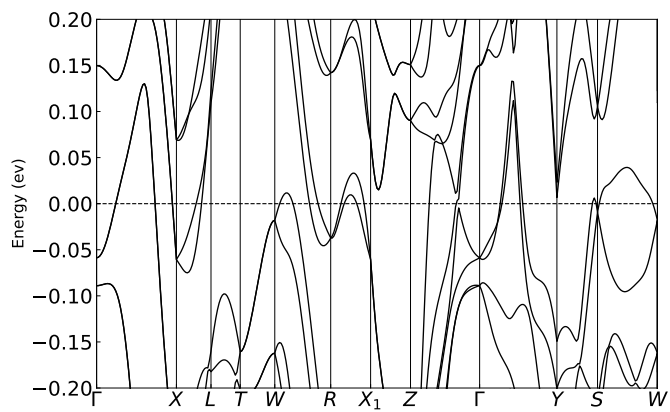

**Supplementary Figure S39.** NbReSi, SG 46, WSM

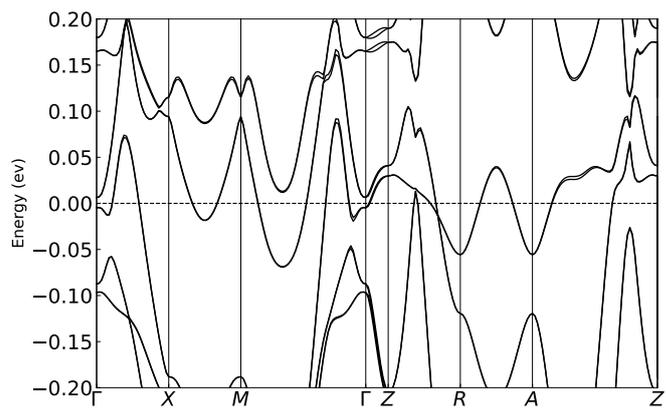

**Supplementary Figure S40.** OTi$_3$, SG 149, WSM

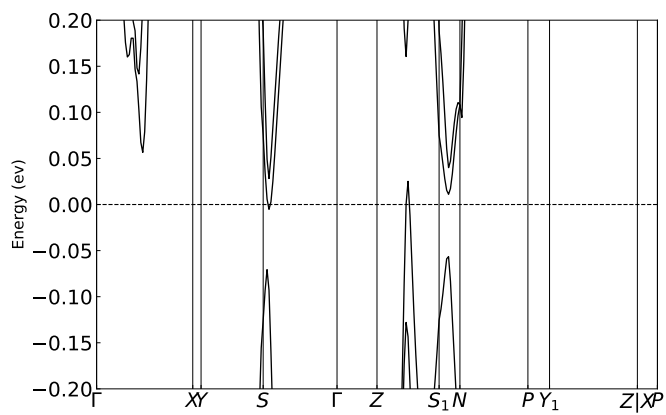

**Supplementary Figure S41.** PTa, SG 109, WSM

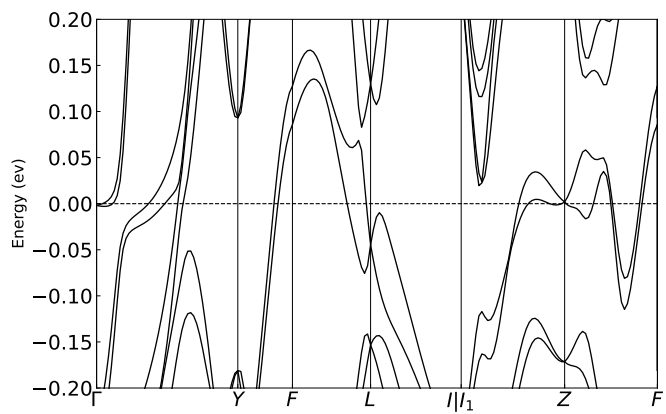

**Supplementary Figure S42.** Pb$_3$Pd$_5$, SG 5, WSM

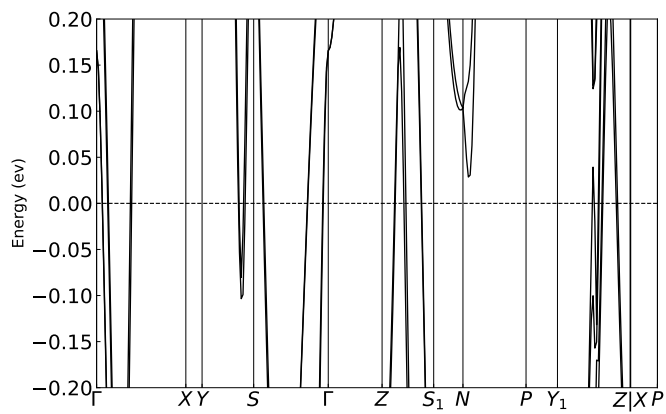

**Supplementary Figure S43.** PdSn$_3$Sr, SG 107, WSM

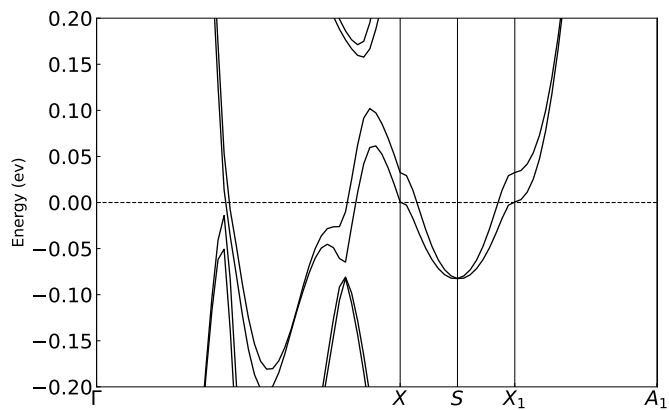

**Supplementary Figure S44.** Pd$_2$Sb, SG 36, WSM

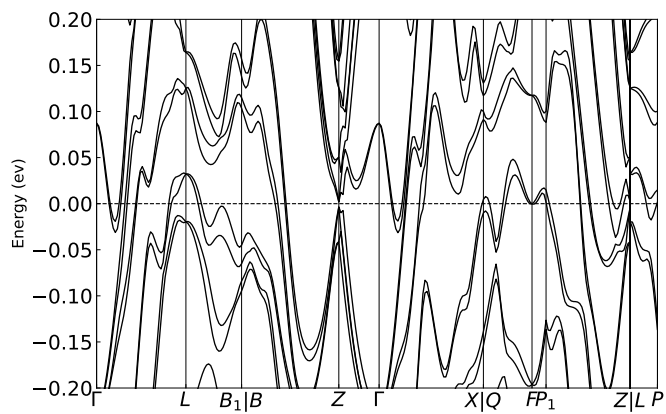

**Supplementary Figure S45.** Pd$_8$Sb$_3$, SG 146, WSM

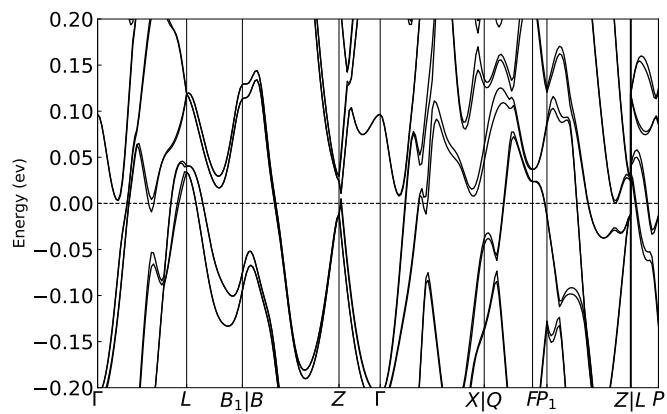

**Supplementary Figure S46.** Pd$_8$Sb$_3$, SG 161, WSM

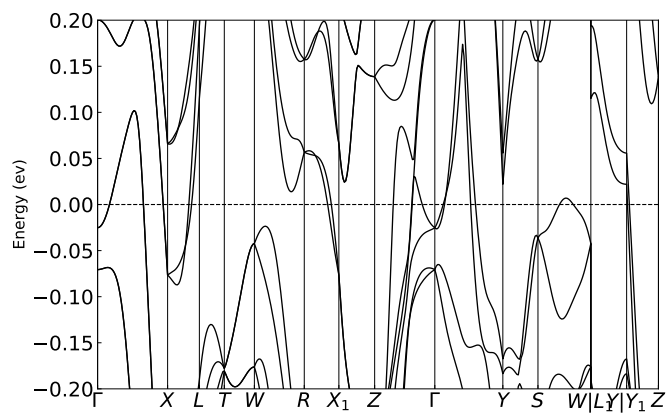

**Supplementary Figure S47.** ReSiTa, SG 46, WSM

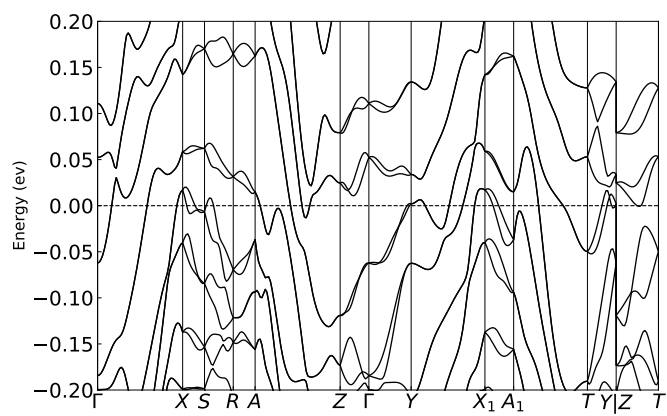

**Supplementary Figure S48.** Re$_2$ScSi$_3$, SG 38, WSM

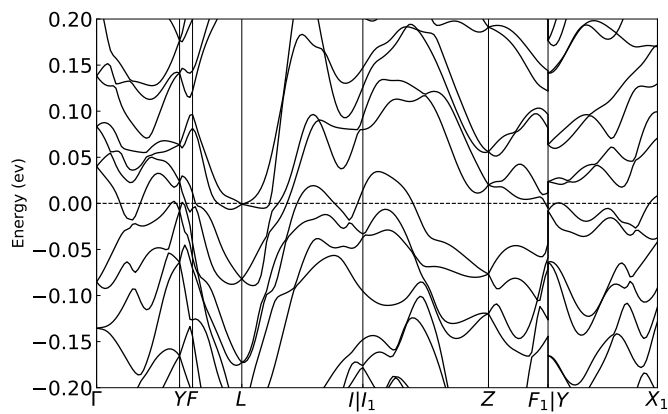

**Supplementary Figure S49.** $Re_2Sc_3Si_3$, SG 5, WSM

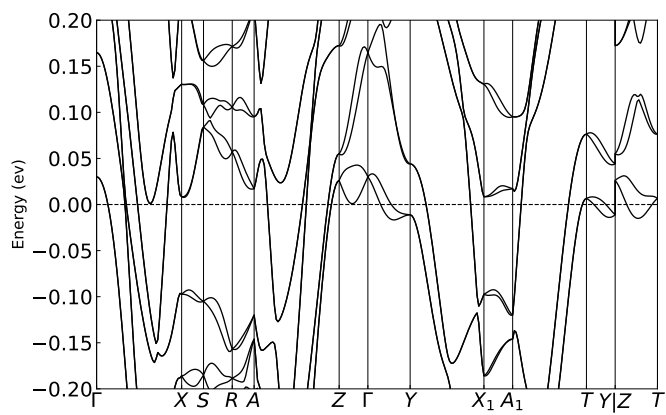

**Supplementary Figure S50.** $Re_8Sc_5Si_{12}$, SG 38, WSM

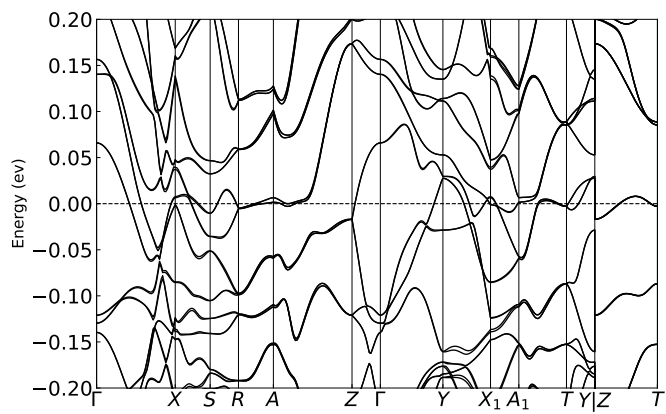

**Supplementary Figure S51.** $Sc_9Te_2$, SG 36, WSM



\end{document}